\documentclass[a4paper,11pt]{article}
\pdfoutput=1 %

\usepackage{jcappub} %
\usepackage{xcolor}
\usepackage{textcomp}
\usepackage{soul}
\usepackage{ulem}
\usepackage[T1]{fontenc} %
\usepackage{comment}
\usepackage{amsmath,amssymb}
\usepackage{subcaption}
\usepackage{xspace}

\usepackage{enumerate}
\usepackage{url}
\usepackage{bm}
\usepackage{hyperref}
\newcommand{\bs}[1]{\boldsymbol{#1}}
\newcommand{\mc}[1]{\mathcal{#1}}

\newcommand{\bv}[1]{\boldsymbol{v}_{ #1}}

\newcommand{\dif}{\mathrm{ d}}
\newcommand{\rmdel}[1]{\delta_{\mathrm{ #1} }}

\newcommand{\bshat}[1]{\hat{\boldsymbol{#1}}}

\newcommand{\relalpha}{\alpha_\mathrm{o}}
\newcommand{\relalphas}{\alpha_\mathrm{s}}
\newcommand{\kalpha}{\alpha_{\rm c}}
\newcommand{\kms}{${\rm km\, s^{-1}}$\xspace}
\newcommand\HI{$\textrm{H}\scriptstyle\mathrm{I}$}
\newcommand{\peuc}{H$\alpha$\xspace}
\newcommand{\pska}{\HI\xspace}
\newcommand{\mcov}{\boldsymbol{\mathsf {C}}}
\newcommand{\legpol}[3]{\mc{L}_{#1}(\bshat{#2}\cdot\bshat{#3})}
\newcommand{\legpolvo}[2]{\mc{L}_{#1}(\bshat{v}_o\cdot\bshat{#2})}

\newcommand{\liger}{\textsc{LIGER}\xspace}
\newcommand{\tj}[6]{ \begin{pmatrix}
  #1 & #2 & #3 \\
  #4 & #5 & #6 
\end{pmatrix}^2}

\newcommand{\snrt}{S$/$N\xspace}
\newcommand{\snrm}{{{\rm S}/{\rm N}}}
\newcommand{\RBFS}{B-FOTO\xspace}

\title{\boldmath Measuring our peculiar  velocity from spectroscopic redshift surveys}
\defcitealias{Elkhashab_2021}{Paper I}
\author[a,b,c,d,e]{ Mohamed Yousry Elkhashab,} %
\author[c,e,f,g]{Cristiano Porciani,}
\author[a,b,h]{and Daniele Bertacca}

\affiliation[a]{Dipartimento di Fisica e Astronomia Galileo Galilei, Universit\`a di Padova, 35131 Padova,Italy}
\affiliation[b]{ INFN Sezione di Padova,  I-35131 Padova, Italy.}
\affiliation[c]{Dipartimento di Fisica – Sezione di Astronomia, Universit\`a di Trieste, Via Tiepolo 11, 34131 Trieste, Italy}
\affiliation[d]{INAF - Osservatorio Astronomico di Trieste, Via Tiepolo 11, I-34131 Trieste, Italy}
\affiliation[e]{IFPU - Institute for Fundamental Physics of the Universe, via Beirut 2, 34151 Trieste, Italy}
\affiliation[f]{Argelander-Institut für Astronomie, Auf dem Hügel 71, D-53121 Bonn, Germany}
\affiliation[g]{SISSA, International School for Advanced Studies, Via Bonomea 265, 34136 Trieste, TS,
Italy}
\affiliation[h]{INAF - Osservatorio Astronomico di Padova, Vicolo dell'Osservatorio 5, I-35122 Padova, Italy.}

\emailAdd{mohamed.elkhashab@inaf.it}

\abstract{
Our peculiar velocity
imprints a dipole
on galaxy density maps derived from redshift surveys.
The dipole gives rise to an oscillatory signal in the multipole moments
of the observed power spectrum which we indicate as the finger-of-the-observer (FOTO) effect. Using a suite of large mock catalogues mimicking ongoing and future \peuc- and \pska-selected surveys,
we demonstrate that the oscillatory features can be measured
with a signal-to-noise ratio of up to 7 (depending on the sky area coverage and provided that observational systematics are kept under control on large scales).
We also show that the FOTO effect cannot be erased
by correcting the individual galaxy redshifts.
On the contrary, by leveraging the power of the redshift corrections, we propose a novel method to determine both the magnitude and the direction of our peculiar velocity.
After applying this technique to our mock catalogues,
we conclude that it can be used to either test the kinematic interpretation of the temperature dipole in the cosmic microwave background or to extract cosmological information
such as the matter density parameter and the equation of state of dark energy.
}

\begin{document}
\maketitle
\flushbottom
\section{Introduction}
The standard $\Lambda$CDM model of cosmology relies on the assumption that 
the Universe can be described as a perturbed Friedmann--Lema\^{i}tre--Robertson--Walker (FLRW) background.
The underlying 
cosmological principle implies that ideal observers comoving %
with the local mean flow of matter (which defines the `matter rest frame' at every location\footnote{This is also known as the `cosmic' or `large-scale structure' rest frame.}) measure the same large-scale properties in all directions.
The isotropy is 
broken if the observer has a peculiar velocity w.r.t. the matter frame, as expected in the actual Universe due to the presence of small-scale inhomogeneities.

Soon after the detection of the cosmic microwave background (CMB), it became evident that the observed CMB temperature should exhibit a kinematic dipole anisotropy due to the Doppler boost associated with the peculiar motion of the Earth \citep{Stewart+Sciama1967,Peebles+Wilkinson_1968,Heer-1968}.
The basic idea is that the CMB has been released in the rest frame of the matter which last scattered the radiation 13.7 Gyr ago (in fact, baryonic matter and radiation were strongly coupled until recombination). Nowadays, the CMB and matter should still share a common rest frame provided that they have been co-moving (on average, as fluids) which is implicitly assumed with the cosmological principle. 
This implies that the mean velocity of matter in the CMB rest frame should vanish as the volume over which the average is taken grows big. In other words, there should be no bulk flows of matter on Hubble-sized scales.

More than 40 years of observations have shown that 
the dipole temperature anisotropy
(obtained after correcting for 
the modulation of the signal due to the orbit of the ground- or space-based detectors around the barycenter of the Solar System)
has an amplitude of $[3361.90 \pm 0.04 \,\mathrm{(stat.)}\pm 0.36\, \mathrm{(syst.)}]\,\mu\mathrm{K}$ and points towards the direction
of Galactic coordinates $l=263^\circ\!.959\pm 0^\circ\!.003 \,\mathrm{(stat.)}\pm 0^\circ\!.017\, \mathrm{(syst.)}$ and $b=+48^\circ\!.260\pm 0^\circ\!.001 \,\mathrm{(stat.)}\pm 0^\circ\!.007\, \mathrm{(syst.)}$ \citep{COBE_CMB_1996, Hinshaw_CMB_2009, planck-dipole-18, Delouis+2021_CMB}. 
Given that the observed dipole anisotropy of the CMB is two orders of magnitude larger than what is measured for higher multipoles, it is usually assumed that the kinematic component dominates over the intrinsic one (i.e. the dipole anisotropy generated on the last-scattering surface\footnote{Note that adiabatic perturbations generated in inflationary models produce no net dipole at last scattering to leading order \cite{Turner91, Erickcek+2008}. The Doppler CMB dipole induced by super-horizon perturbations exactly cancels the intrinsic dipole they generate through the Sachs-Wolfe effect.}). This logical step is known as the kinematic interpretation of the CMB dipole,
and is usually invoked to define the CMB rest frame as the rest frame of a local observer who detects no dipole anisotropy.  Adopting the kinematic interpretation implies that the Solar System has a velocity of nearly $370 \,{\rm km\, s}^{-1}$ relative to the CMB rest frame. 
From this,
we infer that the Local Group of galaxies moves with a velocity of approximately $630 \,{\rm km\, s}^{-1}$ 
relative to the CMB rest frame, in the direction $(l,b)\sim (276^\circ, 30^\circ)$. 
Four decades of studies have been dedicated to check whether this motion could have been gravitationally
induced 
by the surrounding galaxy distribution measured with redshift surveys. The state of the art is that the peculiar velocity generated within $200 \,h^{-1}$ Mpc points only $\sim10^\circ$ away from the direction of
the LG velocity
inferred using the CMB dipole and a residual bulk-flow contribution of $\sim 160$ km s$^{-1}$ should arise from sources lying beyond this volume \cite[][and references therein]{Carrick+15}. 
Finally, it is worth mentioning that
the relativistic aberration and the Doppler effect associated with
the motion of the observer w.r.t. the CMB frame also distorts the pattern
of CMB anisotropies \citep{challinor_2002_l(l+1)}.
These signatures, first detected with the Planck satellite, 
are consistent with the kinematic interpretation of the dipole
\citep{PlanckXXVII} 
and have been used to set an upper limit on the amplitude of the intrinsic dipole \citep{Ferreira_2021l(l+1)_thoery,Ferreira_2021_cmbl(l+1)}. 

If the 
kinematic interpretation of the CMB dipole is correct,
then   
the peculiar velocity of the Solar System should also induce
a kinematic dipole in the observed distribution of galaxies on the sky
via aberration and Doppler effects, as first proposed by  \citet{Ellis_Baldwin_1984}.
The expected amplitude of the signal is of one part in a thousand
w.r.t. the monopole term.
Key to detecting such a small dipole is a galaxy survey with homogeneous selection criteria (to prevent spurious signals) 
and high-density of tracers (to beat shot noise), covering a wide solid angle (to unambiguously measure a dipole anisotropy), and deep enough to overcome intrinsic anisotropies generated by the local large-scale structure which can be as high as the target signal
\citep[e.g.][]{Itoh+2010, Gibelyou_2012_dipole_Multi_survey, Nadolny_2021}. 
Large catalogues of radio-continuum-selected galaxies 
with median redshift $z\sim 1$ (corresponding to a comoving distance of $\sim 3.3$ Gpc in the $\Lambda$CDM model) satisfy all these criteria and are thus suitable samples 
\citep{yoon-huterer-18, Baleisis_1998_radio_dipole}.
As soon as such catalogues including millions of sources became available, it became apparent that results could depend on the method of analysis and on the procedure adopted for handling systematic errors (selection effects, tiling, sky cuts, bias of estimators, limitations of the theoretical models, etc.). For instance, a dipole anisotropy consistent with the CMB one \citep[][see also \citealp{cheng2024radiosourcedipolenvss}]{Blake_2002_radio_NVSS_dipole} or a signal discrepant both in amplitude and direction
\citep{Gibelyou_2012_dipole_Multi_survey} have been extracted from the same catalogue.
More recent studies consistently measure
a dipole anisotropy compatible with the CMB direction but with a substantially larger amplitude 
\citep{Tiwari_2015_NVSS_DIPOLE,Siewert_2021_dipole_radio_measurment}.  
Similar results have been obtained
for quasars selected in the mid-infrared band from the recent data release \citep[CatWISE2020,][]{Eisenhardt_2020} of  the Wide-field Infrared Survey Explorer \citep[WISE,][]{Eisenhardt_2020}. In this case, the discrepancy with the CMB dipole reaches a high confidence level, corresponding to $\sim 5\sigma$ in Gaussian statistics 
\citep{Singal_2021_dipole_quasers,Secrest_2021_dipole_quasers,Secrest_2022_dipole_quasers,Dam+2023}. The discrepancy in the amplitude of the observed (projected) dipole could be partly mitigated by taking into account several redshift-dependent terms which describe
the evolution in the number density and clustering of the tracers 
\citep{Dalang_2022_MAG_BIAS, Guandalin+2022}, {however, \citet{vonHausegger:2024jan}  asserts  that the kinematic dipole measurement is robust against the redshift evolution of these terms.} 
To complicate the picture,
a recent analysis of two new radio-continuum surveys with a combined sky coverage of 90\% finds a dipole signal in good agreement with the CMB \cite{Darling_2022_galaxy_dipole}.

From a theoretical perspective, it is possible to build scenarios in which the galaxy and CMB dipoles differ.
For instance, there could be an 
actual large-scale gradient in the galaxy number density on scales larger than our Hubble patch 
generated in multi-field inflationary scenarios
\cite[e.g.][but see also \citealp{Hirata_2009}]{Erickcek+2008}
or 
the CMB dipole might not be (entirely) kinematic but partially generated by primordial isocurvature perturbations on super-Hubble scales \citep[e.g.][]{Langlois+Piran96, Turner91,Guillem+2022,ebrahimian2023realisticdipolecosmologydipole}.
Yet other possibilities are that the matter and CMB frames do not coincide any longer today \cite{Matzner_1980} or that the CMB dipole appears because dynamic dark energy and matter are not co-moving  \cite[e.g.][]{Maroto-2006, Harko-Lobo_2013}.
In all these cases, a large-scale bulk flow of galaxies should be present in the CMB frame.
Observational evidence for such bulk flow has been claimed based on the kinetic Sunyaev-Zel’dovich effect of galaxy clusters \citep[][see however \citealp{Osborne+2011}]{Kashlinsky+08, Kashlinsky+2010}  %
and the scaling relations of galaxy clusters selected in the X-ray band \cite{Migkas+2020}.

Since a convincing detection of inconsistent CMB and galaxy dipoles would provide smoking-gun evidence for new physics,
in this paper, we introduce a new method to determine the peculiar velocity of the Solar System w.r.t. the matter frame  based on galaxy spectroscopic redshift surveys. 
At variance with previous studies that focused 
on measuring 
the dipole moment of the galaxy distribution projected
on the celestial sphere, we utilise
the simplest observable that quantifies galaxy clustering in three dimensions: the monopole moment of the galaxy power spectrum, $P_0(k)$. This is obtained by expanding the anisotropic power spectrum (which depends on the relative direction of the wavevector and the line of sight)
in Legendre polynomials. 
In an earlier study \citep[][hereafter \citetalias{Elkhashab_2021}]{Elkhashab_2021},
we demonstrated both analytically and numerically 
that the peculiar velocity of the observer adds an oscillatory feature to $P_0(k)$ at very large scales. %
The signal is proportional to the squared magnitude of the velocity of the observer but also depends on the expansion history of the Universe and on the selected galaxy population through several bias parameters. 
Moreover, it is possible to build an estimator for an anisotropic
monopole moment of the power spectrum which retains information on
the direction of the observer velocity.
Drawing an analogy with the so-called `fingers of God'
generated by the peculiar velocities of distant galaxies in clusters, we  
dubbed these characteristic signatures the `Finger Of The Observer' (FOTO) effect. In order to fully exploit the potential of this observable, we devise
a method in which apply different Doppler redshift transformations to  individual redshifts of the observed catalogue, creating multiple galaxy distributions with tuned dipole (and FOTO) signals. We combine these distributions to extract the magnitude and direction of the observer velocity. 
Precise Redshift measurements are necessary for the proposed method. Therefore, our method is well suited for applications to next-generation spectroscopic galaxy redshift surveys that will cover large fractions of the sky out to high redshift and sample high number densities of tracers.

We demonstrate the potential of our method using a  set of mock surveys generated with the \liger (LIght cones with GEneral Relativity) method \citep[][]{MIKO_2017,Elkhashab_2021,euclidcollaboration2024euclidpreparationimpactrelativistic}. The \liger method is a numerical framework designed to construct galaxy catalogues, incorporating relativistic redshift-space distortions (RSD) up to linear order in cosmological perturbations. It starts from Newtonian N-body simulations and accurately models the impact of relativistic RSD on galaxy clustering at large scales. The modelling of relativistic RSD is necessary for our analysis as the kinematic dipole -- and consequently the FOTO effect -- is not determined solely by Doppler and aberration effects but also by flux perturbations. The observed overdensity is thus affected by the expansion history of the Universe, the evolution of the source number density (evolution bias) and the magnification bias that arises from the magnitude limit of a galaxy redshift survey \citep[e.g.][ see also \citealp{Bertacca:2019} for the impact of these terms on the two-point correlation function]{Maartens+2018,Bahr-Kalus:2021jvu,  Elkhashab_2021}. %

We generate two distinct sets of galactic mock catalogues that differ in their covered redshift ranges, as well as in their evolution and magnification biases. The first set of mock catalogues mimics the expected distribution of galaxies that the \textit{Euclid} mission \citep{euclidcollaboration2024euclidiovervieweuclid} will measure, while the second set is based on the 21cm neutral hydrogen galaxy population targeted by the extension of the Square Kilometre Array Observatory %
\citep[\texttt{SKAO2},][]{Yahya_2015}. In our analysis, we assume that these catalogues cover the entire sky, although we demonstrate that the FOTO signal remains measurable even for partial sky coverage. Additionally, we neglect observational systematics, such as those arising from stellar density or galactic extinction, which can be addressed using specialised estimators \citep{Burden_2017, Paviot_2022}.
Using these mocks, we validate our analytical prediction for the FOTO signal,  analyse the impact of the artificial Doppler shifts on the observed power spectrum monopole and finally, extract the velocity of the observer and cosmological parameters.

This paper is structured as follows. We begin by providing a brief overview of the kinematic dipole and derive the FOTO effect for all power spectrum multipoles in Sec.~\ref{sec:FOTO}. In Sec.\ \ref{sec:MOCKS_DESCRIP}, we describe the methodology employed to produce the mock catalogues. Next, in Sec.~\ref{sec:MOCKS_MEASURMENTS_ALL}, we measure the FOTO signal from the mock catalogues and compare it to our analytical predictions. %
In Sec.~\ref{Sec:RED_TRANSF},  we study the effect of different redshift corrections  on the power spectrum monopole.  We then leverage these corrections to extract our peculiar velocity and set constraints on various cosmological parameters. %
 Finally, we summarize our findings in Sec.~\ref{Sec:conc}.

\section{Redshift-space distortions and
the finger-of-the-observer effect}
\label{sec:FOTO}
\subsection{Cosmological redshift and peculiar velocities}
Let us consider a FLRW universe with cosmic time $t$, expansion factor $a$, and Hubble parameter $H$.
A source comoving with the cosmic expansion radiates light at wavelength $\lambda_\mathrm{e}$.
It is standard textbook material that this light is
received at a different wavelength $\lambda_\mathrm{o}$ by a comoving observer. Namely, $\lambda_\mathrm{o}=[a(t_\mathrm{o})/a(t_\mathrm{e})]\,\lambda_\mathrm{e}$
where $t_\mathrm{e}$ and $t_\mathrm{o}$ denote the coordinate time of emission and reception, respectively. 
The ratio 
$a(t_\mathrm{o})/a(t_\mathrm{e})=1+z_\mathrm{cos}$
defines the cosmological redshift $z_\mathrm{cos}$.

If the source has a (proper) peculiar velocity $\bv{\mathrm{e}}$ at emission,
then the observer measures a different wavelength for the light signals. In this case, 
the observed redshift $z_\mathrm{obs}$ is given by the expression
\begin{equation}
1+z_\mathrm{obs}={\lambda_\mathrm{o}\over \lambda_\mathrm{e}}=
{\lambda_\mathrm{o}\over \lambda_\mathrm{e}'}\,{\lambda_\mathrm{e}'\over \lambda_\mathrm{e}}=
\left(1+z_\mathrm{cos}\right)\,\left(1+z_\mathrm{pec, e}\right)\;,
\end{equation}
where $\lambda_\mathrm{e}'$ denotes the wavelength measured by a comoving observer at the event of emission, which can be computed from
$\lambda_\mathrm{e}$ using a simple Lorentz transformation.
This gives
\begin{align}
    z_\mathrm{pec,e} =  \frac{1+(\bv{\mathrm{e}}\cdot\hat{\bs{r}})/{c}}{\sqrt{1-\left(v_\mathrm{e}/c\right)^2}}-1\simeq \frac{\bv{\mathrm{e}}\cdot \hat{\bs{r}}}{c}=\frac{v_{\parallel , \mathrm{e}}}{c}
    \,,
\end{align} 
where $\hat{\bs{r}}$ indicates the direction along the line of sight (from the observer towards the source), $c$ denotes the speed of light in vacuum, and the second equality holds in the limit $v_\mathrm{e}\ll c$.

The peculiar velocity of the observer similarly alters the observed redshift,
\begin{equation}
1+z_\mathrm{obs}={\lambda_\mathrm{o}/\lambda_\mathrm{e}}=
(\lambda_\mathrm{o}'/\lambda_\mathrm{e})\,(\lambda_\mathrm{o}/\lambda_\mathrm{o}')=
(1+z_\mathrm{cos})\,(1+z_\mathrm{pec, o})\,
\end{equation}
where $\lambda_\mathrm{o}'$ denotes the wavelength measured by a comoving observer at the event of detection. In this case,
\begin{align}
    z_\mathrm{pec,o} = \frac{\sqrt{1-\left(v_\mathrm{o}/c\right)^2}}
    {1+(\bv{\mathrm{o}}\cdot\hat{\bs{r}})/c}-1
    \simeq -\frac{\bv{\mathrm{o}}\cdot \hat{\bs{r}}}{c}=-\frac{v_{\parallel, \mathrm{o}}}{c}
    \,
\end{align} 
(the negative sign here ensures that
motion towards the source gives rise to blueshift).
The expression above is written in terms
of the line of sight measured by the 
observer with peculiar velocity.\footnote{Due to relativistic aberration, the comoving  observer  will measure a different
line of sight
\begin{align}
\hat{\bs{r}}'=\left(\frac{\hat{\bs{r}}\cdot \hat{\boldsymbol{v}}_\mathrm{o}+v_\mathrm{o}/c}{1+\hat{\bs{r}}\cdot \boldsymbol{v}_\mathrm{o}/c}\right)\,\hat{\boldsymbol{v}}_\mathrm{o} 
+\sqrt{1-(v_\mathrm{o}/c)^2}\,
\frac{\hat{\bs{r}}-\left(\hat{\bs{r}}\cdot \hat{\boldsymbol{v}}_\mathrm{o}\right)\,\hat{\boldsymbol{v}}_\mathrm{o}}{1+\hat{\bs{r}}\cdot \boldsymbol{v}_\mathrm{o}/c}
\end{align}
such that
$\hat{\bs{r}}'\cdot \hat{\boldsymbol{v}}_\mathrm{o}=\left(\hat{\bs{r}}\cdot \hat{\boldsymbol{v}}_\mathrm{o}+v_\mathrm{o}/c
\right)/(1+\hat{\bs{r}}\cdot \boldsymbol{v}_\mathrm{o}/c )$.
To linear order in $v_\mathrm{o}/c$, these expressions reduce to $\hat{\bs{r}}'\simeq 
\left(1-\hat{\bs{r}}\cdot \boldsymbol{v}_\mathrm{o}/c \right)\,\hat{\bs{r}}+\boldsymbol{v}_\mathrm{o}/c$
and $\hat{\bs{r}}'\cdot \hat{\boldsymbol{v}}_\mathrm{o}\simeq
\hat{\bs{r}}\cdot \hat{\boldsymbol{v}}_\mathrm{o}+[1-(\hat{\bs{r}}\cdot \hat{\boldsymbol{v}}_\mathrm{o})^2]\,v_\mathrm{o}/c$. 
}

Putting everything together, we have
\begin{equation}
1+z_\mathrm{obs}=(1+z_\mathrm{cos})\,(1+z_\mathrm{pec, e})\,(1+z_\mathrm{pec, o})\;.
\label{eq:obsred}
\end{equation}
It is worth stressing that, so far, we adopted a kinematic approach to modelling the impact of peculiar velocities. 
In the standard model, peculiar velocities are generated via gravitational instability, and large-scale cosmological perturbations introduce additional corrections to $z_\mathrm{obs}$ (see Section~\ref{sec:releff}).

\subsection{Radial redshift-space distortions from peculiar velocities}
\label{sec:FOTO-RSD}
It is common practice to use the observed redshift as a distance indicator in order to estimate galaxy positions on our past light cone. The conversion from redshifts to comoving distances is done using an unperturbed FLRW model. Since $z_\mathrm{obs}= z_\mathrm{cos}+\Delta z$, with \begin{equation}
\Delta z\simeq (1+z_\mathrm{cos})\;{(v_{\parallel, \mathrm{e}}-v_{\parallel, \mathrm{o}})\over c}\,,
\end{equation} the resulting galaxy distribution is not a faithful reproduction of the actual one and is affected by the so-called `redshift-space distortions'. 
Neglecting aberration,
the reconstructed
position of a galaxy with peculiar velocity $\bv{}$
is separated from the real one by
a shift along the line of sight of comoving size

\begin{equation}    
\label{eq:deltar}
\begin{split}
\Delta r=&\int_{z_\mathrm{cos}}^{z_\mathrm{obs}} \frac{c}{H(z')}\,\mathrm{d}z'\simeq 
\frac{c}{H(z_\mathrm{cos})}\,\Delta z=
\frac{1+z_\mathrm{cos}}{H(z_\mathrm{cos})}\,(v_{\parallel, \mathrm{e}}-v_{\parallel, \mathrm{o}})\\=&\frac{1+z_\mathrm{obs}}{H(z_\mathrm{obs})}\,(v_{\parallel, \mathrm{e}}-v_{\parallel, \mathrm{o}})=
\frac{v_{\parallel, \mathrm{e}}-v_{\parallel, \mathrm{o}}}{aH}\;,
\end{split}
\end{equation}
where the last expression defines the compact notation we use in the remainder of the paper.

To first approximation, we can neglect velocity dispersion and describe large-scale peculiar motions in terms of a continuous and smooth velocity field $\boldsymbol{v}$. Starting from this assumption,
in a seminal paper \citep{Kaiser87}, Nick Kaiser showed that the overdensity of galaxies
in redshift space (i.e. obtained using the observed redshift as a distance indicator), $\delta_{\rm obs}$, and its counterpart in real space (i.e. the overdensity that would be obtained if we could use  $z_\mathrm{cos}$ as a distance indicator), $\delta$, are related as follows
\textcolor{black}{
\begin{equation}
    \delta_{\rm obs} = \delta - \frac{1}{a\,H}\frac{\partial v_\parallel}{\partial r} -    \frac{\alpha}{r} \,\frac{v_\parallel-v_{\parallel,\mathrm o}}{aH}\,,
    \label{eq:kaiser}
\end{equation}
where $r$ denotes the comoving distance along the line of sight,\footnote{By construction, the redshift space coordinates match the observed redshift $z_{\rm obs} = 1/a-1$ to the comoving distance $r$ via $r = \int_0^{z_{\rm obs}} [c/H(z)]\,\dif z$.  That redshift-distance relation also allows us to denote composite functions, such as $H[z_{\rm obs}(r)]$, simply as $H(r)$.
 } {$v_{\parallel}$ represents the peculiar velocity field, }  
all functions except $v_{\parallel,\mathrm{o}}$} are evaluated at the same position, and  \citep{Kaiser87,Hamilton-Culhane}
\begin{align}
    \label{eq:alpha_class}
    \alpha =\frac{\dif \ln (r^2\,\bar{n})}{\dif \ln r}
    =    2+\frac{\dif \ln \bar{n}}{\dif \ln r}\,,
\end{align}
    with $\bar{n}$ the (comoving) underlying mean number density of galaxies.
 Eq.~(\ref{eq:kaiser}) follows from
 imposing that the number of galaxies is conserved as real space is mapped to redshift space and is obtained assuming  the velocity and density perturbations
grow according to linear theory, the radial derivative of  $(a\,H)^{-1}$ is vanishing and neglecting the temporal evolution of $v_\parallel$. 
In Eq.~(\ref{eq:kaiser}),  the term containing the derivative of $v_\parallel$ quantifies the radial fractional compression/dilation of an infinitesimal volume element since that the velocity changes along the line of sight.
At the same time, the term proportional to $\alpha$ reflects the fractional change
in the expected number of galaxies when the volume element is radially displaced along the line of sight (with two contributions: a geometrical factor and the change in $\bar{n}$).

The standard Kaiser amplification factor for the power spectrum
follows from Eq.~(\ref{eq:kaiser}) by assuming that: (i) the survey is deep enough so that the term proportional to $\alpha$ can be neglected
(as it should dominate only when $r \sim k$, where $k$ is the magnitude of the wavevector, $\bs{k}$, in the plane-wave expansion of the galaxy overdensity.); 
(ii)
all lines of sight point towards the same direction
(distant-observer or plane-parallel approximation). 

Eq.~(\ref{eq:kaiser}) was originally derived to model low-redshift surveys and, as already mentioned, ignores the lightcone evolution of $\Delta r$.  
Before moving on, it is important to understand the consequences of relaxing this assumption.
It follows from Eq.~(\ref{eq:deltar}) that
the Jacobian determinant of the coordinate transformation from real to redshift-space contains the term 
\begin{equation}    
{\dif \Delta r\over\dif r}=
{\dif [(v_{\parallel}-v_{\parallel,\mathrm{o}})/(aH)] \over\dif r}=
(aH)^{-1}{\dif v_\parallel\over\dif r}
+(v_\parallel-v_{\parallel,\mathrm{o}})\,{\dif (aH)^{-1}\over \dif r}\,.
\end{equation}
Ignoring the temporal evolution of $v_{\parallel}$ and the radial dependence of the $aH$ in the previous expression, we can see that ${\dif \Delta r/\dif r} = (aH)^{-1} {\partial v_{\parallel}/\partial r}$, i.e. the second term on the rhs of Eq.~\eqref{eq:kaiser}.  
Considering the previously neglected derivative
gives two additional contributions to 
 $\delta_\mathrm{obs}$, namely the temporal derivative of $v_{\parallel}$ and 
 $-[1-{\dif}\ln H/{\dif} \ln (1+z)]\,(v_\parallel-v_{\parallel,\mathrm{o}})/c$. 
 Thus, Eq.~(\ref{eq:kaiser})
is still valid as long as we replace  the partial derivative of the source peculiar velocity field with the total derivative and 
$\alpha$ with
 \begin{equation}
    \alpha_\mathrm{c} =\alpha +
    \left[1-\frac{{\dif} \ln H}{{\dif}\ln (1+z)} \right]\,\frac{r\, H}{c\, (1+z)}
    \;.
 \end{equation}

\subsubsection{Flux-limited surveys: evolution and magnification bias}
\label{Sec:Mag_evo_bias}
In this paper, we focus on flux-limited surveys and,
for later convenience, we prefer to express $\alpha$ in terms of other functions which are more commonly used in the recent literature. 
Let us denote by $n(L_{\mathrm{min}},z)$ the comoving number density of galaxies with luminosity $L>L_\mathrm{min}$ (in some waveband) 
 at redshift $z$.
A flux-limited survey only selects galaxies with $L>L_\mathrm{lim}(z)$
at every redshift, i.e. $\bar{n}=n(L_\mathrm{lim}(z),z)$.
We define the evolution bias $\mc{E}$ and the magnification bias $\mc{Q}$ of the selected galaxy population as 
\begin{align}
    \mc{E}(z)&=-\left.\frac{\partial \ln n(L_\textrm{min},z)}{\partial \ln (1+z)}\right|_{L_\mathrm{min}=L_\mathrm{lim}(z)} \;,
\end{align}   
 \begin{align}   
    \mc{Q}(z)&=-\left.\frac{\partial \ln n(L_\textrm{min},z)}{\partial \ln L_\textrm{min}}\right|_{L_\mathrm{min}=L_\mathrm{lim}(z)}  \;.
    \label{eq:EANDQ}
\end{align}
In words, $\mc{E}$ quantifies
the logarithmic change in the comoving number density of selected
galaxies due to the redshift evolution of
the amplitude and shape of the luminosity
function at fixed $L_\mathrm{min}$. 
 Instead,
$\mc{Q}$ gives the response of $\ln \bar{n}$
to changes in the limiting luminosity at fixed $z$. 
Note that, if the galaxy luminosity function scales as $L^{-\gamma}$ around $L_\mathrm{lim}$ and $\gamma>1$ (which is usually the case), then $Q\simeq \gamma-1>0$.

The function $\alpha$ can be straightforwardly expressed
in terms of $\mc{E}$ and $\mc{Q}$ \citep[][]{bertacca_2nd_2,Elkhashab_2021}. In fact, 
\begin{align}
\frac{\dif \ln \bar{n}}{\dif \ln r}&=
-\mc{E}\,\frac{\dif \ln (1+z)}{\dif \ln r}-\mc{Q}\,\frac{\dif \ln L_\mathrm{lim}}{\dif \ln r}\;.
\end{align}
In this paper, we consider
surveys that select galaxies based
on the flux of an emission line.
In this case,
the limiting luminosity is linked to
the flux limit of the survey by
$L_\mathrm{lim}(z)=4\pi\, F_\mathrm{lim}\, \bar{d}_\mathrm{L}^2(z)$,\footnote{The observed flux $F$ is measured as follows. First,
the observed spectrum is rescaled to the source rest frame
(multiplying the frequency and dividing the spectrum by the factor $1+z_\mathrm{obs}$).
Second, a line profile is fit to the rescaled data. Third, the line flux is computed by integrating the fitted profile over
the rest-frame frequency. Eventually, possible absorption corrections are applied.}
where $\bar{d}_\mathrm{L}$ denotes the luminosity distance in the unperturbed FLRW universe (redshift space).
Further assuming a flat model universe with luminosity distance $\bar{d}_L=(1+z)\,r$, we get (see also Appendix C in \citetalias{Elkhashab_2021})
\begin{align}
\frac{\dif \ln \bar{n}}{\dif \ln r}&=-\mc{E}\,\frac{\dif \ln (1+z)}{\dif \ln r}-
\mc{Q}\,\frac{\dif \ln \bar{d}_L^2}{\dif \ln r}
=-\mc{E}\,\frac{rH}{c(1+z)}-2\mc{Q}\left[1+\frac{\dif \ln (1+z)}{\dif \ln r}\right]
\nonumber\\
&=
-\mc{E}\,\frac{rH}{c(1+z)}-
2\mc{Q}\,\left[1+ \frac{rH}{c(1+z)}\right]
\end{align}
or, equivalently, 
\begin{align}
    \label{eq:kalpha_mag_def}
       {\kalpha} =2\,(1-\mc{Q}) +\left[  1-2\mc{Q}-\mc{E}- {\frac{\dif \ln H}{\dif \ln (1+z)}}\right]\,\frac{r H}{c\,(1+z)}\,,
\end{align}
which reduces to ${\kalpha} \simeq 2\,(1-\mc{Q})$ for $r\ll c/H$
(i.e., $z\ll 1$, the regime in which Eq.~(\ref{eq:kaiser}) was derived). In this limit, 
where $v_\parallel-v_{\parallel,\mathrm{o}}>0$ so that the shift from real to redshift space points away from the observer,
$2\mc{Q}\,(v_\parallel-v_{\parallel,\mathrm{o}})/(aHr)$ gives a positive contribution to 
$\delta_\mathrm{obs}$ reflecting the fact that
$\bar{n}$ is lower at the redshift-space location
than at the corresponding real-space one (as only intrinsically brighter galaxies satisfy the selection criteria of the survey).

For a typical survey,
$\mc{E}$ and $\mc{Q}$ are expected to be of order unity (see e.g. Section~\ref{sec.mock_specifics}) while, sufficiently far away from the observer,
$(v_\parallel-v_{\parallel,\mathrm{o}})/aH$ is much smaller than $r$ (or the Hubble radius). Therefore,
the contribution to the galaxy overdensity proportional to $\kalpha/r$ should be
small compared to the
other terms in the right-hand side of Eq.~(\ref{eq:kaiser}). This is why it is often neglected when modelling clustering
in the distant-observer approximation. However, this term could give important clustering contributions in 
wide-angle surveys where spatial correlations between
the other terms are small at separations with large components transverse to the line of sight \citep[e.g.][]{szalay+98, Bertacca:2012tp,Raccanelli:2016avd,MIKO_2017}.

\subsubsection{General relativistic effects}
\label{sec:releff}
According to the theory of general relativity, the propagation of radiation from distant galaxies
to our telescopes is affected by
intervening inhomogeneities on top of the peculiar motions of sources and observer. Therefore,
a fully relativistic treatment of redshift-space distortions needs to account for additional effects with respect to Eq.~(\ref{eq:kaiser})
by considering (among others)
the bending of light rays due to intervening density fluctuations (gravitational-lensing
deflection and magnification), 
gravitational redshift, and the difference between the rest frames of the source and the observer.

These phenomena (sometimes collectively called relativistic or lightcone projection effects) not only modify the relation between $z_\mathrm{obs}$ and $z_\mathrm{cos}$ given in Eq.~(\ref{eq:obsred}) through the integrated and non-integrated Sachs-Wolfe effects, but also perturb the luminosity and angular-diameter distances to the sources. 
To first order in the perturbations, the magnification 
$1+\mu=(\bar{d}_\mathrm{L}/d_\mathrm{L})^2$
(where $d_\mathrm{L}$ denotes the luminosity distance to a source and both numerator and denominator are evaluated at $z_\mathrm{obs}$), receives the following contributions from the peculiar velocities
\cite[e.g.][]{Sugiura+99, Pyne+2004, Hui+Greene06}
\begin{align}
\label{eq:magnif}
\mu_{v}&= 2\left[1-\frac{c(1+z)}{Hr} \right]\,\Delta \ln z + 2 \frac{v_{\parallel,\mathrm{o}}}{c} =
2\left[1-\frac{c(1+z)}{Hr} \right]\,
\frac{v_{\parallel, \mathrm{e}}-v_{\parallel, \mathrm{o}}}{c}+ 2 \frac{v_{\parallel,\mathrm{o}}}{c} \nonumber\\
&=2\frac{v_{\parallel, \mathrm{e}}}{c}-
2\,\frac{(1+z)}{Hr}\,(v_{\parallel, \mathrm{e}}-v_{\parallel, \mathrm{o}})
\;,
\end{align}
where $\mu_{v}\subset \mu$. %
The first term in the first line follows from the redshift perturbation while the second comes from special-relativistic aberration.
Note that the latter 
cancels out one of the
contributions to $\mu$
coming from the redshift correction
and proportional to $v_{\parallel,\mathrm{o}}$.
It is instructive to relate the magnification perturbation to the radial shift $\Delta r$ given in 
equation~(\ref{eq:deltar}). 
The last term in Eq.~(\ref{eq:magnif}) coincides with the `geometrical magnification' $-2\Delta r/r$ generated by the shift \citep{Kaiser+2015}. At redshifts $z\ll 1$, when $H\simeq H_0$ and $r\simeq cz/H_0$, the first term can be neglected and  $\mu_{v}$ reduces to $-2 (v_{\parallel, \mathrm{e}}-v_{\parallel, \mathrm{o}})/(cz)$.
At finite redshifts, however, the correction $2v_{\parallel,\mathrm{e}}/c$ in Eq.~(\ref{eq:magnif}) becomes appreciable.

The perturbation in the luminosity distance directly impacts the observed density contrast. Since
the limiting luminosity of a flux-limited survey varies as
$L_\mathrm{lim}(z)=4\pi\, F_\mathrm{lim}\, d_\mathrm{L}^2=
4\pi\, F_\mathrm{lim}\, (1+\mu)^{-1}\,\bar{d}_\mathrm{L}^2
$, the additional contribution 
$-\mc{Q}\,[(1+\mu)^{-1}-1]\simeq Q\,\mu$
must be added to
$\delta_\mathrm{obs}$ as the number
of galaxies that satisfy the selection
criteria of the survey are influenced by magnification \citep{Broadhurst_1995}.
Eq. ~(\ref{eq:magnif}) ensures that the contribution of the peculiar velocities, in particular the term sourced by the observer, is accounted for in the magnification correction. 

Taking into account all relativistic corrections to linear order, 
one obtains
\citep{Yoo:2009au, Bonvin-Durrer2011, Challinor:2011bk, Jeong:2011as, Bertacca_2015_mag}  %
\begin{equation}
\label{eq:delta_velocities_rel_org}
    \delta_{\rm obs} = \delta- \frac{1}{aH}\,\frac{\partial v_{\parallel}}{\partial r} -     \frac{\relalphas}{r}\,\frac{v_\parallel}{aH} +\frac{\relalpha}{r}\,\frac{v_{\parallel,\mathrm{o}}}{aH}+\dots
\end{equation}
where
 \begin{align}
     \relalphas&= 
      2\,(1-\mc{Q})  +\left[ 1+2\mc{Q}-\mc{E}-  \frac{\dif \ln H}{\dif \ln (1+z)}\right]\frac{r \, H}{c\, (1+z)}\,,%
     \label{eq:alphaGRsource}
\end{align}
\begin{align}\label{eq:alphaGRobs_full}
     \relalpha&=   2\,(1-\mc{Q}) +\left[ 3-\mc{E}- \frac{\dif \ln H}{\dif \ln (1+z)}\right]\frac{r\, H}{c\, (1+z)}\,,
\end{align}
and the ellipses indicate several other terms that do not need to be written down explicitly for understanding this work.\footnote{In compact notation, they can be expressed as the sum of $2\,(\mc{Q}-1)\,\kappa$ (where $\kappa$ denotes the gravitational-lensing convergence) with a handful of terms proportional to the Bardeen gravitational potentials.}  Note that $\relalphas\simeq \relalpha \simeq {\kalpha} \simeq 2\,(1-\mc{Q})$ when $r\ll c/H$ but these coefficients are all different otherwise,
reflecting our considerations on the magnification above.

\subsection{The finger-of-the-observer (FOTO)  effect}
In order to emphasize the impact of $\boldsymbol{v}_{\mathrm{o}}$ 
we rewrite Eq.~(\ref{eq:delta_velocities_rel_org}) as
\begin{equation}
\label{eq:delta_velocities_rel_obs}
    \delta_{\mathrm{obs}} = \delta_{\mathrm {com}} +\relalpha\,\frac{\boldsymbol{v}_\mathrm{o}  \cdot \hat{\bs{r}}}{aH r}\,,
\end{equation}
where $\delta_{\mathrm {com}}$ collects all the other terms (including those that were represented by the ellipses) and represents the galaxy overdensity in redshift-space measured by an observer who comoves
with the Hubble expansion.
Eq.~(\ref{eq:delta_velocities_rel_obs}) shows that
an observer with a peculiar velocity measures an additional deterministic contribution to $\delta_\mathrm{obs}$,
\begin{equation}
\label{Eq:DIP_DELTA}
\delta_{\mathrm {dip}}=    \relalpha\,\frac{\boldsymbol{v}_\mathrm{o}  \cdot \hat{\bs{r}}}{aHr}\,,
\end{equation}
which imprints a characteristic dipole pattern on the sky 
and displays a variable amplitude as a function of radial distance
reflecting the properties of the selected galaxy population and the
expansion history of the Universe.
The dipole term proportional to $\boldsymbol{v}_\mathrm{o}$ generates
contributions to 
the summary statistics of $\delta_{\mathrm{obs}}$ which then do not 
coincide with those of $\delta_{\mathrm {com}}$.
This is what we refer to as the ``finger-of-the-observer'' {(FOTO)} effect. An intriguing question is whether this effect could be used to measure
$\boldsymbol{v}_\mathrm{o}$.

In \citetalias{Elkhashab_2021}, 
we investigated how $\delta_{\mathrm {dip}}$ alters the
monopole moment of the galaxy power spectrum, $P_0(k)$, measured with traditional estimators. We found that \citep[see also][]{Bahr-Kalus:2021jvu}, in an ideal full-sky survey, 
\begin{align}
  P_{0,\mathrm{obs}} (k) = P_{0,{\mathrm{com}}} (k) +P_{0,\mathrm{dip}}(k)\;,
\end{align}
where the correction due to the observer's peculiar velocity is
\begin{align}
\label{eq:fing}
  P_{0,\mathrm{dip}}(k)
  =  \frac{16\pi^2}{3} \,\frac{v_{\mathrm{o}}^2}{H_0^2}\,\frac{I_1^2(k)}{\int \bar{n}^2\,\dif^3 r}\;,
\end{align}
with
{\begin{align}
I_\ell(k)&=\int
\frac{r\,\bar{n}\,\relalpha}
{a\, H/H_0 }\,j_\ell(k r)\,\dif r\;,\label{eq:Idef}
\end{align}
and $j_\ell$ denotes the spherical Bessel function of order $\ell$. }
If one considers a redshift bin with sharp boundaries,
then
$\boldsymbol{v}_\mathrm{o}$ leaves an imprint in $P_{0,{\mathrm{com}}} (k)$ in terms of an oscillatory feature which becomes prominent at small wavenumbers. 
For a narrow shell of width $\delta r \ll r$ within which all functions can be treated as constants, 
\begin{equation}
\label{eq:thin_limit}
P_{0,\mathrm{dip}}(k)\simeq  \frac{4\pi}{3}\,\relalpha^2 \,\left[\left(\frac{v_{\mathrm{o}}}{aH}\right)^2\, \delta r \right]\,j_1^2(k r)\;,
\end{equation}
which is independent of the number density of galaxies.
Here, $v_{\mathrm{o}}/(aH)$ gives the comoving separation at which the Hubble flow generates a velocity of amplitude $v_{\mathrm{o}}$.
Assuming the CMB dipole is of kinematic origin gives
$v_{\mathrm{o}}/(aH)\simeq 4\,h^{-1}$ Mpc for $0.5 \lesssim z \lesssim 2$. 
In \citetalias{Elkhashab_2021}, we have shown that this characteristic feature can be detected with high confidence by using galaxy redshift surveys of the current generation. In this paper,
we investigate what information can be extracted from the measurements. 

\subsubsection{Multipoles of the power spectrum}
\label{sec:FOTO_MULTI}

 We extend the analytical treatment of \citetalias{Elkhashab_2021} to all the multipoles of the power spectrum. Our starting point is the Yamamoto  estimator in the local plane parallel approximation \citep{Yamamoto_2006,bianchi_measuring_2015,scoccimarro_fast_2015}  
\begin{equation}
    \label{eq:YAMMATO_ESTIMATOR}
    \begin{split}
    \hat{P}_{\ell,{\rm obs}}&(k) \equiv \\
    &\frac{2\ell +1}{\int \bar{n}^2\,\dif^3 r}\iiint\bar{n}(r_1)\,\bar{n}(r_2)\,\delta_{\rm obs}(\bs{r}_1)\,\delta_{\rm obs}(\bs{r}_2) \mathrm{e}^{i\bs{k} \cdot (\bs{r}_2  - \bs{r}_1)}\mc{L}_\ell(\bshat{k}\cdot \bshat{r}_2)\,\dif^3 r_1\; \dif^3 r_2  \;\frac{\dif^2\Omega_k}{4\pi}\,,
    \end{split}
\end{equation}
where $\mc{L}_\ell$ denotes the Legendre polynomial of  order $\ell$. Similar to our analysis in \citetalias{Elkhashab_2021}, we assume the cross-correlation terms are negligible. Therefore, we compute the impact of the FOTO signal by replacing $\rmdel{obs}$ in Eq.~\eqref{eq:YAMMATO_ESTIMATOR} with the function $\rmdel{dip}$, defined in Eq.~\eqref{Eq:DIP_DELTA}. The final result reads as (see App.~\ref{APP:FOO_MULTI} for derivation details)
\begin{align}
    \label{eq:Yam_dip_result_TOTAL}
P_{\ell,\mathrm{dip}}(k)&=
    \frac{16\pi^2\,(2\ell +1)}{3}\,\frac{v^2_{\rm o}}{H^2_0}\;\frac{I_1}{\int \bar{n}^2\;\dif^3 r}\;
    \left[\,\sum^{\infty}_{\ell'=0}\tj{\ell}{{\ell'}}{1}{0}{0}{0}  (2{\ell'}+1)(-1)^{\ell'}({\rm i})^   {{\ell'}+1}\;I_{\ell'} \right]   \,,
\end{align}
where $\tj{\ell}{{\ell'}}{n}{0}{0}{0}$ denotes the square of the Wigner-3j symbol. 

\begin{figure}
    \centering
    \includegraphics[width=1.0\textwidth]{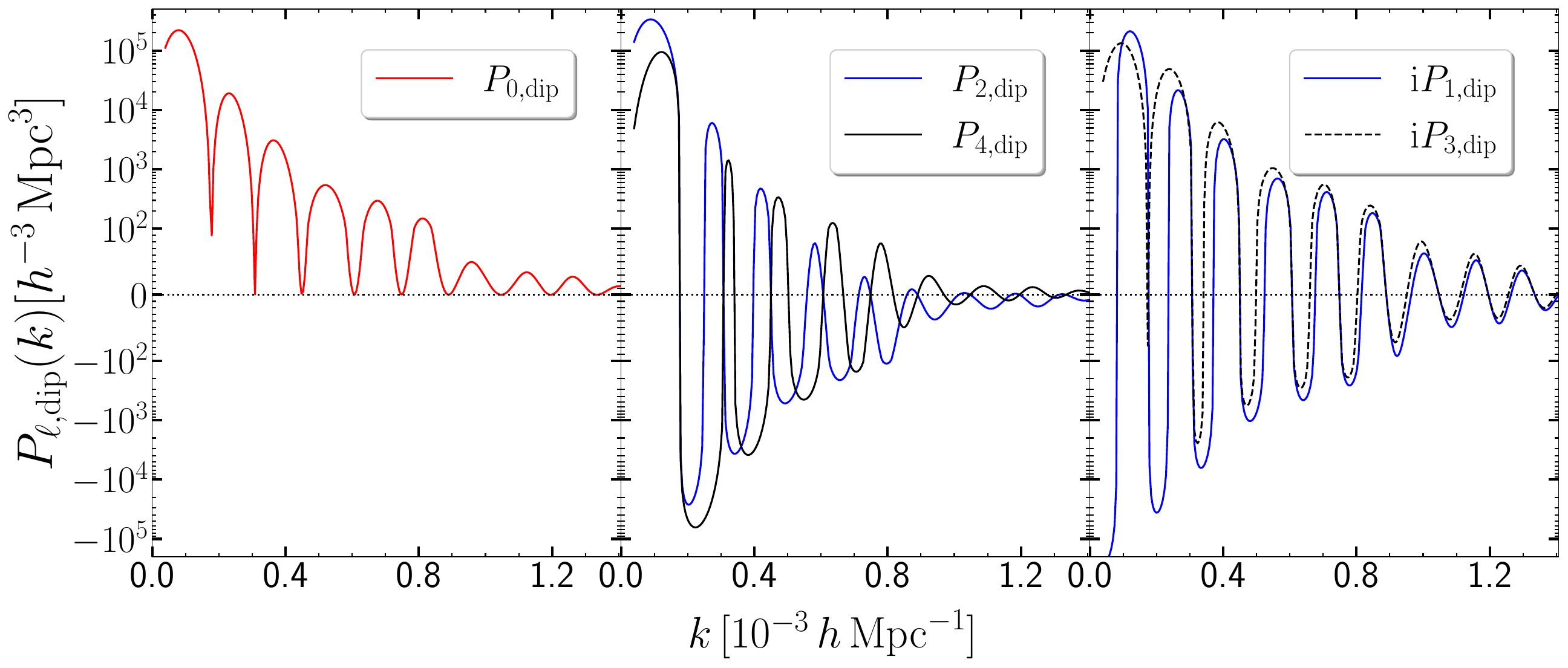}

    \caption{
    The theoretical prediction of the FOTO signal (see Eq.~\ref{eq:Yam_dip_result_TOTAL}) for the first five multipoles of the power spectrum in the redshift bin $z\in(0.9,1.8)$. The monopole is shown on the left panel, while the higher-even and odd multipoles are plotted in the middle and right panels, respectively. This figure is generated assuming that $v_{\rm o} \approx 369\,{\rm km\, s}^{-1}$ \citep{planck-dipole-18} , $\relalpha=4$ and that $\bar{n}$ matches the expected distribution of  \peuc galaxies. We compute the latter using  the  luminosity function model 3 in \citep[][see Sec.~\ref{sec.mock_specifics} for more details]{pozzetti16}.}
    \label{fig:ALL_MULTI}
\end{figure}

We can directly see that setting \(\ell = 0\) in Eq.~\eqref{eq:Yam_dip_result_TOTAL} yields Eq.~\eqref{eq:fing}. It is also clear that the monopole is the only positive-definite contribution, whereas the sign of the FOTO signal for higher multipoles varies based on the combined impact of the $I_{\ell'}$ functions (see Eq.~\ref{eq:Idef}). This behaviour is illustrated in Fig.~\ref{fig:ALL_MULTI}. In addition to the positivity of the monopole signal, the figure shows that the signal decays more strongly in the even multipoles compared to their odd counterparts. 

From Eq.~\eqref{eq:Yam_dip_result_TOTAL}, we also observe that even contributions are real, while odd contributions are purely imaginary. This stems from the asymmetry of the estimator in Eq.~\eqref{eq:YAMMATO_ESTIMATOR}, which uses the end-point convention, defining the line of sight as the direction pointing to one of the two position vectors. Under this convention, the first spatial integral is symmetric under wavevector reflection $(\mathbf{k} \rightarrow -\mathbf{k})$, whereas the second spatial integral is symmetric only for even multipoles and antisymmetric for odd ones. Consequently, the real component averages to zero for odd multipoles, while the imaginary part averages to zero for even multipoles.\footnote{We note that selecting a different line-of-sight convention, such as the mid-point or median methods, eliminates these imaginary contributions \citep{Reimberg+2016}. Similarly, using an alternative basis, such as spherical Bessel functions for the overdensity expansion, can also remove these contributions \citep[e.g.][]{semezato2024}.}

\section{Mock galaxy redshift surveys}
\label{sec:MOCKS_DESCRIP}
We build a large suite of mock galaxy redshift surveys
to assess what constraints the {FOTO} effect could place
on either the peculiar velocity of the Sun or on the cosmological parameters appearing within $I_1$ (see Eq.~\ref{eq:fing}).
To represent the state of the art in galaxy redshift surveys, 
we consider two types of spectroscopic surveys where galaxies are selected either based on their H$\alpha$ emission line 
or the 21cm emission from neutral hydrogen (\HI).\footnote{These surveys are flux-limited only to first approximation as their selection function also depends on the source size (for H$\alpha$) and the line profile (for \HI).}
However,
as this paper serves as a proof of concept
and we are interested in assessing the best possible constraints that can be set,
we only consider full-sky surveys in our main study. 
In this case, Eq.~(\ref{eq:fing}) provides an exact model for the signature of the FOTO effect
and we do not need to model the imprint of an angular window function.
Although we make this assumption for simplicity,
it is not unreasonable to imagine that redshift
surveys will soon cover major fractions of the sky.
For instance, forecasts for
a nearly all-sky \HI\ survey detecting one billion galaxies
out to $z\simeq 2$ with the
Square Kilometre Array Observatory (SKAO) have been presented in the literature \cite[e.g.][]{Yahya_2015}.
We anyway discuss the impact of partial sky coverage in Section~\ref{sec:S_N}.
\subsection{Heliocentric rest frame}
In Sec.~\ref{sec:FOTO}, we have considered a general, abstract observer. From now on, we focus on galaxy spectra taken from the Earth or nearby space.
Published redshifts are usually given in the heliocentric frame, after correcting for the motion of the telescope around the Sun
with velocities $v_\mathrm{tel}\lesssim 30$ km s$^{-1}$ ($v_\mathrm{tel}/c\lesssim 10^{-4}$).\footnote{Telescopes have a time-varying peculiar velocity and
sometimes we combine observations taken at different epochs in order to measure redshifts. In these cases, the correction needs to be performed statistically.}
We neglect this small correction in this work.

Eq.~(\ref{eq:obsred}) shows that
the heliocentric redshift $z_\mathrm{hel}$ 
and 
the redshift in the comoving (or CMB) frame differ by 
a direction-dependent term generated by the peculiar motion of the Sun,
 \begin{align}
\label{eq:redshift_correction}
    1+z_{\rm cmb} := 
    \frac{1+z_\mathrm{hel}} {1+z_{\mathrm{pec}, \odot}}\simeq
    \frac{1+z_\mathrm{hel}} {1-v_{\parallel, \odot}/c}
    \,.
\end{align} 
Assuming that the CMB dipole is of kinetic origin gives $|v_{\parallel, \odot}/c|<1.2\times 10^{-3}$ 
which is comparable with the uncertainty of
redshift measurements performed with
slitless spectroscopy from space \cite[e.g.][]{euclidcollaboration2024euclidiovervieweuclid} and
substantially larger than the redshift error in current
and future ground-based cosmological surveys \cite[e.g.,][]{Lan_2023}. 
Therefore $\boldsymbol{v}_\odot$
should leave an imprint on the statistics we use to measure 
galaxy clustering.%

\subsection{The \liger method} 

Since we want to study 2-point clustering at very
large scales, 
we use the new implementation of \liger  method \cite{MIKO_2017} presented in \citetalias{Elkhashab_2021}
to generate mock galaxy catalogues accounting for relativistic redshift-space distortions to first order in the cosmological perturbations.\footnote{The \liger code is available at \url{https://astro.uni-bonn.de/~porciani/LIGER/}}
Starting from the sequence of snapshots produced by
a Newtonian N-body simulation,
\liger determines the intersection between the perturbed null worldlines of the photons emitted by a source located within the domain of the simulation
and the backward light-cone of an observer. 
As an output, \liger provides the redshift-space position of the source and its magnification.

It is important to note that, contrary to other techniques, \liger does not replicate the simulation box in order to build a lightcone, as this would introduce spurious artefacts 
at the large scales we are interested in studying.
Therefore, the box size of the input N-body simulation must be larger than the transverse size of the portion of the light cone created in the output.
As we want to analyse sections of full-sky light cones with sizes that are comparable with the Hubble volume, it is not possible to use N-body simulations that resolve the dark-matter haloes hosting the galaxies. A special implementation of \liger
has been developed for applications of this kind 
(see \cite{MIKO_2017} and \citetalias{Elkhashab_2021} for further details).
In this case, the basic \liger algorithm is first applied to the dark-matter distribution and the mock galaxy catalogues are produced in a second step using the \textsc{buildcone} toolbox (see \citetalias{Elkhashab_2021}).
This step requires specifying four functions of the background redshift
that characterise the galaxy population under study, namely the observed number density of galaxies $\bar{n}$, the evolution bias $\mc{E}$, the magnification bias $\mc{Q}$, and the linear bias factor $b$ which relates the galaxy and dark-matter overdensities in real space.
The final output is a mock catalogue listing
the observed angular positions and redshifts for the galaxies in the simulated survey.

\subsubsection{Specifics of the mock catalogues}
\label{sec.mock_specifics}

We consider a standard $\Lambda$CDM cosmological model based on the results of the Planck mission \cite{planck18}. The background is defined by  
the matter density parameter
$\Omega_\mathrm{m,0} = 0.3158$, the baryon density parameter $\Omega_\mathrm{b,0} = 0.0508$, and the present-day dimensionless Hubble constant $h = 0.673$.
The scalar perturbations are described by a power-law power spectrum with a 
primordial spectral index $n_\mathrm{s} = 0.966$ and an amplitude $A_{\rm s} = 2.1 \times 10^{-9}$ at wavenumber $k = 0.05\,{\rm Mpc}^{-1}$.
We compute the corresponding linear matter power spectrum by using the Boltzmann solver \textsc{camb} \citep{CAMBS}.

We simulate the distribution
of dark matter within 35 periodic cubic boxes, each with a comoving side of $L_{\rm box} = 12\,h^{-1}$ Gpc.
Since we are only interested in very large scales, 
we use the  \textsc{music} code \citep{music} 
to evolve the distribution of $1024^3$ particles
of mass $m_\mathrm{p}=1.4\times 10^{14}\,h^{-1}$ M$_\odot$
using second-order Lagrangian perturbation theory.
In \citetalias{Elkhashab_2021}, we have demonstrated that these simulations are sufficiently accurate for our goals.

We use 21 snapshots (linearly spaced in the scale factor, $a$, from $a=1/3$ to $a=1$) of the \textsc{music} simulations as an input to \liger for building the full-sky dark-matter lightcones. Note that the lightcone section at $z=1.8$ (the maximum redshift we consider) has a radius
of $3.37\,h^{-1}$ Gpc, which means we could extract multiple lightcones from the same simulation box.  
We thus use hexagonal close packing %
to extract four non-overlapping light-cones from each simulation box. Eventually, we employ the \textsc{buildcone} toolbox to generate 140 mock galaxy distributions.

In order to model the selection function for the sample
H$\alpha$ survey (\peuc), we rely on our previous forecasts
for the wide spectroscopic survey of the Euclid mission \citep{euclidcollaboration2024euclidiovervieweuclid}. Following \citetalias{Elkhashab_2021}, we compute $\bar{n}$,
$\mc{Q}$ and $\mc{E}$ in the redshift range $0.9<z<1.8$
using model 3 in \cite{pozzetti16} for the H$\alpha$ luminosity function and assuming a flux limit of $2\times 10^{-16}$ erg cm$^{-2}$ s$^{-1}$ 
together with a uniform $70\%$ completeness.  For $b$,
 we adopt the linear relation $b(z)=1.46+0.68(z-1)$ from \cite{EuclidVII}.
 These specifics are meant to represent a generic state-of-the-art survey targeting emission-line galaxies from space.

\begin{figure}
    \centering
    \includegraphics[width=0.8\textwidth]{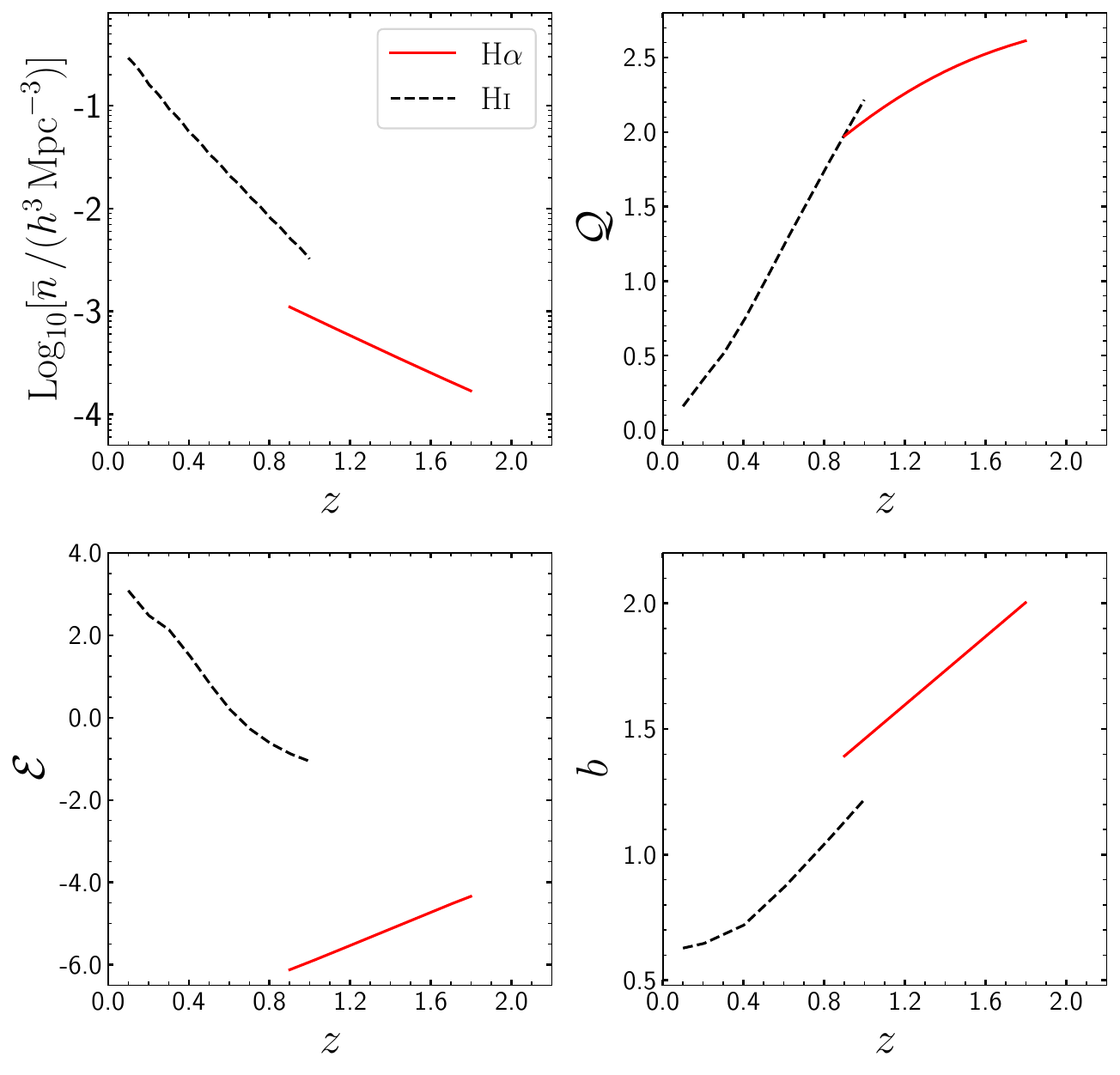}
    \includegraphics[width=.56\textwidth]{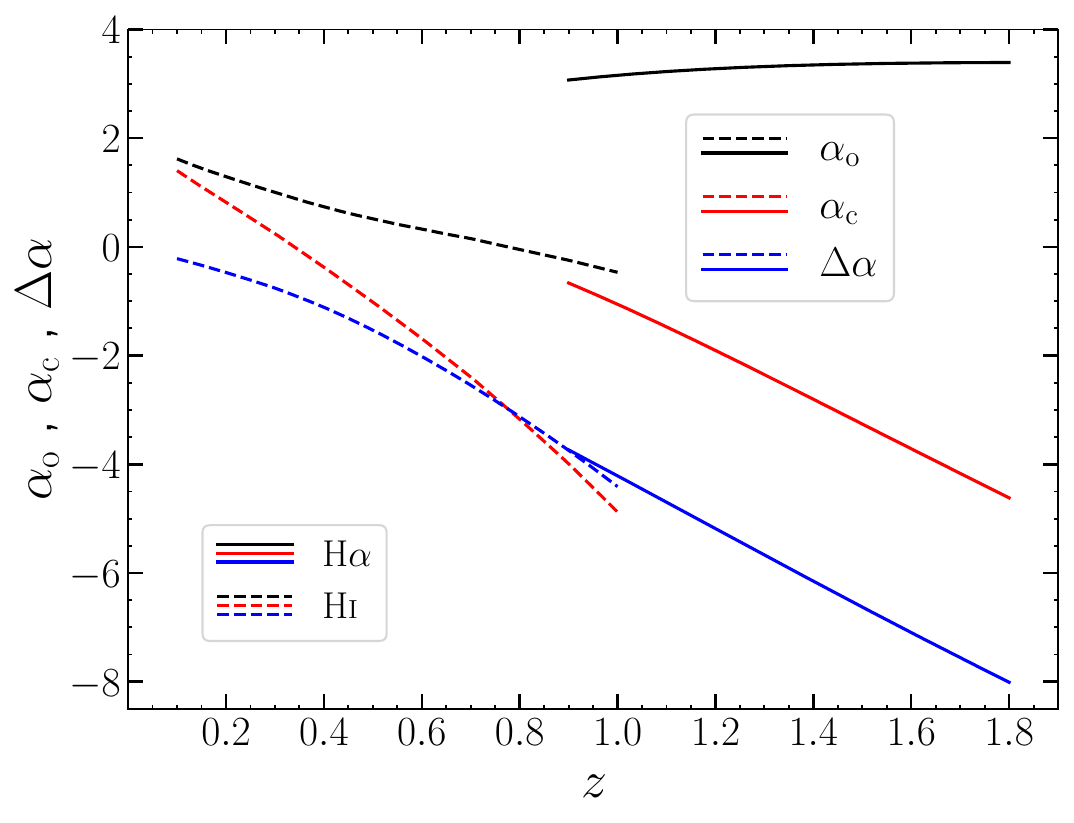}
    \caption{The top set of figures shows the functions $\bar{n}$, $\mc{Q}$, $\mc{E}$ and $b$ used to build the mock galaxy catalogues.  Solid and dashed lines refer to the sample \peuc and \pska surveys, respectively. The bottom figure displays the functions $\relalpha$, $\kalpha$ and $\Delta \alpha$ in black, red, and blue, respectively.}
    \label{fig:SURV_FUNCS}
\end{figure}

As a reference case for the selection function of the sample \HI\ survey, 
we adopt the \HI\ galaxy survey planned with the Square Kilometre Array Observatory \cite{Abdalla+2015}.
We thus use a cubic spline to interpolate between the tabulated values 
for $\bar{n}$,  $\mc{E}$ and $\mc{Q}$
given in \cite{Maartens_2021} 
(based on a semi-analytic model of galaxy formation built upon an N-body simulation [\citealp{Obreschkow+09}]). 
For the linear bias, we adopt the values in \citet{Yahya_2015} 
obtained from the same model. 
We consider the redshift interval $0.1<z<1$. A complete summary of the adopted functions is shown in Fig.~\ref{fig:SURV_FUNCS}.
We note that the $b$ and $\mc{Q}$ show similar trends with redshift for the two sample surveys
while this similarity is not present for
$\mc{E}$ \citep[{see}][]{Maartens_2021}.

In order to study the impact of the FOTO effect on $P_0$,
we generate two sets of mock catalogues (composed of 140 realisations each) that differ only because of the peculiar velocity of the observer.
In the first collection (hereafter HRF, short for `in the heliocentric rest frame'), we make sure that $\boldsymbol{v}_\mathrm{o}$ coincides with
the peculiar velocity that generates the full CMB dipole as measured by the Planck mission \citep{planck-dipole-18}, i.e., $\bv{\rm o} = \bv{\odot}$. In the second (hereafter CRF, short for `in the CMB -or comoving- rest frame'), we consider a comoving observer with no peculiar velocity ($v_{\rm o}=0$).

\begin{figure}[tbp]
\centering 
\includegraphics[width=0.9\textwidth]{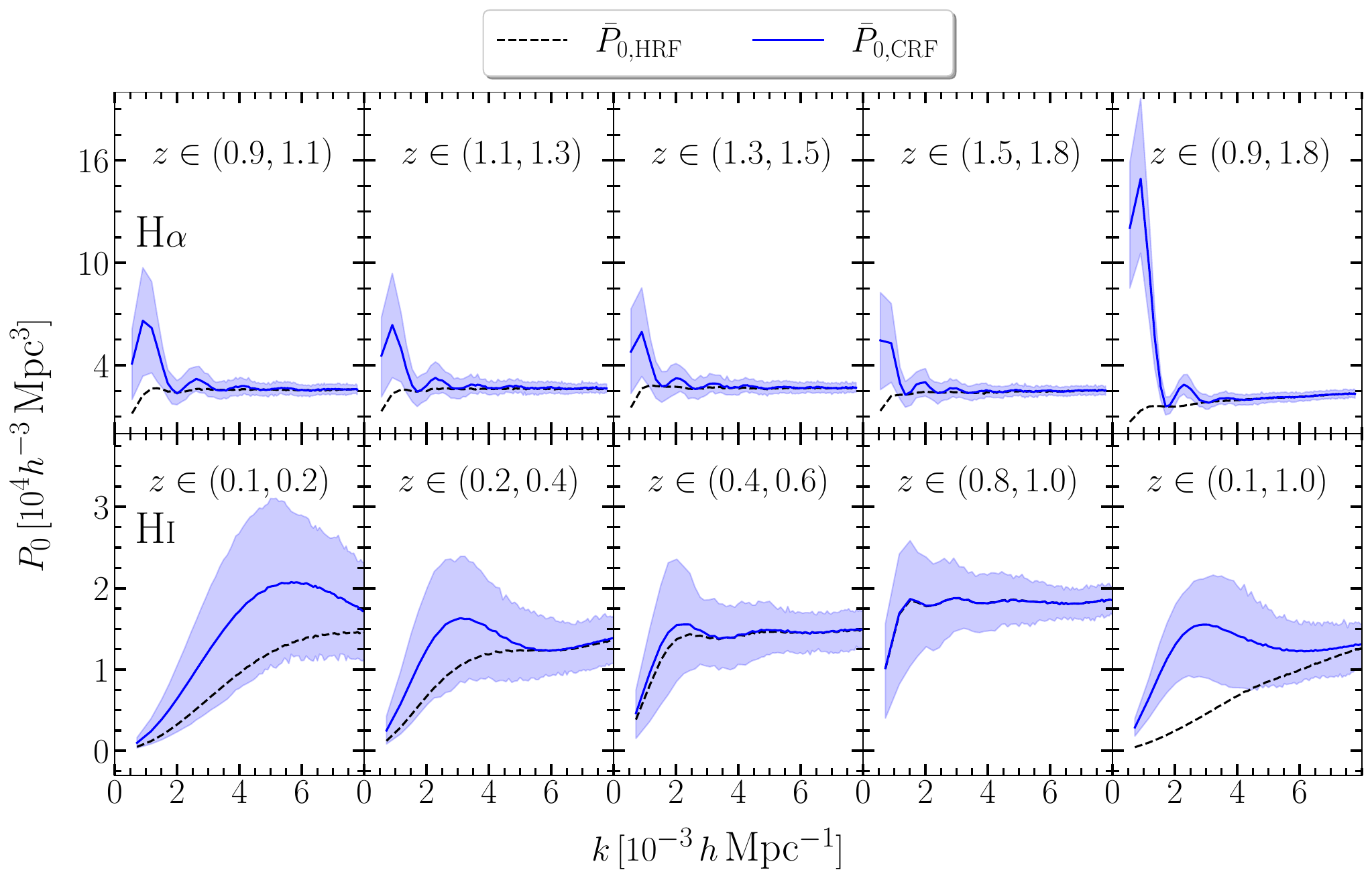}
\includegraphics[width=0.9\textwidth]{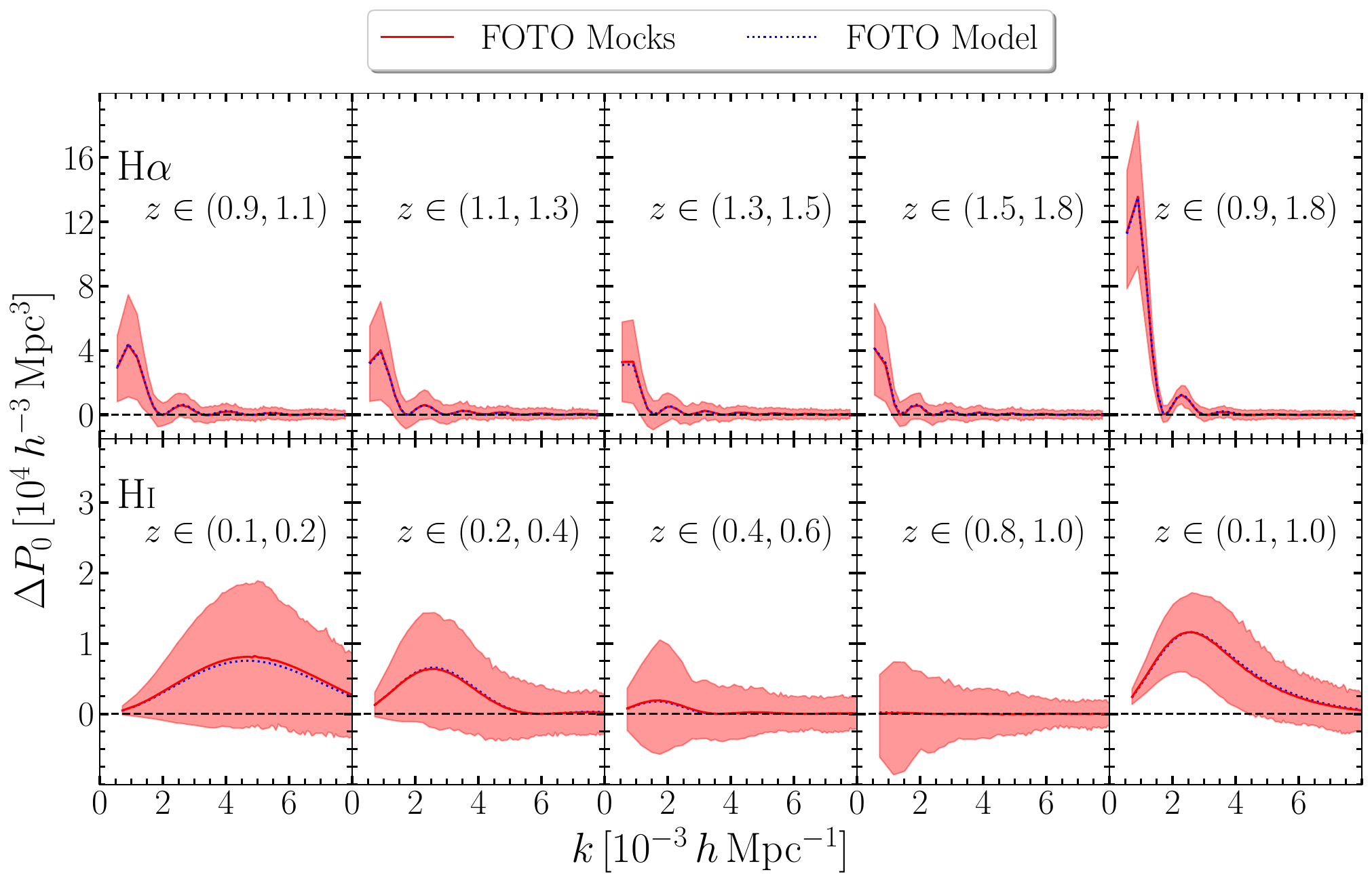}
\caption{\label{fig:SPECTRA_ORG} 
Top: Measurements of the power spectrum monopole in the HRF (blue solid) and CRF (black dashed) averaged over the 140 mocks. 
The shaded regions denote the root mean square (rms) scatter in the HRF. The higher and lower sub-panels display the results for the  \peuc and \pska surveys, respectively.
Bottom: As in the top figure, but for
the FOTO signal (solid) obtained by subtracting the measurements of $P_0(k)$ in the HRF and CRF. 
The dotted blue line (which almost perfectly coincides with the solid red line) represents the theoretical model in  Eq.~\eqref{eq:fing}.}
\end{figure}

\section{Measuring the FOTO effect}
\label{sec:MOCKS_MEASURMENTS_ALL}

We employ the Feldman-Kaiser-Peacock (FKP)  estimator \cite{fkp}  to measure the power spectrum monopole
from the mock catalogues. Fourier transforms are computed with the FFT algorithm \citep{Heideman_1985_FFT} 
after embedding the sections of the light cones within a cubic box  of side  $L_{\rm FFT} = 16 \,h^{-1}\, {\rm Gpc}$ %
and sampling the density on a Cartesian grid with $1024^3$ voxels.
This setup allows us to measure $P_0$ for $k\geq k_\mathrm{f}=2\pi/L_{\rm FFT} = 3.9\times10^{-4}\,h\,\rm{Mpc}^{-1}$ and thus cover the range  of wavenumbers for which the FOTO effect imprints a detectable signal.
In order to minimise the variance of the FKP estimator, we weigh galaxies based on the mean local density, $\hat{n}$, evaluated  in each catalogue within radial shells of thickness $\delta r = 10 \,h^{-1}$ Mpc. Namely, we use the weights
$w(r) = [ 1 + \hat{n}(r)\, \mc{P}_0]^{-1}$ with $\mc{P}_0 = 20\,000\,h^{-3}$ Mpc$^{3}$ which represents a typical value for the power spectrum at the scales of interest.

In the top panel of Fig.~\ref{fig:SPECTRA_ORG}, we show the mean $P_0$ obtained by separately averaging the measurements over the 140 HRF (solid blue lines) and the 140 CRF (dashed black lines) catalogues.  We consider four relatively narrow redshift bins and a broader one for both example surveys.  As expected, in the HRF, $P_0$ shows evident oscillations 
on large scales, which are absent in the CRF. The wavenumbers at which the peaks and valleys appear change with the characteristic redshift, $z_{\rm c}$,  of the galaxy samples.

In %
the bottom panel of Fig.~\ref{fig:SPECTRA_ORG},
we focus on the FOTO signal itself. We show the difference between the  $P_0$ obtained in the HRF  and its counterpart in the CRF.  The blue dotted line denotes the analytical results computed using Eq.~\eqref{eq:fing}. The analytical prediction agrees extremely well with the mock estimates for both example surveys.\footnote{The model can be extended to account for Alcock--Paczyński distortions as shown in appendix~\ref{Sec:AP}.} The amplitude of the FOTO signal and its redshift evolution differ  between the two example surveys reflecting 
the variations in 
$\relalpha^2$ for the different galaxy populations
(bottom panel of Fig.~\ref{fig:SURV_FUNCS}).
Moreover, the wavenumber at which the FOTO signal peaks --hereafter $k_{\rm p}$-- 
is determined by the redshift bin under consideration %
as the spherical Bessel function in Eq.~\eqref{eq:Idef} reaches a local maximum when $k\sim r(z_\mathrm{c})^{-1}$.

We quantify the detectability of the FOTO signal using the likelihood-ratio test for comparing two
simple hypotheses (namely, the observer is at rest
either in the CRF or in the HRF). This frequentist method measures the separation between the probability distribution functions of the likelihood ratios obtained under the different null hypotheses
 and thus quantifies the possibility to discern between the two models using a measurement of $P_0(k)$
(for further details, see Sec.~4.4 in \citetalias{Elkhashab_2021}, Sec.~4 in \cite{euclidcollaboration2024euclidpreparationimpactrelativistic} as well as Sec.~3.3 and  Appendix~A in \cite{MIKO_2017}).  The outcome of the test is presented in terms
of a signal-to-noise ratio (\snrm) for the detection of the FOTO effect. Given a data vector $\bs{d}$ and
the covariance matrix of the measurement errors $\mcov$, the result of the likelihood-ratio test coincides
with the usual estimate $\snrm=\bs{d}^T\, \mcov^{-1} \bs{d}$ when $\mcov$ is the same under the two null hypotheses (but are more general if $\mcov$ changes).
Using the FOTO signal as the data vector, estimating $\mcov$ numerically from the variations of the measurements in our mock catalogues, and assuming Gaussian measurement errors,
we obtain the highest detectability score of $\snrm\simeq 6.8$ for the \peuc survey in the redshift bin $z\in(0.9,1.8)$. 
\subsection{Is the FOTO signal detectable for realistic survey geometries?}
\label{sec:S_N}

\begin{figure}
    \centering
    \includegraphics[width=0.55\linewidth]{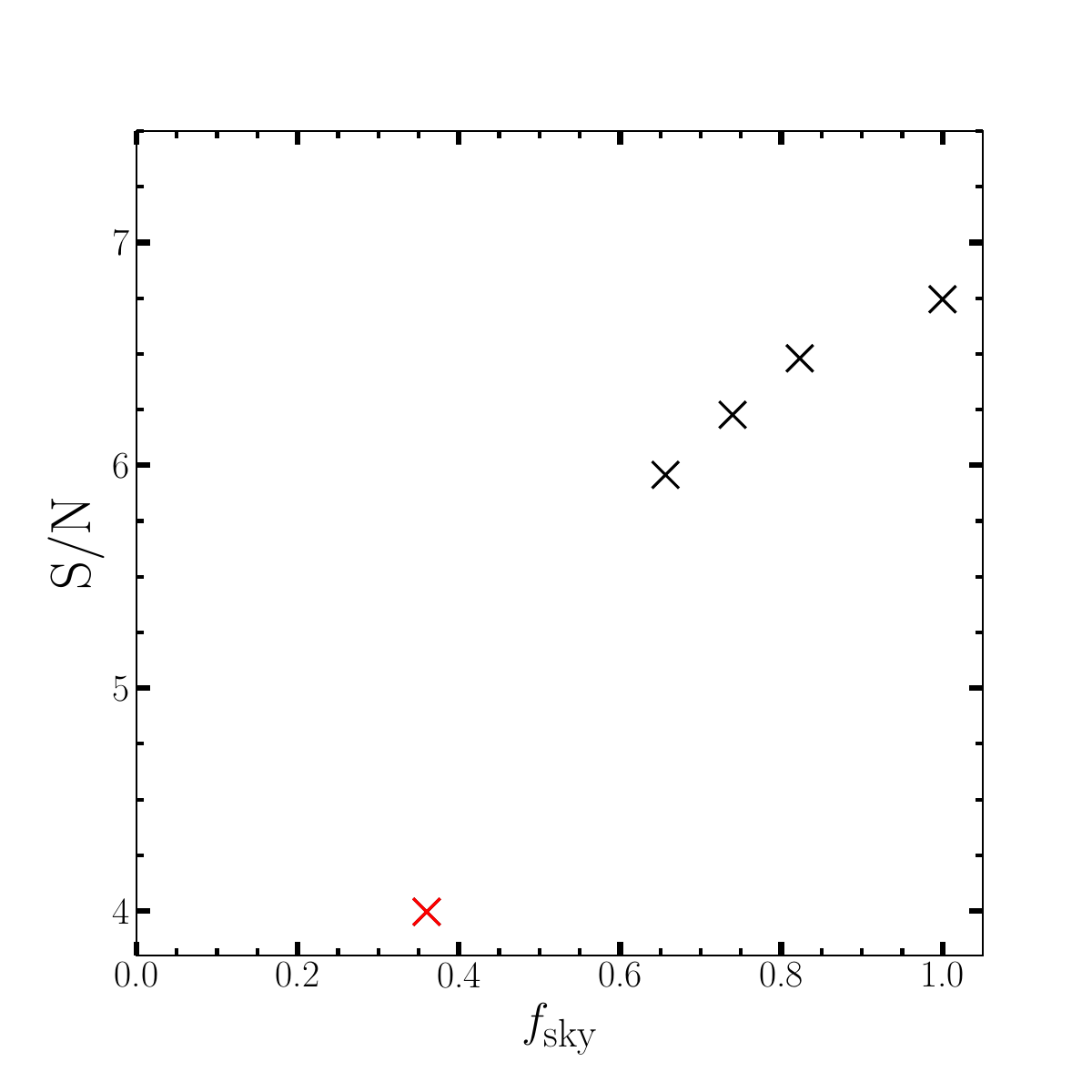}
    \caption{The signal-to-noise ratio of the FOTO signal as a function of the sky fraction covered by our example \peuc survey for galaxies in the redshift interval $z\in (0.9, 1.8)$.
    }
    \label{fig:S_N}
\end{figure}

Thus far, our investigation has focused on the ideal scenario where the entire sky is observable. However, in actual surveys, significant portions of the celestial sphere are inaccessible to the observer
and this alters the estimates of the power spectrum on large scales. %
It is thus important to 
test whether or not the FOTO signal is detectable  in these cases. 
To that end, we repeat the likelihood-ratio test considering different survey geometries.
Our results for the \peuc survey  in the redshift bin $z\in(0.9,1.8)$ are shown in Fig.~\ref{fig:S_N}. From right to left, we consider different footprints: \textit{i}) full-sky, \textit{ii}) masking the regions within $10^\circ,15^\circ,$ and $20^\circ$ from the Galactic plane, and \textit{iii}) masking the regions within $20^\circ$ from the Galactic and  Ecliptic planes (red symbol), similar to some on-going surveys. 
Although, as expected, the $\snrm$ increases with the fraction of the sky covered, the FOTO effect remains detectable with $\snrm \approx 4$ even when the most severe cuts are applied. 

We finally investigate the impact of redshift measurement errors on the \snrt. To this end, we add Gaussian errors with zero mean and a standard deviation of $10^{-3}(1+z)$ to the \peuc mocks.
These figures mirror the specifications of current redshift surveys from space \citep{euclidcollaboration2024euclidiovervieweuclid}.
We find that redshift errors induce negligible corrections to the \snrt ($\approx 0.1\%$). 

{
 A  partial-sky coverage also introduces mixing between different $k$-modes, thereby altering the signal's shape \citepalias[see Fig.~12 in][]{Elkhashab_2021}. This effect must be taken into account in the model for the FOTO signal. There are two approaches available in the literature; the mixing matrix formalism \citep[e.g.][]{wilson_rapid_2017,Beutler+14} and the spherical Fourier Bessel method \citep[see][]{Wen_2024}. We leave the implementation of these techniques to future work. }

\subsection{
Constraining cosmological parameters }
\label{Sec:MCMC_FOTO}

The FOTO signal is sensitive to the expansion history of the Universe in multiple ways.
The most evident one is perhaps
the logarithmic derivative ${\mathrm{d} \ln H}/{\mathrm{d} \ln (1+z)}$ appearing in the definition of $\relalpha$. On top of that, the $I_1(k)$ integral is influenced by
the time evolution of the ratio $H/H_0$ and variations in the extremes of integration.
It thus makes sense to test whether this sensitivity could be used to constrain some cosmological parameters  from observations. As a proof of concept, we consider two idealized scenarios.

In the first, we set the cosmological parameters to their true values %
and see what constraints we can place
on the magnitude of $\bv{\odot}$ (hereafter, velocity inference). 
This setting represents the situation
in which one wants to test the kinematic interpretation of the CMB dipole using the FOTO signal combined with tight constraints on the
cosmological parameters coming from other probes.

Alternatively,
in the second scenario, we aim to measure the value
of some cosmological parameters while
adopting a prior on the peculiar velocity of the Sun
based on the Planck mission
\cite{planck-dipole-18}.
In this case, one
adheres to the kinematic interpretation of the CMB dipole
and uses the FOTO effect as a cosmological probe.
To start with, we consider a flat $\Lambda$CDM model
so that $H(z)/H_0=[\Omega_{\mathrm m,0}(1+z)^3+1-\Omega_\mathrm{m,0}]^{1/2}$ 
and we only need to perform inference on the present-day value of the matter density parameter (hereafter, density inference).%

In both scenarios, we assume to know the evolution and magnification biases exactly.
As a case study, we use the full-sky \peuc example survey within the redshift range \(z \in (0.9, 1.8)\) for which we obtained the highest \snrm\ for the FOTO signal.

\begin{figure}
    \begin{subfigure}[b]{0.45\textwidth}
    \centering
    \includegraphics[width=1\textwidth]{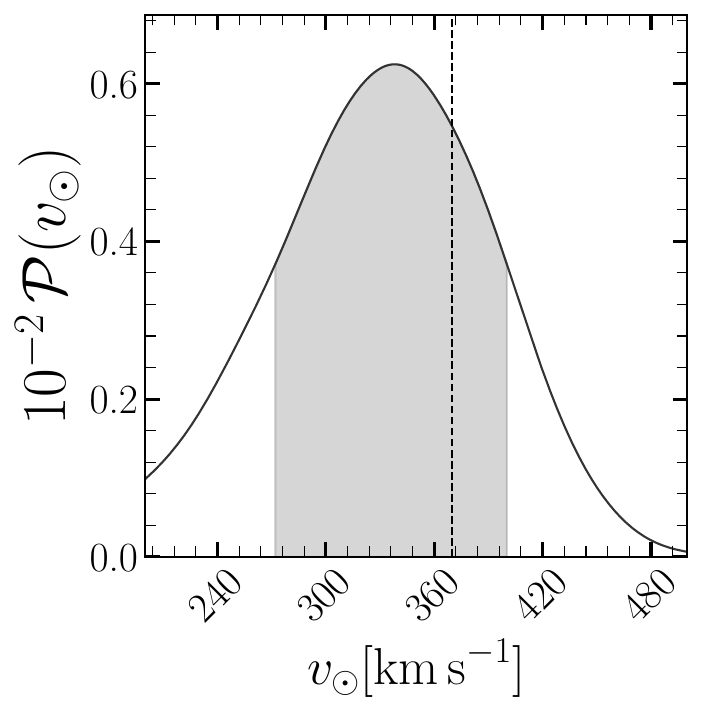}%
    \end{subfigure}
    \begin{subfigure}[b]{0.45\textwidth}
        \flushright
    \includegraphics[width=1\textwidth]{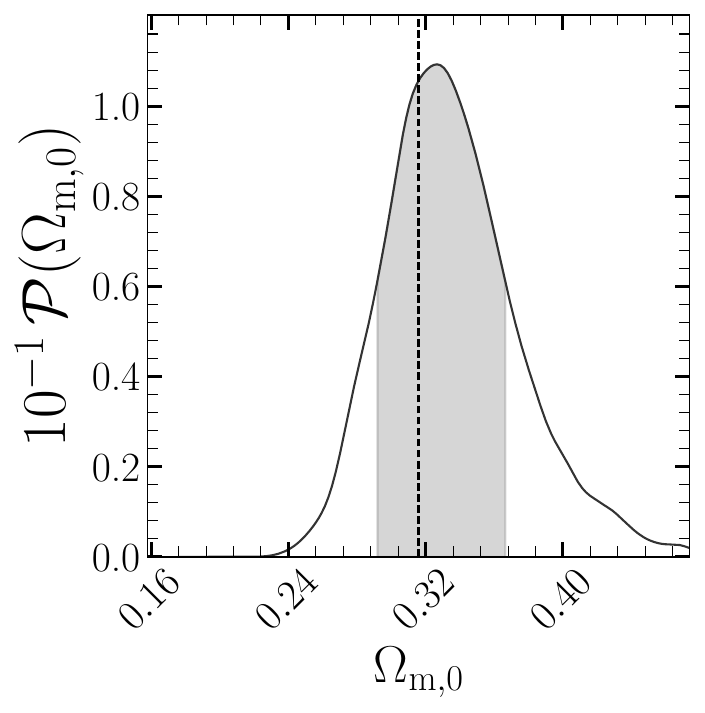}%
    \end{subfigure}%
    \caption{ The posterior distributions of the magnitude of the observer velocity (left) and the present-day value of the matter density parameter (right) extracted from one of the \peuc HRF mocks. The dashed vertical line indicates the true value of the parameters while the shaded regions denote the 68$\%$  HPDI. 
    } 
    \label{fig:Corner_V_NO_SHIFT}
\end{figure}

\subsubsection{Bayesian inference}

Given a model $\bs{m}$ depending on a set of tunable parameters $\bs{\theta}$ and some data $\bs{d}$, we compute the posterior distribution of the parameter values $\mathcal{P}(\bs{\theta}|\bs{d})$ by updating the assumed prior distribution $\mathcal{P}(\bs{\theta})$ with the likelihood function of the observed data, 
$\mc{P}(\bs{d}|\bs{\theta})$.
We assume Gaussian measurement errors and write
\begin{equation}
    \mc{P}(\bs{d}|\bs{\theta})\propto \frac{\exp\left\{-\frac{1}{2}[\bs{d}-\bs{m}(\bs{\theta)}]^T\;  \mcov^{-1} \;[\bs{d}-\bs{m}(\bs{\theta)}]\right\}}{(2\pi)^{n/2} \sqrt{\det{\mcov}}}\,,%
\end{equation}
where $\mcov$ is the covariance matrix of the errors and $n$ denotes the size of the dataset.

In our investigation, $\bs{d}$ represents the difference between the power-spectrum monopole extracted from one of our HRF mock catalogues and the mean of the $P_0$ estimates from the CRF mocks,  $\bar{P}_{0,\rm CRF}(k)$, which is a proxy for an exact theoretical model that does not include the FOTO effect.
We compute the monopole of the power spectrum in five bins spanning the range $k\in(0.39, 5.34) \times 10^{-3}\, h \,\mathrm {Mpc}^{-1}$.

For the theoretical model, we use Eqs.~(\ref{eq:fing}) and (\ref{eq:Idef})
that accurately reproduce the mean signal from the mocks.
As often done in cosmology, we do not vary $\mcov$
with $\bs{\theta}$ and estimate it numerically from the variations of the measurements in the 140 HRF mocks. 
Moreover, we sample the posterior distribution using the 
Markov chain Monte Carlo (MCMC) method as implemented
in the \texttt{emcee} package \cite{EMCEE}. 

For the inference of $\Omega_\mathrm{m,0}$,
we adopt a flat uniform prior in the range $[0,1]$.
Choosing a prior for $v_{\odot}$ is less straightforward.
Numerical simulations in the $\Lambda$CDM scenario show that the peculiar velocity of dark-matter halos with fixed mass and local environmental density approximately follows a Maxwellian distribution
\cite{Sheth_2001_prior,Dam+2023},
\begin{equation}
\label{eq:prior}
\mc{P}(v_{\odot}) = \sqrt{\frac{2}{\pi}}\frac{v^2_{\odot}}{\sigma^3}\exp{\left(-\frac{v^2_{\odot}}{2\sigma^2}\right)}\,,
\end{equation}
(i.e. a chi distribution with three degrees of freedom
and scale parameter $\sigma$). For halo masses comparable to the Local Group ($\sim 3\times 10^{12}$ M$_\odot$, \cite{Benisty_2022}) %
or the Milky Way ($\sim 1.3\times 10^{12}$ M$_\odot$, \cite{Watkins+2019}),
one expects values of $\sigma$ of the order of a few hundred km s$^{-1}$.
However, it is necessary to 
account also for the motion of the Galaxy within its halo and for the fact that the Sun orbits the centre of the Milky Way at a speed of about 250 km s$^{-1}$.
For simplicity, we use a slightly-informative proper prior on $v_{\odot}$ in the form of a  Maxwell distribution with $\sigma=300$ km s$^{-1}$.
This prior distribution has mode $\sqrt{2}\sigma\simeq 420$ km s$^{-1}$, expectation
$2\sigma \sqrt{2/\pi}\simeq 480$ km s$^{-1}$, and standard deviation $\sqrt{(3\pi-8)\sigma^2/\pi}\simeq 200$ km s$^{-1}$. The 68\% highest prior density interval extends from 240 to 640 km s$^{-1}$.
{In Appendix~\ref{sec:Vel_prior}, }we {show} that our final results are dominated by the likelihood and are little influenced by the adoption of this particular prior.

In Fig.~\ref{fig:Corner_V_NO_SHIFT}, we present the 
posterior distributions for $v_\odot$ and $\Omega_\mathrm{m,0}$ derived from one of the mock
catalogues.
In both cases,
the posterior nicely peaks around the true value
and decays exponentially fast in the tails. 
Averaging the results over the 140 mocks, we do not see evidence of biased estimates. Moreover,
by taking the half width of the 68\% highest posterior density interval (HPDI) as a measure of
uncertainty, we find 
$\overline{\Delta v}_\odot =58$ \kms ($16\%$ relative error) and
$\overline{\Delta \Omega}_{\rm m, 0}=  0.036$ ($11\%)$ which
are both substantially narrower than the adopted priors.
This demonstrates that the FOTO signal has constraining power and can be used to measure either $v_\odot$ or $\Omega_\mathrm{m,0}$, provided that observational systematic effects are kept under control and corrected for on very large scales.
%
%
%

{So far, we have assumed to know $\mc{Q}(z)$ and $\mc{E}(z)$ precisely, as it is difficult to gauge how well stage IV surveys will measure the luminosity function of their target galaxies. However, we can  estimate how much  uncertainties in these functions impact the velocity measurement. Assuming  $(1\%,10\%)$ errors on $\relalpha$, we find that the velocity HPDI increases to $\overline{\Delta v}_\odot = (62,71)$ \kms.}\footnote{{ The measurements of  $\mc{Q}$ and $\mc{E}$ in the eBOSS survey \citep{Measurment_S_MAG} suggest that  $\relalpha$ of the \peuc survey could be determined with 2\% precision. This shows that assuming a  10\% error is a fairly pessimistic scenario.}}

\section{The impact of redshift transformations}
\label{Sec:RED_TRANSF}
\subsection{Can we cancel the FOTO signal?}
\label{Sec:Cancellation}
Naively, one might think that correcting the observed
redshifts to the CMB frame as in Eq.~(\ref{eq:redshift_correction})
removes any dependence of the galaxy distribution on the peculiar velocity of the observer.
However, the redshift correction only shifts galaxies along the radial direction while it does not consistently adjust their angular positions and apparent flux.
As derived in Appendix~\ref{app:alpha_deriv}, 
the overdensity of the redshift-corrected galaxies is
 \begin{equation}\label{eq:corrected_overdns}
 \delta_\mathrm{corr}=
 \delta_{\rm obs} -\,\alpha_\mathrm{c}\,\frac{\bv{\odot} \cdot {\hat{\bs{r}}}}{a\,H\, r}= \delta_{\mathrm {com}} +(\relalpha-\alpha_c)\,\frac{\bv{\odot}  \cdot \hat{\bs{r}}}{a\,H\,r}\;,
\,,
 \end{equation}
and  still depends on the observer velocity. 
In this case, the amplitude of the dipole term scales
 proportionally to
 \begin{equation}
     \Delta \alpha \equiv \alpha_c -\relalpha=
     -\frac{2\,r H}{c(1+z)}\left(  \mathcal{Q}  +1\right)\,.\label{eq:delta_alpha}
 \end{equation}
 In Fig.~\ref{fig:SURV_FUNCS}, we show the evolution of $\Delta\alpha$ with redshift for both our example surveys (blue curves). 
 We note that $\Delta \alpha$ is always negative as
$\mc{Q}\geq0$ (see Sec.~\ref{Sec:Mag_evo_bias}). Its magnitude increases (almost) linearly with $z$ and
approaches zero only when $z\to 0$.

 It follows from the discussion above that, after
 the redshift correction is applied to the galaxies, the power-spectrum monopole presents an oscillatory term as in Eqs.~\eqref{eq:fing} and
\eqref{eq:Idef} but where $\relalpha$ is replaced by $\Delta \alpha$. We dub this term the C-FOTO effect (with C short for `corrected redshifts') and indicate
the equivalent of the function $I_\ell(k)$ with
the symbol
\begin{equation}
J_{\ell}(k)=\int^{r(z_f)}_{r(z_i)}  \frac{r\,\bar{n}\,\Delta\alpha}
{a\, H/H_0 }\,j_\ell(k r)\,\dif r\;.\label{eq:Idef_corrected}    
\end{equation} 

\begin{figure}[tbp]
\centering 
\includegraphics[width=1\textwidth]{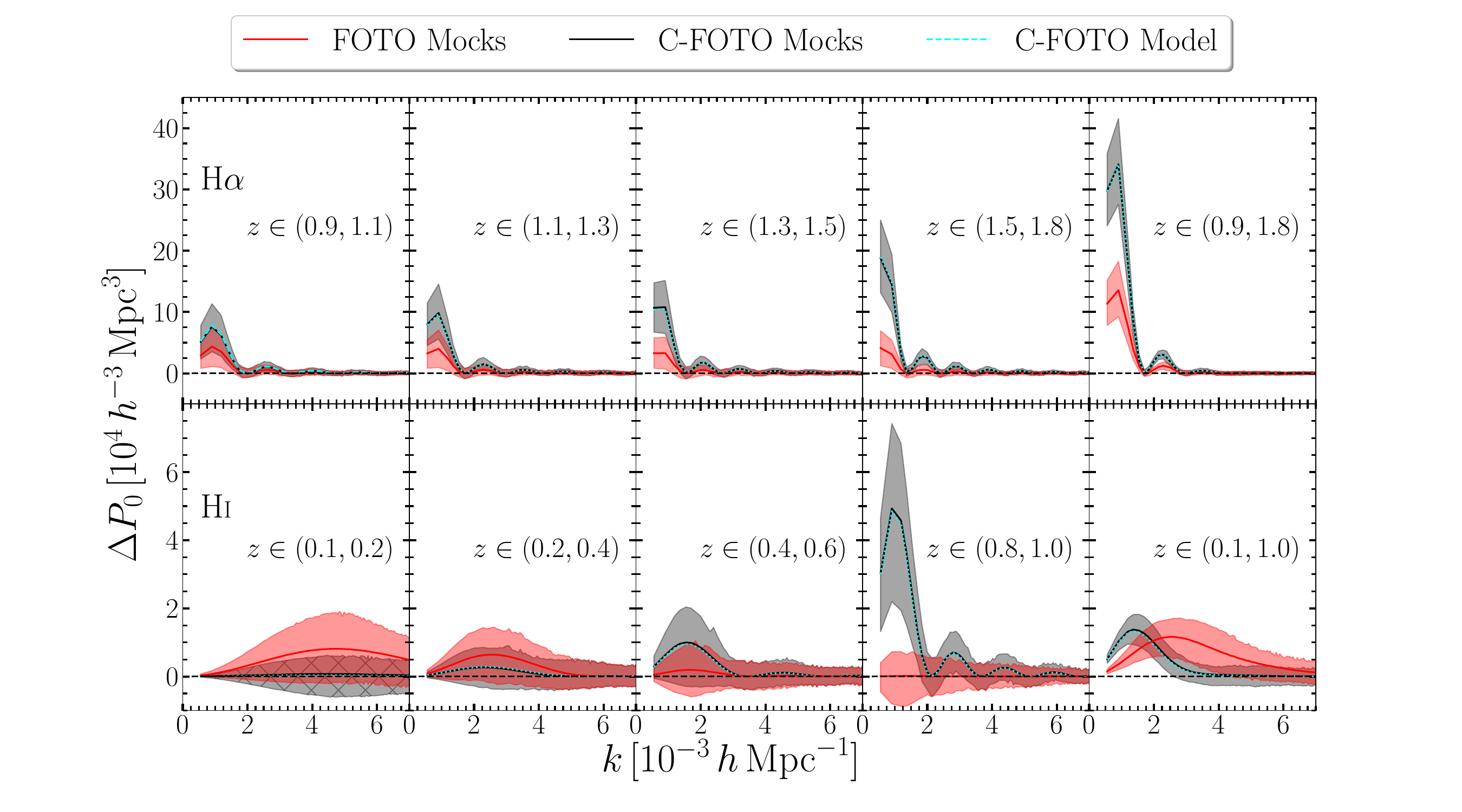}
\caption{\label{fig:corrected_spectra}
The mean C-FOTO signal (black) and its scatter (grey shade) from the mocks are overplotted to the bottom panel of Fig.~\ref{fig:SPECTRA_ORG}
showing the corresponding quantities for the FOTO signal.
The cyan dashed lines (which almost perfectly overlap with the black lines) show the theoretical predictions based on Eq.~\eqref{eq:Idef_corrected}}
\end{figure}

In Fig.~\ref{fig:corrected_spectra}, we compare
he C-FOTO signal derived from our set of HRF mocks
(black lines and gray shaded regions indicate the mean and rms scatter among the catalogues, respectively)
with the original FOTO signal (red lines and salmon shaded regions). As expected, the redshift correction
to the CMB frame cancels the FOTO signal at low redshifts for the \pska example survey (where $\Delta \alpha$ approaches zero). However, it significantly amplifies the signal in all the other cases.
Additionally, the redshift correction alters the wavenumber at which the signal peaks (as $\relalpha$ and $\Delta \alpha$
follow different trends with redshift).
Finally, we highlight the remarkable agreement between the analytical predictions (cyan dashed lines) and the measured spectra.

In summary, the FOTO signal is not erased by correcting the galaxy redshifts to the CMB frame. The persisting C-FOTO signal combines the contributions from relativistic aberration and magnification. While the former can be easily corrected for (see appendix~\ref{app:aberration}), the latter 
is more difficult to handle. 
A possible option to completely cancel the FOTO effect could be to adopt a field-based approach in which the (redshift-dependent) evolution and magnification biases are first measured \citep[see e.g.][]{mag.bias_meas,Measurment_S_MAG} and then $\delta_\mathrm{com}$ is obtained by inverting Eq.~\eqref{eq:delta_velocities_rel_obs}.

\subsection{Can we boost the FOTO signal?}
\label{sec:boost}

Fig.~\ref{fig:corrected_spectra} shows that, in many cases, 
consistently transforming the galaxy redshifts to the CMB frame leads to an enhancement of the oscillatory features in $P_0(k)$ generated by the peculiar velocity of the observer.
This result suggests that similar transformations could
be used to enhance some features that are not apparent in the raw data. 
The idea goes as follows.
We pick an arbitrary velocity vector
$\bv{\rm art}$ (with modulus $v_\mathrm{art}\ll c$) and we transform all galaxy redshifts (but not their angular positions and luminosities) to
 the rest frame of
a fictitious observer that moves with peculiar
velocity $\bv{\rm art}+\bv{\odot}$ (e.g. the CMB rest frame considered in Eq.~(\ref{eq:redshift_correction}) and Sec.~\ref{Sec:Cancellation}
 is obtained by setting $\bv{\rm art}=-\bv{\odot}$):
\begin{equation}
\label{eq:redshift_boost}
    1+z_{\rm art} = \frac{1+z_{\rm hel}} {1+v_{\parallel, \rm art}/c}\,,
\end{equation}
where $v_{\parallel, \rm art} = \bv{\rm art}\cdot \bshat{r}$ (i.e. the redshift correction is direction dependent). 
At linear order, the galaxy overdensity of the redshift-corrected galaxies is
\begin{align}
    \delta_{\mathrm{B}}   &= \rmdel{com}  + \frac{\relalpha}{a\,H\,r} \,(\bv{\odot} \cdot \bshat{r})+  \frac{\kalpha}{a\,H\,r}(\bv{\rm art} \cdot \bshat{r}) \,,\label{eq:ART_DELTA_alpha}
\end{align}
which can be recast as
\begin{align}
    \rmdel{\mathrm{B}} - \rmdel{com} = \frac{\relalpha}{a\,H\,r}[(\bv{\rm art} + \bv{\odot})\cdot\bshat{r}] + \frac{\Delta \alpha}{a\,H\,r}(\bv{\rm art}\cdot\bshat{r})\,. 
\end{align}
Now, following  the procedure delineated in \citetalias{Elkhashab_2021}, we compute the power-spectrum monopole of $\delta_{\mathrm{B}}$ and we subtract $P_0(k)$ evaluated in the CRF, to obtain
 \begin{equation}
 \begin{split}
     &P_{0, \rm \mathrm{B}dip}(k;\bv{\rm art}) \equiv  P_{0,\mathrm{B}} (k;\bv{\rm art}) - P_{0,\rm com} (k) \approx \\  &\frac{16 \pi^2}{3H^2_0\int \bar{n}^2\,\dif^3 r}
     \left\{{\left|\bv{\odot}+\bv{\rm art}\right|^2}I_1^2(k) + 2{\left[(\bv{\odot}+\bv{\rm art})\cdot \bv{\rm art}\right]} \,
I_{1}(k)\,J_{1}(k) + {v^2_{\rm art}}\,
J_{1}^2(k)\right\}=\\
&\frac{16 \pi^2}{3H^2_0\int \bar{n}^2\,\dif^3 r}
\left\{v_\odot^2 \,I_1^2(k)+
2\,\bv{\odot}\cdot \bv{\rm art}\,I_1(k)[I_1(k)+J_1(k)]
+v_\mathrm{art}^2 [I_1(k)+J_1(k)]^2\right\}
\,,
 \end{split}
\label{eq:shifted_power_ang}
\end{equation}
that we denote as the boosted FOTO signal (B-FOTO in short). Three things are noteworthy. First, the term
proportional to $v_\odot^2$ in the last row of
Eq.~(\ref{eq:shifted_power_ang}) coincides with the original FOTO effect.
Second, due to the presence of the term $\propto  \bv{\odot}\cdot \bv{\rm art}$, the B-FOTO signal depends on both
the magnitude and the direction of $\bv{\odot}$. Third, the signal is influenced by the background cosmology, which determines the values of the integrals $I_1$ and $J_1$. Therefore, at least in principle,
the B-FOTO signal gives us a handle to measure
the peculiar velocity vector of the Sun and/or the
cosmological parameters by combining results obtained with different $\bv{\rm art}$. We will investigate this possibility in the remainder of this paper.

\begin{figure}%
\centering 
\includegraphics[width=1\textwidth]{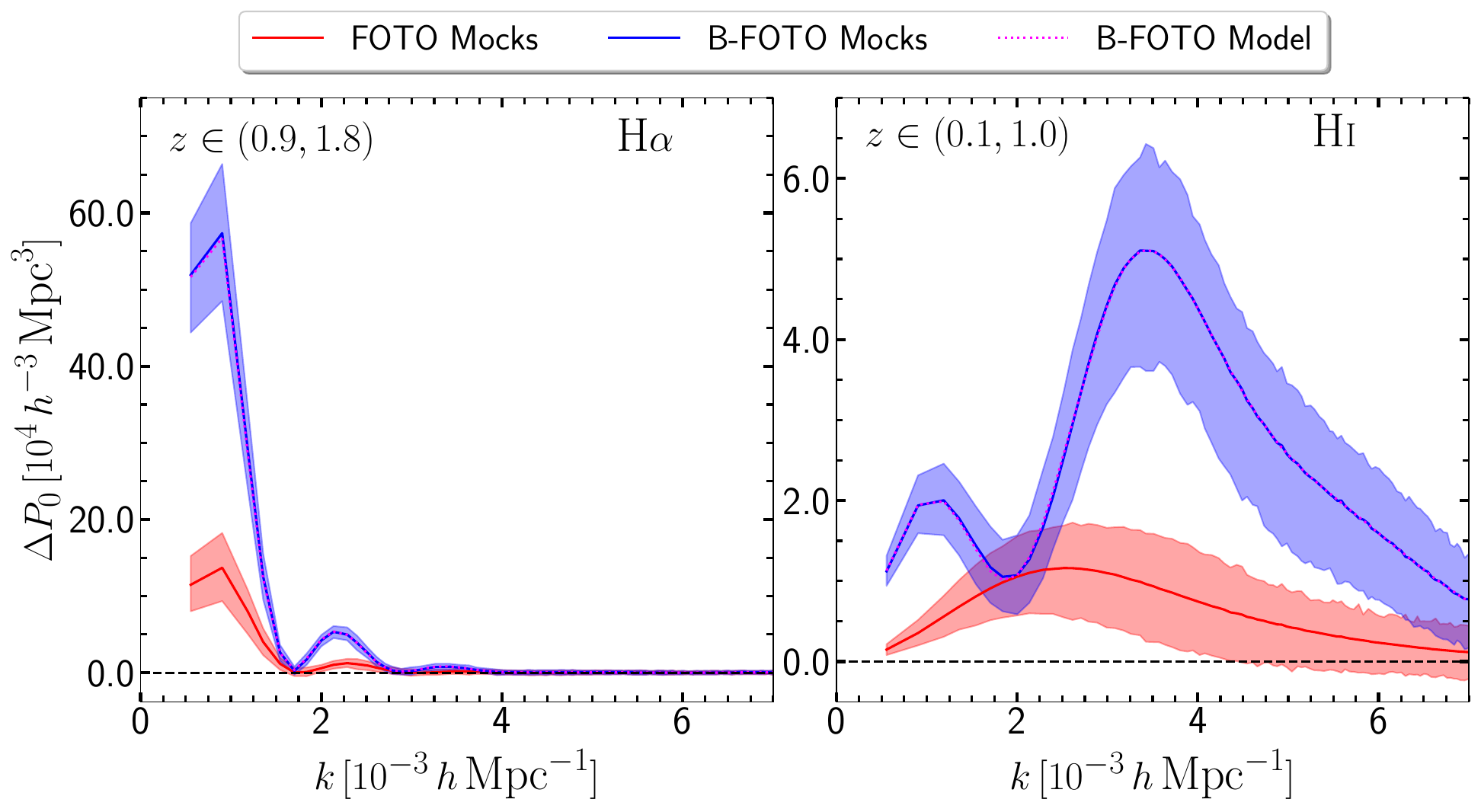}
\caption{\label{fig:BFOTO}%
Two examples of the B-FOTO signal obtained by
applying the redshift correction given in Eq.~(\ref{eq:redshift_boost}) to all the galaxies in a survey. Specifically, we present results for the widest redshift bins considered in Fig.~\ref{fig:SPECTRA_ORG}
for the \peuc and \pska example surveys 
and take $v_{\rm art} = 3 v_{\odot}$ with
$\bv{\mathrm{art}}\perp \bv{\odot}$.
The resulting mean B-FOTO signal (blue) and its scatter (light-blue shade) are overplotted to the corresponding sub-panels of Fig.~\ref{fig:SPECTRA_ORG}.
The magenta dotted lines (which almost perfectly overlap with the blue lines) display the theoretical predictions given in Eq.~\eqref{eq:shifted_power_ang}.} %
\end{figure}

For illustrative purposes,  in Fig.~\ref{fig:BFOTO}, 
we consider the widest
redshift bins of the \peuc and \pska
example surveys and compare the original FOTO effect (red) with the B-FOTO signal (blue) we obtain 
after picking $\bv{\odot}\cdot \bv{\rm art }= 0$ and $v_{\rm art} = 3 v_\odot$.
It is evident that the redshift correction
changes both the amplitude and the $k$-dependence of the oscillations. 
In this specific realisation,
the \snrt of $\Delta P_0$ increases from 6.8 (FOTO) to 22.6 (B-FOTO) in the \peuc case and from 3.4 to 9.1 for the \pska data. This means that the contribution
proportional to $v^2_\mathrm{art}$ in the last row of Eq.~(\ref{eq:shifted_power_ang}) is clearly distinguishable from the noise.
Since this term is sensitive to the cosmological parameters,
the B-FOTO effect should help us setting tighter constraints on $\Omega_{\mathrm m,0}$ as we will discuss in the next section.
The figure also shows that Eq.~(\ref{eq:shifted_power_ang}) provides an extremely
accurate model for the B-FOTO signal in a full-sky survey as the corresponding magenta dotted line
is hardly distinguishable from
the mean signal extracted from the mock catalogues (blue solid line).

\subsection{Can we derive cosmological constraints from the boosted FOTO signal?}
\label{sec:constraints}

In this section, we test the feasibility of extracting cosmological information from the \RBFS signal.
To this purpose, we extend the analysis performed in Sec.~\ref{Sec:MCMC_FOTO} and consider the same
two scenarios.
The main difference here is that measuring the \RBFS effect allows us to perform inference 
on both the magnitude and the direction of $\bv{\odot}$.
The central idea is as follows.
The last row of Eq.~(\ref{eq:shifted_power_ang}) shows that the B-FOTO signal receives three distinct contributions. The first term on the right-hand side is the standard FOTO effect whose constraining power we have explored in Sec.~\ref{Sec:MCMC_FOTO}.
The third term is not sensitive to $\bv{\odot}$ and only depends on the cosmological
parameters.
The key ingredient in our discussion is the second term which can be used to set constraints on  the scalar product $\hat{\bs{v}}_{\odot}\cdot
    \hat{\bs{v}}_\mathrm{art}$. Given a credible level,
    we can thus confine $\hat{\bs{v}}_{\odot}$ within a spherical cone around $\pm\hat{\bs{v}}_\mathrm{art}$.
    Obviously,
this localisation becomes more precise if one combines 
results obtained using $\bv{\rm art}$ vectors pointing
in different directions. 
We characterize
unit velocity vectors
using galactic coordinates with latitude $b$ and longitude $l$ so that
\begin{equation}
    \hat{\bs{v}}_{\odot}\cdot
    \hat{\bs{v}}_\mathrm{art}\equiv \cos \psi =
    \cos b_\odot \cos b_\mathrm{art}
\cos (l_\odot-l_\mathrm{art})+\sin b_\odot \sin b_\mathrm{art}\,.\end{equation}
Since the B-FOTO signal depends on $\cos \psi$,
it is necessary to vary $b_\mathrm{art}$ and $l_\mathrm{art}$ to constrain $\hat{\bs{v}}_{\odot}$.

Contrary to Sec.~\ref{Sec:MCMC_FOTO}, where we focused on the \peuc mocks, here we also consider the
\pska survey.
In particular, we set constraints on the peculiar velocity of the Sun
using the redshift range $z\in (0.1, 1.0)$ and employing six bins covering the interval $k\in(0.39, 8.9) \times 10^{-3}\, h\, \mathrm {Mpc}^{-1}$.
On the other hand, we constrain the cosmological parameters using the narrower redshift range
$z\in (0.8, 1.0)$ and five bins within $k\in(0.39, 5.1) \times 10^{-3}\, h\, \mathrm {Mpc}^{-1}$. 
For the \peuc survey, we use the same redshift range and binning strategy
as in Sec.~\ref{Sec:MCMC_FOTO}.

\begin{table}[tbp]
\centering
\begin{tabular}{|c| c c c c c|}
\hline
&%
$\bs{v}_{1,\rm art}$&$\bs{v}_{2,\rm art}$&$\bs{v}_{3,\rm art}$&$\bs{v}_{4,\rm art}$&$\bs{v}_{5,\rm art}$
\\
\hline
$v_{\rm art}/v_{\odot,{\mathrm{true}}}$& $3$&$1$&$2$&$2$&$3$\\
$b_{\rm art}$ & $+40^\circ$ & $-23^\circ$ & $-48^\circ$  & $+48^\circ$  &  $-17^\circ$ \\
$l_{\rm art}$ & $304^\circ$& $154^\circ$ & $84^\circ$& $264^\circ$ & $333^\circ$ \\
$\psi$&${30^\circ}$&${120^\circ}$&$180^\circ$&$0^\circ$&${90^\circ}$\\
\hline
\end{tabular}
\caption{\label{tab:vel}The five velocity vectors we used to measure the B-FOTO signal and perform Bayesian inference of the model parameters.}
\end{table}
We present results obtained combining mock data obtained with five distinct $\bv{\rm art}$ vectors pointing in different directions and having different magnitudes (see Table~\ref{tab:vel}). This gives us a total of 25 or 30 data points for each inference run.
We have tested that using fewer shifts degrades the final constraints. On the other hand, 
 we do not use more shifts in order to avoid generating substantial biases and random errors in the numerically estimated inverse covariance matrix (i.e. to make sure
 that the dimension of $\mcov$ is substantially smaller than the number of HRF catalogues used to estimate it).
Our tests also reveal that adopting values of $v_{\mathrm{art}}\gg v _{\odot,\mathrm{true}}$ weakens the constraints on the model parameters (probably because the right-hand side  of Eq.~(\ref{eq:shifted_power_ang}) is dominated by the last term in that case). 
For this reason, we only consider $v_{\mathrm{art}}\leq 3\, v_{\odot,\mathrm{true}}$ in our example.
Finally, we point out that we do not try to optimize the batch of $\bv{\mathrm{art}}$ vectors so that to minimize the size of the
resulting credible intervals, although this is likely possible (for instance with a sequential procedure in which the next $\bv{\mathrm{art}}$ to use is picked based on the results obtained that far). In fact, we use the very same set of $\bv{\mathrm{art}}$
for both inference scenarios.

\subsubsection{Velocity inference}
\begin{figure}
    \centering
    \includegraphics[width=.8\textwidth]{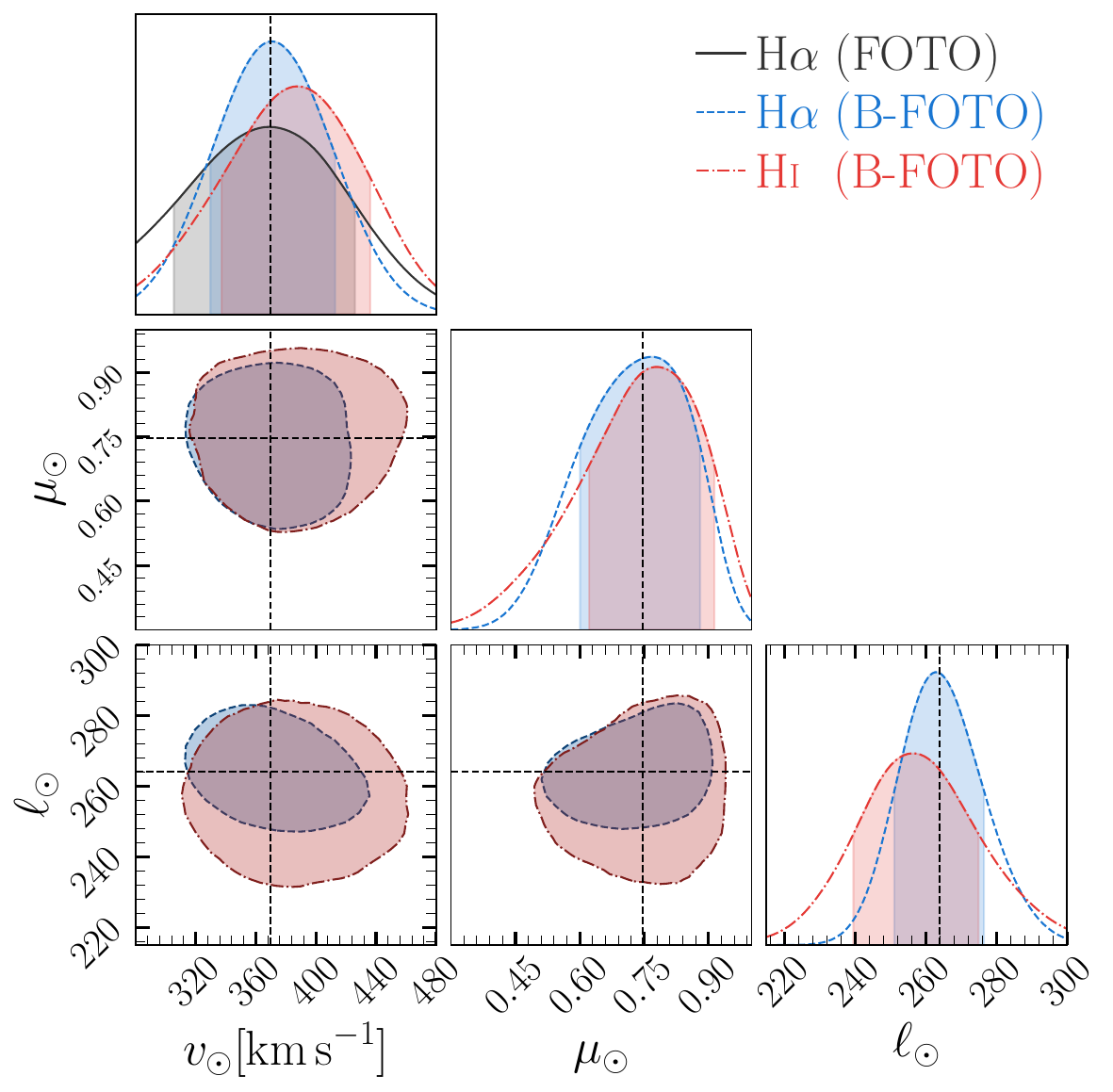}
    \caption{Posterior distribution for the velocity inference based on the B-FOTO effect {from individual realisations of the \peuc (blue, obtained using the same mock catalogue as in Fig.~\ref{fig:Corner_V_NO_SHIFT}) and \pska (red) surveys.} The off-diagonal panels show the contour levels for the marginalized 2D posterior corresponding to the 68\% credible regions for the model parameters. The diagonal panels display the 1D distributions (lines) and the corresponding 68\% HDPIs (shaded areas). In the top-left panel, we also reproduce the posterior for $v_\odot$ from the (unboosted) FOTO signal (already shown in the left panel of Fig.~\ref{fig:Corner_V_NO_SHIFT}).}
    \label{fig:POST_ONE}
\end{figure}
To set constraints on the peculiar velocity of the Sun, we use three model parameters
$\bs{\theta} = \{v_{\odot},\, \mu_\odot ,\,l_{\odot}\}$
with $\mu_\odot = \cos (90^\circ-b_\odot)$.
For the angular variables, we adopt uniform priors within the ranges $\mu_\odot \in [-1,1]$ and $l_{\odot}\in[0^\circ,360^\circ)$. For $v_\odot$, we use the same prior as in Sec.~\ref{Sec:MCMC_FOTO}.

\begin{figure}
    \centering
    \includegraphics[width=.7\textwidth]{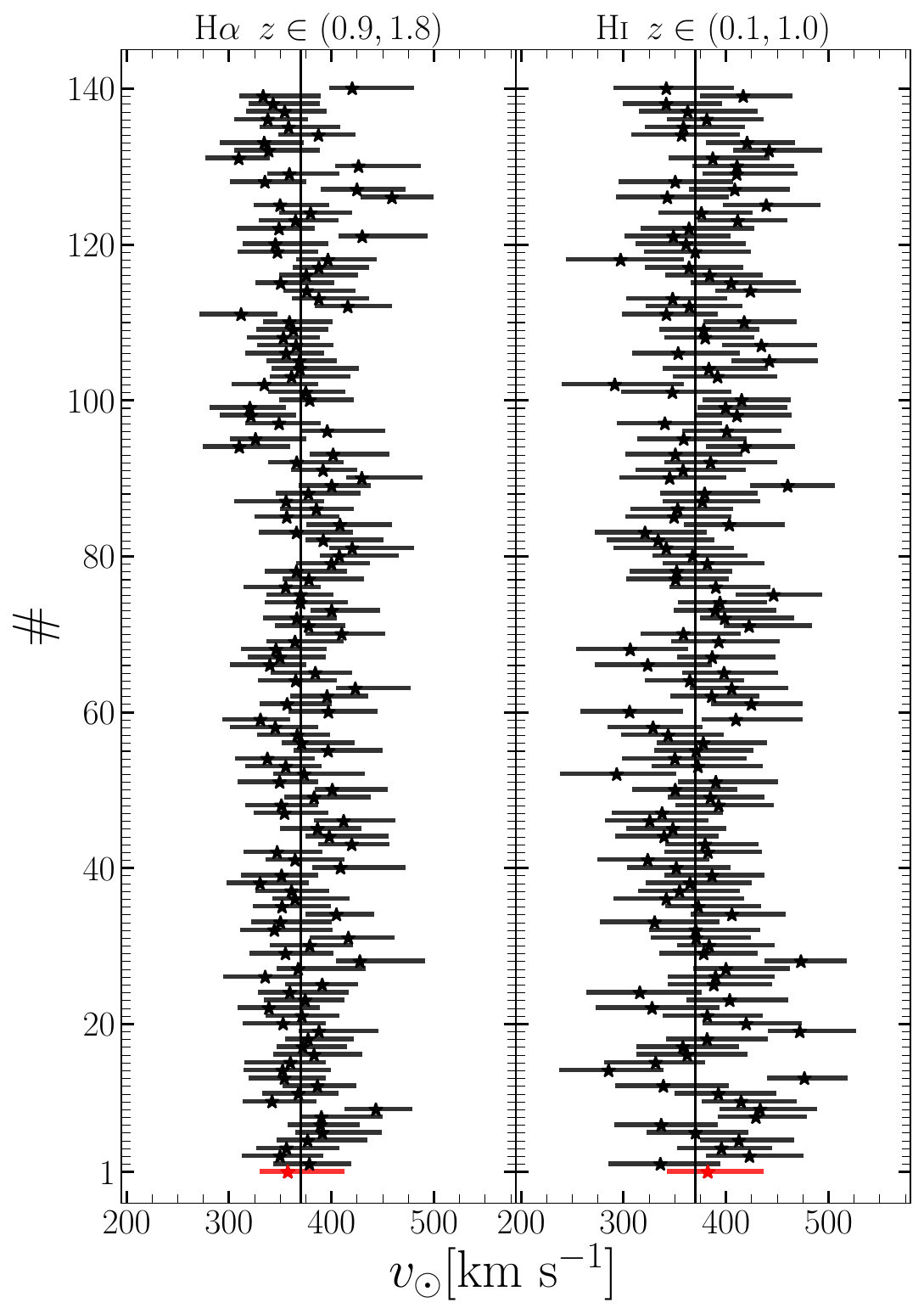}
    \caption{The 68\%  HPDIs for $v_\odot$ (horizontal bars) extracted from the \RBFS data are shown for all  the HRF mock
    surveys. The left and right panels refer to 
    the \peuc and \pska surveys, respectively. The stars indicate the different posterior modes and the vertical black line shows the underlying true value. The red bars highlight the realizations used in Fig.~\ref{fig:POST_ONE}.
}
    \label{fig:BARV}
\end{figure}
\begin{figure}
    \centering
    \includegraphics[width=0.73\textwidth]{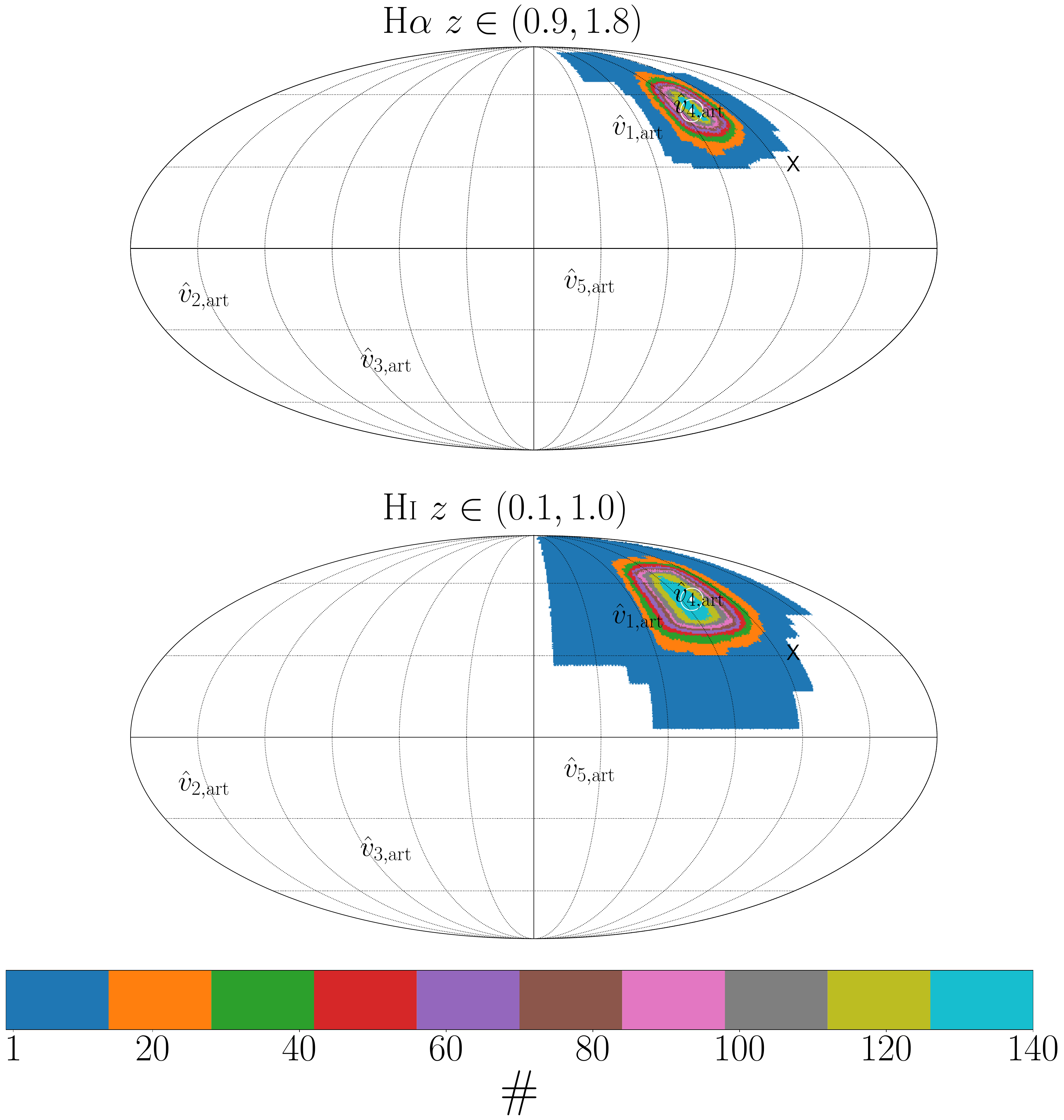}
    \includegraphics[width=0.8\textwidth]{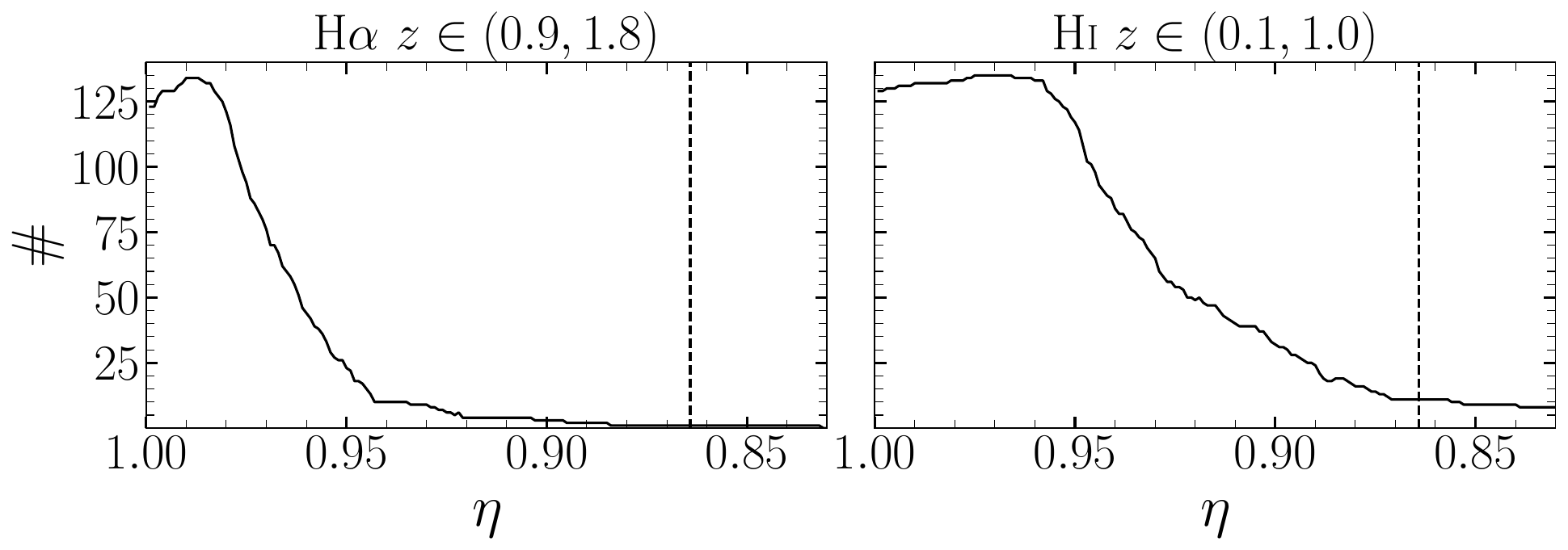}
    \caption{Top: Mollweide projection of the celestial sphere in galactic coordinates illustrating the constraints that the B-FOTO effect can set on $\hat{\bs{v}}_\odot$. The colour coding indicates
    the number of realisations in which a given direction
    lies within the 68\% credible region. The empty circle displays $\hat{\bs{v}}_{\odot,\mathrm{true}}$ while 
    the best-fit direction inferred with the CatWISE quasar catalogue  is marked with a cross.
    The directions in which the five velocity vectors listed in Table~\ref{tab:vel} point are indicated by the corresponding labels.
    The top and bottom sub-plots refer to the mock \peuc and \pska surveys, respectively. 
    Bottom: The number of realisations that include a particular value of $\eta$ in the 68\% HPDI regions for the 
 derived parameter $\bshat{v}_{\odot,\text{true}}\cdot\bshat{v}_{\odot}(\mu_\odot, l_\odot)$. 
 The CatWISE value is indicated with the dashed line.
Note that this plot compresses the information in the maps by averaging over spherical circles.
    }
    \label{fig:ANGLE_PLOTS}
\end{figure}

As an example,
in Fig.~\ref{fig:POST_ONE}, we present the one- and two-dimensional marginal posterior distributions obtained
analysing one of our mock HRF catalogues for both the
\peuc (blue) and \pska (red) surveys.
The 68\% credible regions nicely encompass the true values and show that the \RBFS effect can
constrain $v_\odot$ to better than 11\% (\peuc) and
13\% (\pska). \footnote{For the \peuc survey, we pick the same realisation used in Fig.~\ref{fig:Corner_V_NO_SHIFT}. To ensure a fair comparison, for the \pska survey, we consider the galaxy catalogue associated with the same dark-matter lightcone.} %
For reference, 
we also show the posterior probability of the velocity magnitude from the  (un-boosted) FOTO signal for the \peuc survey (grey, already shown in  
Fig.~\ref{fig:Corner_V_NO_SHIFT}).
It is evident that using the \RBFS technique not only allows
us to determine the direction of $\bv{\odot}$ but
also slightly enhances the constraints on its magnitude.

In Fig.~\ref{fig:BARV}, we plot the 68\% HDPI for $v_\odot$ obtained from the individual 140 mocks. 
There is no evidence of any important bias and
the average
half-size of the HDPI is $\overline{\Delta v_\odot}\simeq 39$ (10.5\% relative error) and 50 \kms (13.5\%)
for the \peuc and \pska example surveys, respectively.
This difference is related to the variations of $\relalpha$ between the selected galaxy populations that determines the strength of the FOTO signal.

In Fig.~\ref{fig:ANGLE_PLOTS},  we present the constraints we obtain on $\hat{\bs{v}}_{\odot}$ from the 140 HRF realisations in a combined way. The empty circle indicates
the true velocity used to build the mock catalogues and
the colour coding represents the number
of realisations in which a given pixel lies within
the 68\% credible region of the Bayesian inference based on the B-FOTO effect. 
Clearly, nearly all realisations include the true
$\hat{\bs{v}}_{\odot}$ in their highly-credible region. 
For reference, we also show with a black cross
the best-fit direction of the dipole in the number counts of the CatWISE quasar catalogue measured
by \citet{Secrest_2021_dipole_quasers}.
The fact that not a single realisation includes the cross in their 68\% credible region shows
that the B-FOTO method can undoubtedly tell it apart from the actual peculiar velocity of the Sun (at least in a full-sky survey). 
We finally note that the \peuc survey has a stronger constraining power, as we already pointed out for $v_\odot$.

In order to present our results in a simpler and more quantitive way, we look at the offset between the direction of the true
peculiar velocity of the Sun and that 
characterised by the model parameters $(\mu_\odot, l_\odot)$. We thus use the variable
\begin{equation}
\label{eq:eta_off}
\eta =\bshat{v}_{\odot,\text{true}}\cdot\bshat{v}_{\odot}(\mu_\odot, l_\odot)\,,
\end{equation}
as a derived variable in our MCMC runs
and obtain posterior samples for it.
In the bottom panel of  Fig.~\ref{fig:ANGLE_PLOTS},
we plot the number of realisations which include
a given value of $\eta$ within the 68\% HDPI.
The counts peak 
at $\eta=0.987$ (corresponding to an angle of approximately $9^\circ$) for the \peuc survey 
and at 
$\eta=0.966$ ($15^\circ$) for the \pska survey. 
These values give an estimate of the typical uncertainty 
which is achieved in the measurement of $\hat{\bs{v}}_\odot$.

In summary, our analysis indicates that we can measure both the magnitude and the direction of $\bv{\odot}$ using the \RBFS technique. The constraining power of the method increases with the amplitude of the FOTO signal,
which scales as the second power of the parameter $\relalpha$\ for the galaxy population under study. 

\subsubsection{Density inference}\label{subsec:matter}
\begin{figure}
    \centering
    \includegraphics[width=0.5\textwidth]{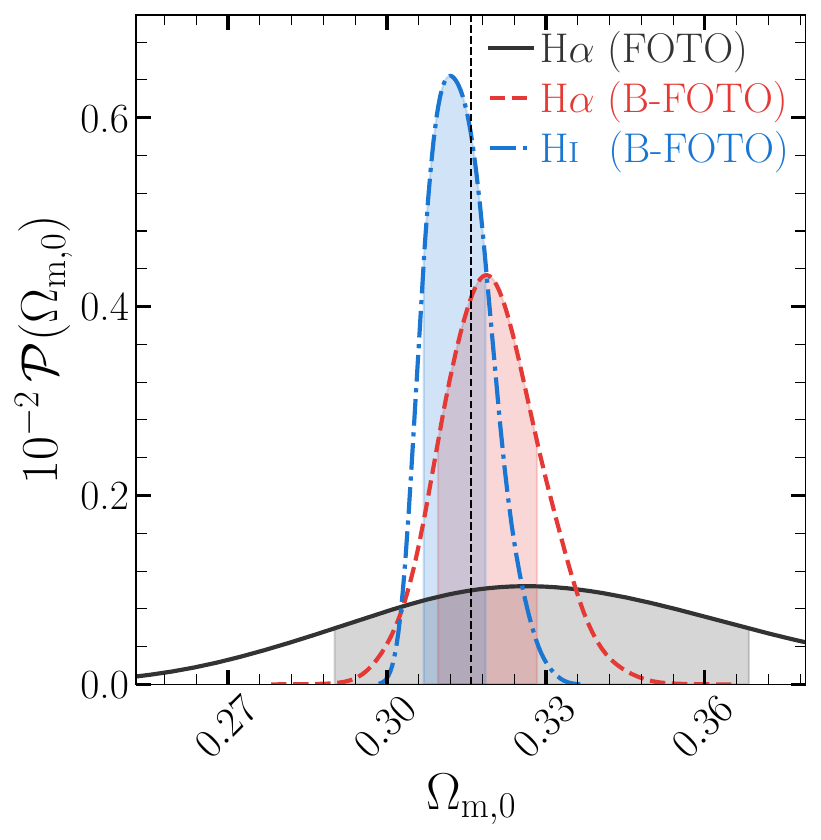}%
    \caption{ {The marginalised posterior distribution of $\Omega_{\rm m,0}$ obtained by fitting the \RBFS signal from one realisation 
    of the \pska (blue, same mock as in Fig.~\ref{fig:Corner_V_NO_SHIFT}) and  \peuc (red) surveys. The posterior derived from the (unboosted) FOTO signal is also reproduced from Fig.~\ref{fig:Corner_V_NO_SHIFT} (grey).} 
    } 
    \label{fig:Corner_m}
\end{figure}

\begin{figure}
    \centering
    \includegraphics[width=0.7\textwidth]{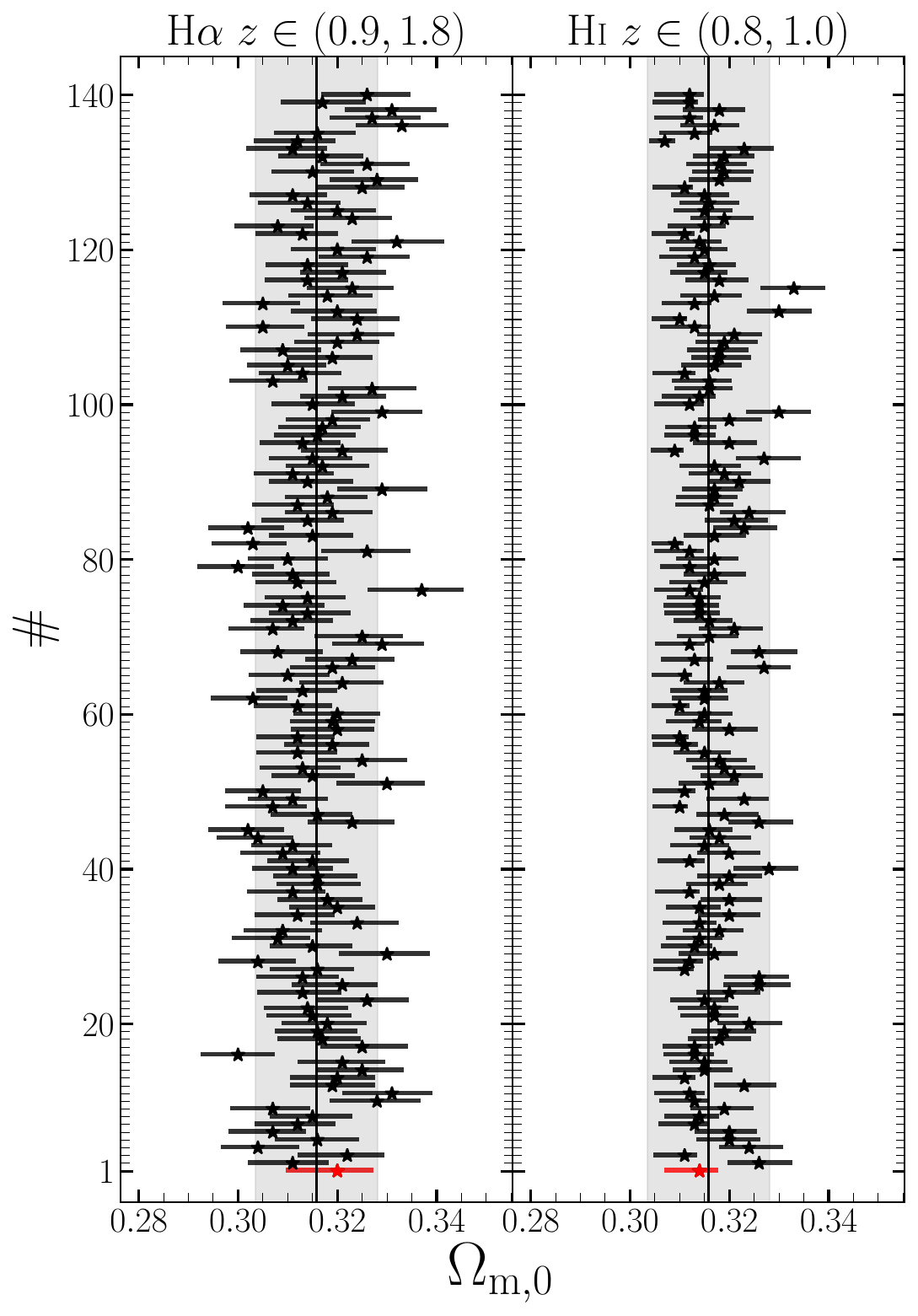}%
    \caption{As in Fig.~\ref{fig:BARV} but for $\Omega_{\rm m,0}$. For reference, we also show  the current CMB constraints \citep{planck18} as a grey band centred around the true value. The red bars emphasize the realisations presented in Fig.~\ref{fig:Corner_m}.
    } 
    \label{fig:HUBB}
\end{figure}

 \begin{figure}
    \centering
    \includegraphics[width=0.6\textwidth]{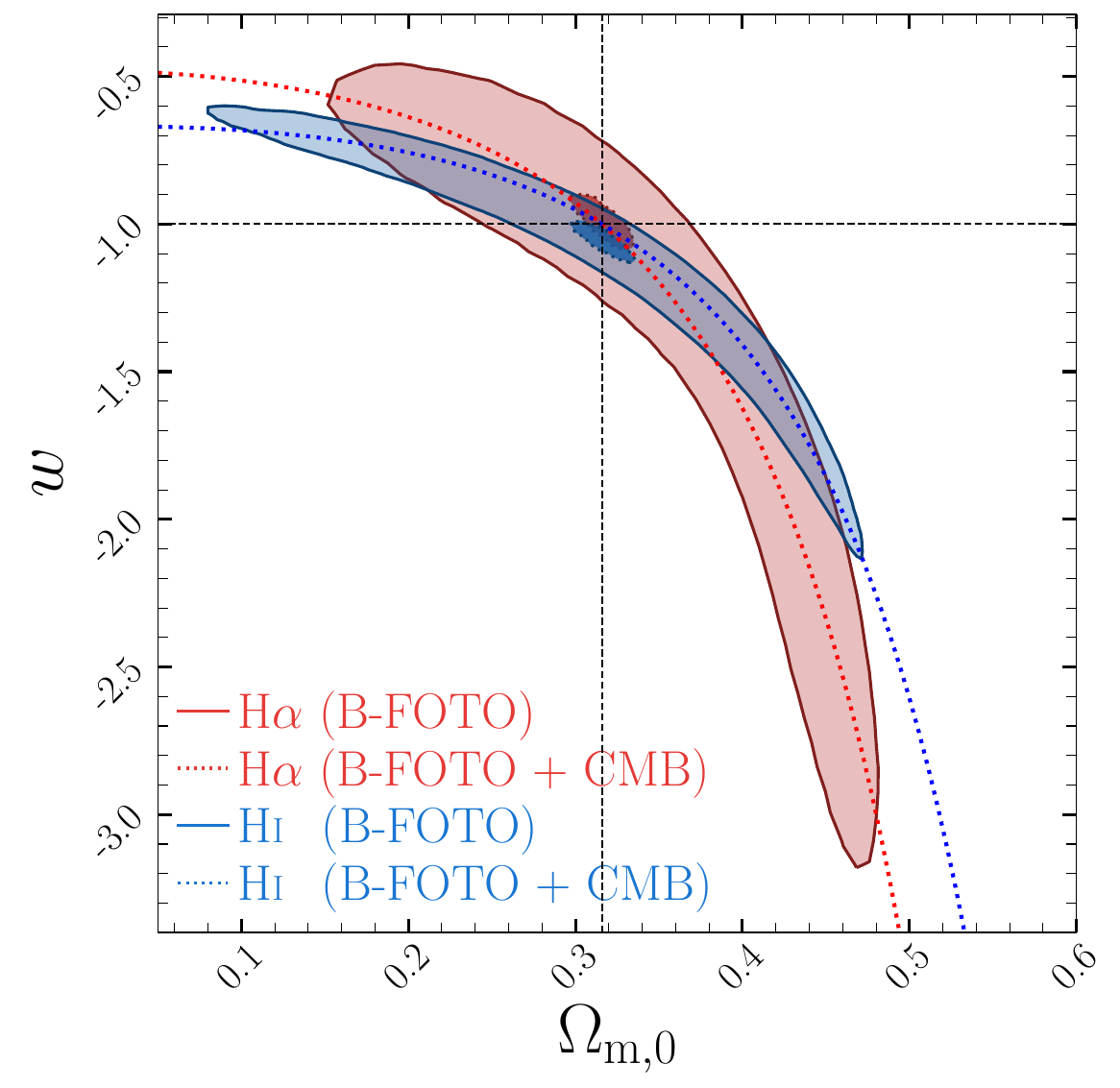}%
    \caption{ The marginalised posterior distribution for $\Omega_{\rm m,0}$ and $w$ for the realizations of the \peuc (light red) and \pska (light blue) example surveys that we  that have used in Fig.~\ref{fig:Corner_m}.
    The 68\% joint credible regions from the B-FOTO
    signal are indicated with solid contours. The darker shaded regions show how they shrink if we adopt a Planck prior on $\Omega_\mathrm{m,0}$.
    The equation $w = \Sigma\,\{\,[1-(0.3)^{\gamma}]/[1-\Omega_{\rm m,0}^{\gamma}]\}^7$ with $\gamma=2$ and 2.5 (which approximates the degeneracy direction of the contours) is plotted with dotted lines.      
    }
    \label{fig:Corner_m_w}
\end{figure}
\begin{figure*}[]

    \centering
    \begin{subfigure}[b]{0.45\textwidth}
        \centering
    \includegraphics[width=1\textwidth]{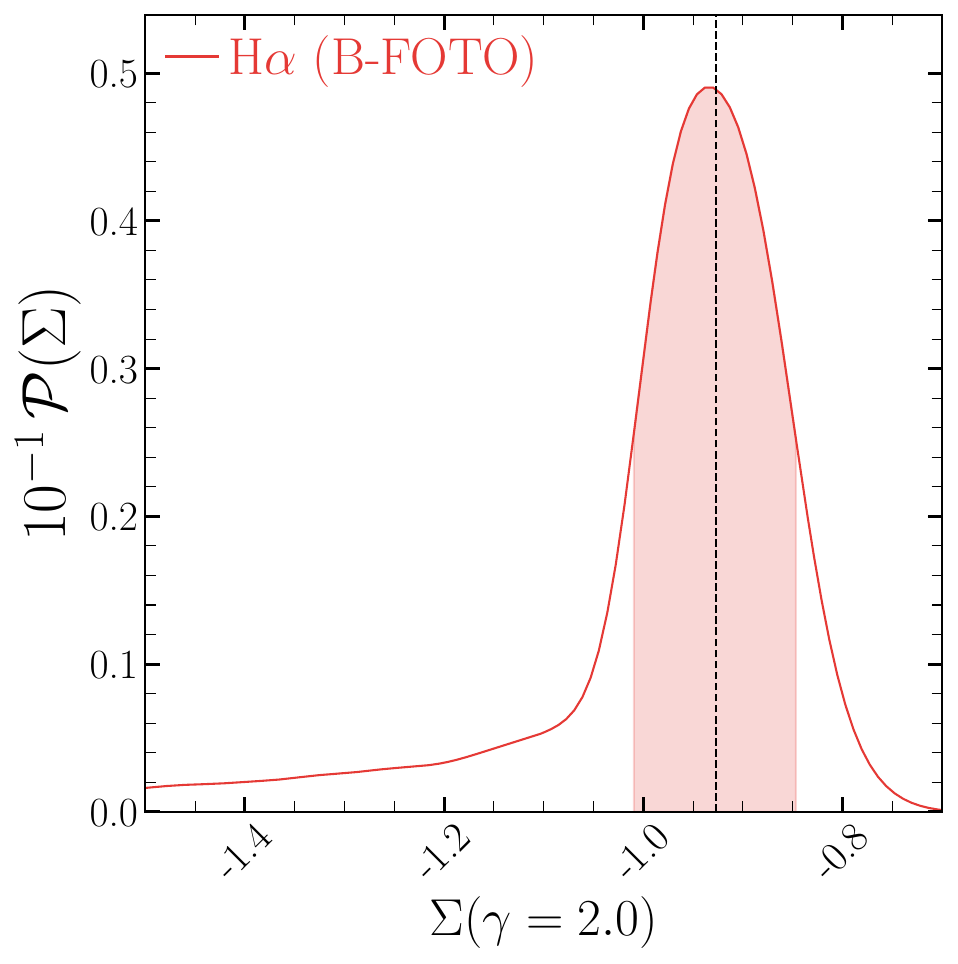}%
    \end{subfigure}%
    ~ 
    \begin{subfigure}[b]{0.45\textwidth}
        \centering
    \includegraphics[width=1\textwidth]{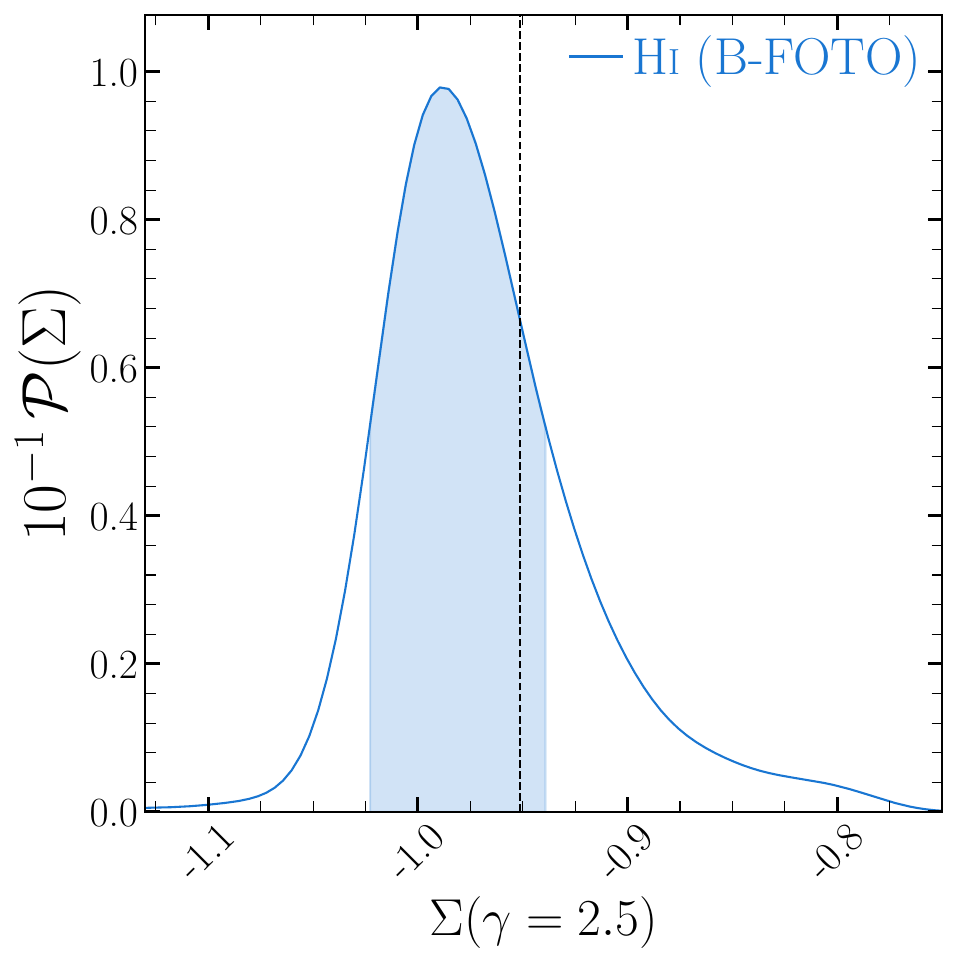}%
    \end{subfigure}%
    \caption{\label{fig:corner_gamma} The posterior distribution for the derived parameter $\Sigma = \{[1-\Omega_{\rm m,0}^{\gamma}]/[1-(0.3)^{\gamma}]\}^{7}\,w$}. 
\end{figure*}

We repeat the density inference described in Sec.~\ref{Sec:MCMC_FOTO} using the B-FOTO data.
We start by considering the same mock catalogues 
used to prepare  Fig.~\ref{fig:POST_ONE} 
(and, for the \peuc survey only, also Fig.~\ref{fig:Corner_V_NO_SHIFT})
and present the resulting marginalised posterior distribution for $\Omega_{\rm m,0}$ in
Fig.~\ref{fig:Corner_m}.
Comparing it to what we have obtained from the FOTO effect, it is evident that the redshift corrections lead to notably tighter constraints.
This conclusion is confirmed by the analysis of 
the 68\% HDPIs from all realisations displayed
in Fig.~\ref{fig:HUBB}. On average, we get
$\overline{\Delta \Omega}_{\rm m, 0}=  0.0084$ from the \peuc survey ($2.7\%$ relative error, a factor of 4.3 better than obtained from the FOTO signal)  and 
$\overline{\Delta \Omega}_{\rm m, 0}=  0.0057$ ($1.8\%)$ for the \pska survey.
Taken at face value, these figures
are competitive with the
current constraints coming from CMB studies \citep{planck18} (grey shaded band). However,
it is important to mention once again that the constraints we derived here are certainly optimistic, as they:
(i) assume a full-sky survey, (ii)
rely on measuring clustering at very large scales
which is prone to systematic errors 
\citep[e.g.][]{Paviot_2022}, and (iii) presume that we perfectly know the
evolution and magnification bias parameters of
the galaxy population under study as a function of redshift, %
On the other hand, the constraining power could
be further improved by considering higher-order multipoles -- see Eq.~(\ref{eq:Yam_dip_result_TOTAL}).

\subsubsection{Constraints on the dark-energy equation of state.}
\begin{figure}
    \centering
    \includegraphics[width=0.72\textwidth]{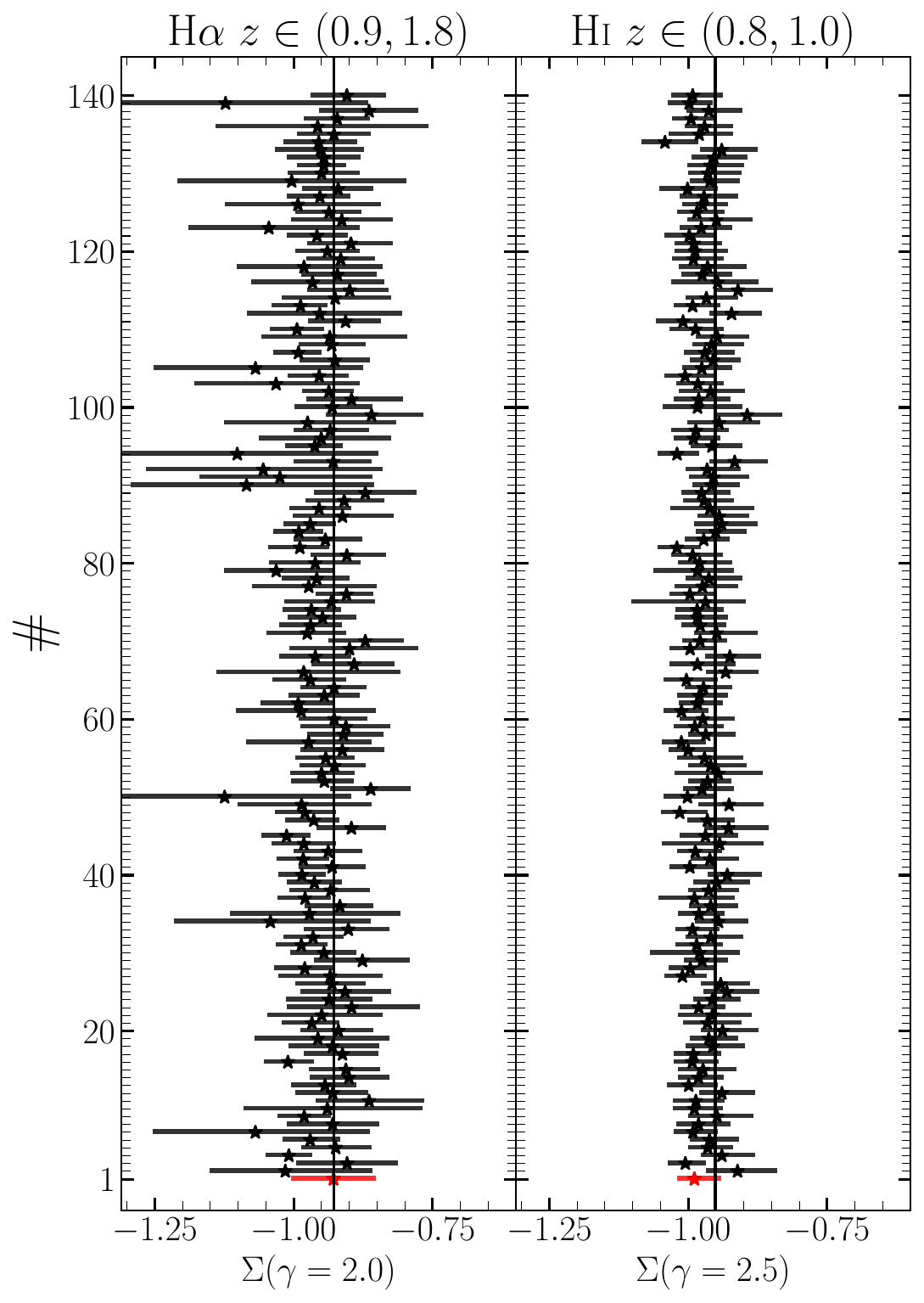}%
    \caption{ Similar to Fig.~\ref{fig:BARV}, but for the constraints on the derived parameter  $\Sigma = \{[1-\Omega_{\rm m,0}^{\gamma}]/[1-(0.3)^{\gamma}]\}^{7}\,w$. The red bars represent the \peuc and \pska survey constraints shown in Fig.~\ref{fig:corner_gamma}.
         }
    \label{fig:GAMMA}
\end{figure}
Given the impressive constraints on $\Omega_{\mathrm m,0}$ we obtained  in the previous section, we relax one of our assumptions
and allow the ratio $H(z)/H_0$ to depend on an additional cosmological parameter.
Namely, we consider flat $w$CDM models in which the
equation-of-state parameter for dark energy $w$
does not evolve with time so that
$H(z)/H_0=[\Omega_{\mathrm m,0}(1+z)^3+(1-\Omega_\mathrm{m,0})^{3(1+w)}]^{1/2}$.
In this scenario, we perform inference on the
model parameters
$\bs{\theta}=\{\Omega_{\rm m,0},w,v_{\odot},\mu_\odot,l_{\odot}\}$. We adopt the same priors as in the density inference, combined with a uniform prior for $w$ in the range $[-6,-1/3]$.

In Fig.~\ref{fig:Corner_m_w},
we present the joint marginalised posterior distribution on $\Omega_{\rm m,0}$ and $w$ extracted from the two realisations we have constantly used as an example in the previous sections. 
Results from the \peuc survey are illustrated in red, while those from the \pska survey are shown in blue. 
It is evident that the B-FOTO signal sets constraints
on a degenerate combination of the parameters.
The banana-shaped iso-probability contours in Fig.~\ref{fig:Corner_m_w} 
approximately stretch along the curves of equation
$w = \Sigma\,\{\,[1-(0.3)^{\gamma}]/[1-\Omega_{\rm m,0}^{\gamma}]\}^7$ with $\gamma = 2.0$ for the \peuc survey ($\Sigma = -0.9274$) and $2.5$ for the \pska survey ($\Sigma = -0.9512$).
\footnote{Note that 
the \pska sample better constrains $w$ (relatively
to the \peuc one)
while the latter better constrains  
$\Omega_{\rm m,0}$. This difference (which reflects the distinct values of $\gamma$) arises 
from the redshift ranges covered by the two data sets.
The \peuc sample contains galaxies that emitted their light when the Universe was still matter-dominated, whereas the specific \pska sample we are considering here lies beyond
the matter-dark energy equality (which takes place
at $z\simeq 1.29$ in our mocks).
In general, we expect $\gamma$ to decrease as the characteristic redshift of a data set grows.}
Therefore, the B-FOTO effect should be able to set tight constraints on the derived parameter, $\Sigma = \{[1-\Omega_{\rm m,0}^{\gamma}]/[1-(0.3)^{\gamma}]\}^{7}\,w$, %
which varies in the
orthogonal direction.
In Fig.~\ref{fig:corner_gamma}, we present the posterior distributions for $\Sigma$ obtained from our MCMC runs.  As expected, $\Sigma$ is  more tightly constrained from the mock data than the original
parameters $\Omega_\mathrm{m,0}$ and $w$. The 68\% credible regions include the true values in each survey and show that the \pska and \peuc surveys  result in half-width HPDIs of $\Delta \Sigma =0.077$ (8.2\%) and $0.040$ (4.3\%), respectively. 

We present the results for all our realizations in Fig.~\ref{fig:GAMMA}. %
 We find that the average half-size of the HPDI is $\overline{\Delta \Sigma} = 0.09$ (9.1\%)  for the \peuc  survey and 0.05 (5.2\%) for the \pska. The posterior mode is always rather
close to the true value which, in most cases (96\% for the \peuc survey and 84\% for \pska), lies
within the 68\% HDPI. 

Obviously, the degeneracy between $\Omega_\mathrm{m,0}$ and $w$ can be broken by combining the B-FOTO measurements with other data. For instance, in Fig.~\ref{fig:Corner_m_w}
we also show the 68\% credible region obtained
by considering the Planck prior on $\Omega_\mathrm{m,0}$ which leads to 
rather tight constraints on $w$ (i.e. $\Delta w\simeq 0.15)$.%

\section{Summary}
 \label{Sec:conc}

Our peculiar motion with respect to the Hubble flow imprints a distinctive dipolar pattern on the galaxy densities we construct from redshift surveys. This feature receives contributions from
(i) the Doppler shift due to the radial component of the velocity, (ii) the relativistic aberration sourced by the transversal component, and (iii) Doppler magnification (i.e. the velocity contribution to the weak-lensing convergence appearing when the observed redshift is used as a distance indicator).
The resulting dipole superimposes an oscillatory component on the observed power spectrum multipoles at ultra-large scales, which we term the FOTO effect. 
In Section~\ref{sec:FOTO} and  Appendix \ref{APP:FOO_MULTI}, we derive the exact 
expression for the oscillatory terms 
contributing to each multipole
in the case of a full-sky survey,
see Eq.~(\ref{eq:Yam_dip_result_TOTAL}).

Using the \liger method, in Sec.~\ref{sec:MOCKS_DESCRIP}.
we build different suites of mock full-sky galaxy catalogues
that incorporate relativistic redshift-space distortions at linear order. 
The galaxy populations in the mocks mimic those expected
in ongoing and future \peuc- and \pska-selected surveys.
For each type of survey, we build 
two suites of 140 mocks each that
differ only in the peculiar velocity of the observer. In one case, the observer has no peculiar motion while, in the other, lies in the heliocentric rest frame (as determined by CMB studies).

In Sec.~\ref{sec:MOCKS_MEASURMENTS_ALL}, we 
validate our mocks by showing that
the average FOTO signal in their power-spectrum monopoles
is in spectacular agreement with the exact theoretical
prediction (Fig.~\ref{fig:SPECTRA_ORG}).
We then move on to investigating the detectability of the FOTO signal in the individual \peuc mocks.
Assuming that observational systematics can be controlled on the largest scales,
we find that, on average, the oscillatory FOTO signal can be identified with 
a signal-to-noise ratio of 6.8 in a full-sky survey
which drops to 4 for currently realistic sky cuts (Fig.~\ref{fig:S_N}).
We also verify that redshift measurement errors
do not degrade the statistical significance of the detection. 

The FOTO effect depends on three factors:
our peculiar velocity, the expansion history of the Universe (more precisely, the ratio $H(z)/H_0$), and
the galaxy population under study (in particular,
its magnification- and evolution-bias parameters).
Since we have demonstrated that the FOTO signal is (at least potentially) detectable with ongoing redshift surveys, it makes sense to investigate what cosmological
information we could extract from it.
To this end, we devise two scenarios. In the first, 
we use the FOTO signal to test the
kinematic interpretation of the CMB dipole by measuring $v_\odot$ with Bayesian inference. This requires setting tight priors on the remaining cosmological parameters from other probes. In the second, we stick to the kinematic interpretation and use the CMB measurements as a prior for $v_\odot$ in order to
constrain the expansion history of the Universe, which, in a $\Lambda$CDM model, only depends on the
present-day value of the matter density parameter.
As a proof of concept, we simplify both scenarios
by assuming to know precisely the
parameters that describe the galaxy populations.
The resulting posterior probabilities for the \peuc full-sky survey are shown in 
Fig.~\ref{fig:Corner_V_NO_SHIFT}.

In Sec.~\ref{Sec:Cancellation}, we demonstrate both analytically and numerically that it is not possible to cancel the FOTO effect by correcting the individual galaxy redshifts to the matter rest frame (as done, for instance, in studies of the Hubble diagram with type Ia supernovae). This is because the redshift correction
only takes care of one of the three contributions to the dipole, leaving relativistic aberration and Doppler magnification unaddressed.
Actually, an approximate cancellation takes place for the mock \pska data at redshifts $z<0.4$ as the residual dipole has a very low amplitude in this case.  For higher redshifts (and for the \peuc mocks), however, the FOTO signal is actually enhanced by the redshift correction (Fig.~\ref{fig:corrected_spectra}).
The only option we see to fully subtract the FOTO effect is to operate at the field level.

Capitalizing on these findings, 
in Sec.~\ref{sec:boost}, we present a method 
that uses different redshift corrections to
simultaneously boost the oscillatory signal and introduce a dependence on the direction of the observer velocity. The basic idea is to transform the individual galaxy redshifts of a catalogue (but not their angular positions and luminosities) to the rest frame of a fictitious observer that moves with a pre-selected peculiar velocity vector.
The resulting power-spectrum monopole
differs from the one measured by a comoving observer because of three additive corrections -- see Eq.~(\ref{eq:shifted_power_ang}) -- whose sum we 
dub the \RBFS signal (short for boosted FOTO).
Two of them are the standard FOTO signal and an analogous term depending on the square modulus of the fictitious velocity, the third one, instead, scales with the scalar product of the true observer velocity and the fictitious one.

 Finally, in Sec.~\ref{sec:constraints}, we investigate
the constraining power of the \RBFS signal. We first consider the same two scenarios presented in Sec.~\ref{Sec:MCMC_FOTO} but with the additional possibility of setting constraints on the direction of our peculiar velocity.
 For this study, we combine the measurements obtained with five different redshift corrections.
 Our results for the first scenario, presented in Figs.~\ref{fig:BARV} and~\ref{fig:ANGLE_PLOTS}, show that a full-sky survey free of systematics can determine the direction of our peculiar velocity to $10^\circ$ and its amplitude to $10\%$ accuracy.
 In the second scenario, we find that the constraints on $\Omega_{\rm m,0}$ are competitive with the CMB as shown in Fig.~\ref{fig:HUBB}.
 At last, we show that, in a $w$CDM model, the \RBFS effect can set tight constraints on a non-linear combination of the matter-density parameter and the equation-of-state parameter for dark energy.

\section*{Acknowledgments}
This paper is supported by the PRIN 2022 PNRR project "Space-based cosmology with Euclid: the role of High-Performance Computing" (code no. P202259YAF), funded by European Union – Next Generation EU. MYE and DB acknowledge support from the COSMOS network (www.cosmosnet.it) through the ASI (Italian Space Agency) Grants 2016-24-H.0, 2016-24-H.1-2018 and 2020-9-HH.0. CP is grateful
to SISSA, the University of Trieste, and IFPU, where part of this work was carried out, for
hospitality and support. We use the healpy software package \citep[available at \url{http://healpix.sourceforge.net},][]{Gorski+05,Zonca2019} to produce the Mollweide projection plots.

\appendix

\section{FOTO multipoles}
\label{APP:FOO_MULTI}
In this section, we derive the FOTO signal  for all multipoles of the observed power spectrum, assuming a full-sky survey. %
To that end, we compute %
\begin{equation}
    \label{eq:Yam_dipole}
    P_{\ell,{\rm dip}}(k) = \frac{2\ell +1}{N}\iiint A_1 A_2 \left({\bs{v}_{\rm o}\over H_0}\cdot \bshat{r}_1\right)\left({\bs{v}_{\rm o}\over H_0}\cdot \bshat{r}_2\right) \mathrm{e}^{i\bs{k} \cdot (\bs{r}_1  - \bs{r}_2)}\mc{L}_\ell(\bshat{k}\cdot \bshat{r}_2)\,\dif^3 r_1 \,\dif^3 r_2 \, \frac{\dif^2\Omega_k}{4\pi}\,,
\end{equation}
where 
\begin{equation}
    A_i = \frac{\bar{n}(r_i)\,\relalpha(r_i)\,}{r_i\, a(r_i)\,H(r_i)/H_0}\,,
\end{equation}
and    $N = \int \bar{n}^2(r)\,\dif ^3\,r$.
We split the integrals into two parts
\begin{align} 
    P_{\ell,{\rm dip}}(k) = \frac{2\ell +1}{N}\int&\left[ \int A_1  \left({\bs{v}_{\rm o}\over H_0}\cdot \bshat{r}_1\right)  \mathrm{e}^{i\bs{k} \cdot  \bs{r}_1}\,\dif^3 r_1 \right]\times\nonumber\\
    &\left[ \int A_2  \left({\bs{v}_{\rm o}\over H_0}\cdot \bshat{r}_2\right)  \mathrm{e}^{-i\bs{k} \cdot  \bs{r}_2}\mc{L}_\ell(\bshat{k}\cdot \bshat{r}_2)\,\dif^3 r_2\right]\frac{\dif^2\Omega_k}{4\pi}\,.
    \label{eq:Yam_dip}
\end{align}
In evaluating the previous integrals, we  use the following identities 
\begin{align}
\label{eq:expand_exp}&\mathrm{e}^{i\bs{k} \cdot  \bs{r}} = \sum_{\ell} (2{\ell} + 1)\,i^{\ell}\, j_{\ell}(k r)\, \mc{L}_{\ell}(\bshat{k}\cdot \bshat{r})\,,\\\label{eq:ortho_legend}
    &\int \legpol{\ell}{q}{r}\,\legpol{{\ell'}}{k}{r}\,\dif ^2\Omega_r = \frac{4\pi}{2{\ell'} +1} \legpol{{\ell'}}{q}{k} \,\delta^{\mathrm{K}}_{\ell{\ell'}}\,, \\
    &\legpol{\ell}{q}{r}\legpol{{\ell'}}{q}{r} = \sum_{n=\ell-{\ell'}}^{\ell+{\ell'}}\tj{\ell}{{\ell'}}{n}{0}{0}{0} (2n+1)\legpol{n}{q}{r}\label{eq:Legendmultiply}\,,
\end{align}
where $\delta^{\mathrm{K}}_{\ell{\ell'}}$ is the Kronecker delta.

For the first square bracket in Eq.~\eqref{eq:Yam_dip}, we expand the exponential and split the integral into radial and angular parts
\begin{align}
    \int A_1  \left({\bs{v}_{\rm o}\over H_0}\cdot \bshat{r}\right)  \mathrm{e}^{i\bs{k} \cdot  \bs{r}}&\,\dif ^3 r = \sum_{\ell'=0}^\infty (2{\ell'}+1)i^{\ell'} \iint  A  \left({\bs{v}_{\rm o}\over H_0}\cdot \bshat{r}\right)j_{\ell'}(k r) \mc{L}_{\ell'}(\bshat{k}\cdot \bshat{r})\,\dif ^2 \Omega_{r}  \,\dif  r\,,\nonumber\\
    &={{v}_{\rm o}\over H_0} \,\sum_{\ell'=0}^\infty (2{\ell'}+1)i^{\ell'} I_{\ell'} \int \legpolvo{1}{r}\legpol{{\ell'}}{k}{r} \,\dif ^2 \Omega_{r}\,,\nonumber\\
    &={{v}_{\rm o}\over H_0} \sum_{\ell'=0}^\infty (2{\ell'}+1)i^{\ell'} I_{\ell'} \frac{4\pi}{2{\ell'} +1}  \legpolvo{{\ell'}}{k} \delta^{\mathrm{K}}_{{\ell'}1}\,,\nonumber\\
    &=4\pi i\,{{v}_{\rm o}\over H_0}  \,   I_1\, \legpolvo{1}{k}\,.
\end{align}
For the second square bracket in Eq.~\eqref{eq:Yam_dip}, we find
\begin{align}
     \int A_2  &\left({\bs{v}_{\rm o}\over H_0}\cdot \bshat{r}\right)  \mathrm{e}^{-i\bs{k} \cdot  \bs{r}}\mc{L}_\ell(\bshat{k}\cdot \bshat{r})\,\dif ^3 r =\nonumber\\ &\sum_{\ell'=0}^\infty(2{\ell'}+1)\,(-i)^{\ell'} \iint A \,\left({\bs{v}_{\rm o}\over H_0}\cdot \bshat{r}\right)\,j_{\ell'}(k r) \mc{L}_{\ell'}(\bshat{k}\cdot \bshat{r})\,\legpol{l}{k}{r}\,\dif ^2 \Omega_{r}  r^2\,\dif  r\,,\nonumber\\
     =&{{v}_{\rm o}\over H_0} \,\sum_{\ell'=0}^\infty(2{\ell'}+1)(-i)^{\ell'} I_{\ell'} \int   (\bshat{v}_o\cdot \bshat{r})\sum_{n=\ell-{\ell'}}^{\ell+{\ell'}}\tj{\ell}{{\ell'}}{n}{0}{0}{0} (2n+1)\legpol{n}{k}{r} \,\dif ^2 \Omega_{r} \nonumber\\
     =&{{v}_{\rm o}\over H_0} \,\sum_{\ell'=0}^\infty\sum_{n=\ell-{\ell'}}^{\ell+{\ell'}}\tj{\ell}{{\ell'}}{n}{0}{0}{0}  (2{\ell'}+1)(2n+1)(-i)^{\ell'} I_{\ell'} \int   \legpolvo{1}{r}\legpol{n}{k}{r} \,\dif ^2 \Omega_{r}\,, \nonumber\\
     =&{{v}_{\rm o}\over H_0} \,\sum_{\ell'=0}^\infty\sum_{n=\ell-{\ell'}}^{\ell+{\ell'}}\tj{\ell}{{\ell'}}{n}{0}{0}{0}  (2{\ell'}+1)(2n+1)(-i)^{\ell'} I_{\ell'}\left[\frac{4\pi}{2n +1}  \legpolvo{n}{k} \delta^{\mathrm{K}}_{n1}\right]\,, \nonumber\\
     =&4\pi\,{{v}_{\rm o}\over H_0}\sum_{\ell'=0}^\infty\tj{\ell}{{\ell'}}{1}{0}{0}{0}  (2{\ell'}+1)(-i)^{\ell'} I_{\ell'}  \legpolvo{1}{k} \,. 
\end{align}
Plugging in both results in Eq.~\eqref{eq:Yam_dip},
we finally obtain
\begin{align}
    \label{eq:Yam_dip_result}
    &P_{\ell,{\rm dip}}(k)=& \nonumber\\\nonumber&\frac{2\ell +1}{N}\,{{v}^2_{\rm o}\over H^2_0} \int\left[ 4\pi i   I_1 \legpolvo{1}{k} \right]\left[ 4\pi\sum_{\ell'=0}^\infty\tj{\ell}{{\ell'}}{1}{0}{0}{0}  (2{\ell'}+1)(-i)^{\ell'} I_{\ell'}  \legpolvo{1}{k}\right]\frac{\dif^2\Omega_k}{4\pi}\\
    &=\frac{16\pi^2(2\ell +1)}{3}\,{v^2_{\rm o}\over H_0^2}\, \frac{I_1} {N}\sum_{\ell'=0}^\infty\tj{\ell}{{\ell'}}{1}{0}{0}{0}  (2{\ell'}+1)(-1)^{\ell'}(i)^{{\ell'}+1} I_{\ell'}   \,.
\end{align}
For first few multipoles, this reduces to
\begin{align}
&P_{0,{\rm dip}}(k) = \frac{16\pi^2}{3} {v^2_{\rm o}\over H_0^2} \,\frac{I_1^2(k)}{N}\,,\\
&P_{1,{\rm dip}}(k) = \frac{16\pi^2 i }{3 } {v^2_{\rm o}\over H_0^2} \,\frac{ I_1(k)\left[I_0(k) - 2I_2(k) \right]}{N}\,,\\
&P_{2,{\rm dip}}(k) = -\frac{16\pi^2 }{5 } {v^2_{\rm o}\over H_0^2} \, \frac{I_1(k)\left[2I_1(k) + 3I_3(k) \right]}{N}\,,\\
&P_{3,{\rm dip}}(k) = \frac{16\pi^2 i }{7 } {v^2_{\rm o}\over H_0^2} \, \frac{I_1(k)\left[-3I_2(k) + 4I_4(k) \right]}{N}\,,\\
&P_{4,{\rm dip}}(k) = -\frac{16\pi^2  }{9 } {v^2_{\rm o}\over H_0^2} \, \frac{I_1(k)\left[4I_3(k) + 5I_5(k) \right]}{N}\,.
\end{align}

\begin{figure*}[]

    \centering
    \begin{subfigure}[b]{0.45\textwidth}
        \centering
    \includegraphics[width=1\textwidth]{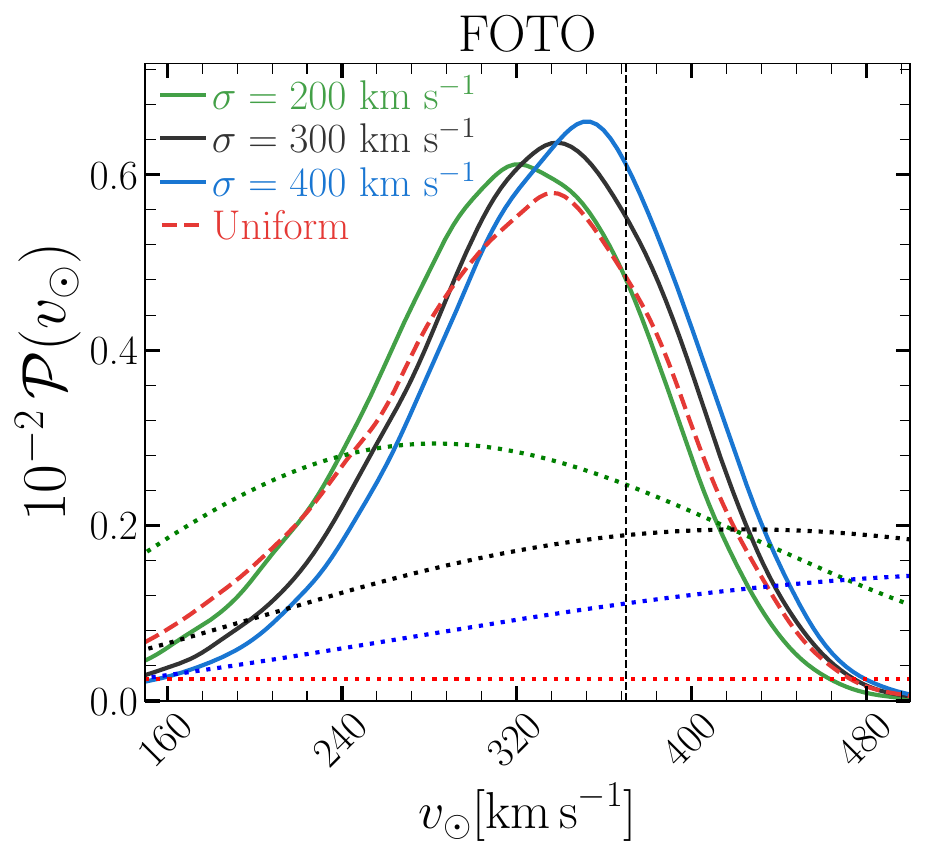}%
    \end{subfigure}%
    ~ 
    \begin{subfigure}[b]{0.45\textwidth}
        \centering
    \includegraphics[width=1\textwidth]{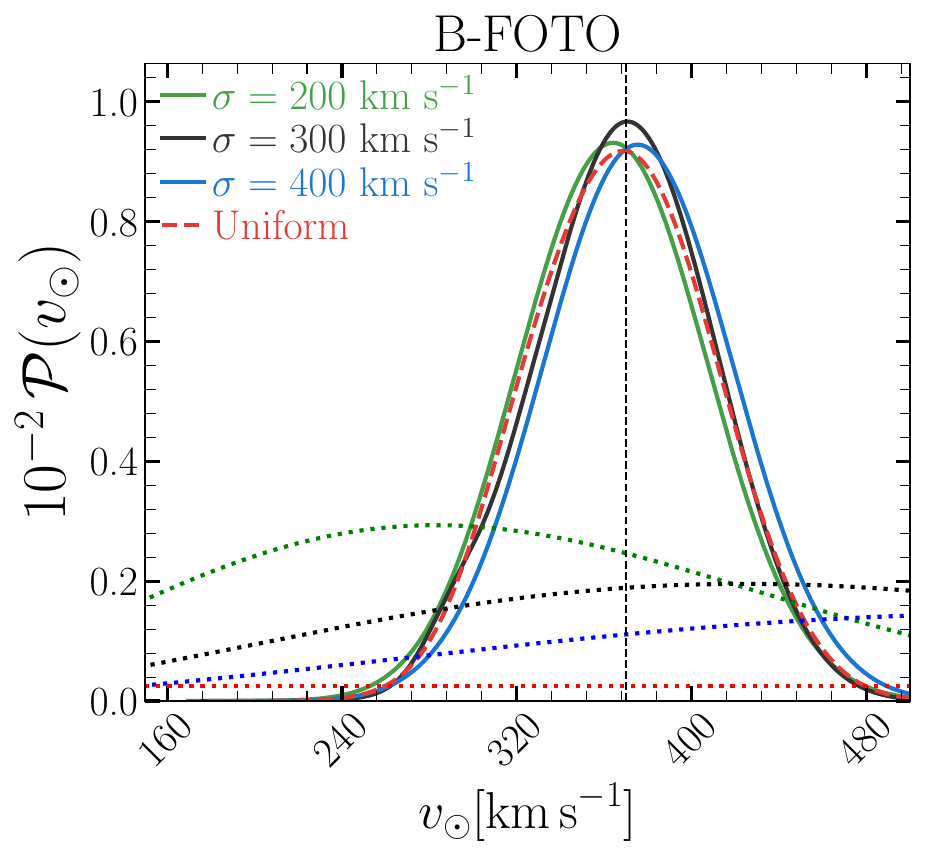}%
    \end{subfigure}%
    \caption{\label{fig:CORNER_PRIOR} {The posterior distributions of $v_\odot$ are shown for different prior assumptions in the velocity inference using the FOTO (left panel) and B-FOTO (right panel) signals. The dotted lines represent the Maxwellian prior distributions corresponding to each posterior of the same colour. The dashed red curves show the inferred posteriors using a uniform prior distribution.}}
\end{figure*}

\section{Velocity prior}
\label{sec:Vel_prior}
{In the velocity inference, we used Eq.~\eqref{eq:prior} with $\sigma = 300$ \kms as the  prior distribution for $v_\odot$. To explore the impact of that assumption, we infer the  posterior distributions for the FOTO (see Fig.~\ref{fig:Corner_V_NO_SHIFT}) and B-FOTO (see Fig.~\ref{fig:POST_ONE}) signals using three additional prior distributions, which are shown in Fig.~\ref{fig:CORNER_PRIOR}. The first two also employ a Maxwellian distribution, but with different $\sigma$ values of 200 and 400 \kms, resulting in the green and blue curves, respectively. Lastly, we use a uniform prior over the range \(v_\odot \in [0, 4000]\) \(\text{km\,s}^{-1}\), which results in the red-dashed curves. 
The  resulting distributions exhibit slight variations in the FOTO case, whereas the impact on the B-FOTO constraints is minimal, highlighting that the posterior is largely driven by the likelihood.}

\section{Alcock--Paczyński distortions}
\label{Sec:AP}
In the analysis of the mock catalogues, we have used the cosmological
parameters of the numerical simulations to transform
galaxy redshifts into distances and then estimate
the power spectrum.
In actual surveys, however, we do not know the
underlying cosmological parameters and some fiducial
values are used to convert redshifts into distances. This leads to the Alcock--Paczyński  \citep[AP,][]{Alcock_Paczynski+79} effect which creates anisotropic distortions in the galaxy density field.
The AP distortions affect the orthogonal and parallel modes of the power spectrum differently. We can relate the  elements $\{k_{\parallel}\, , k_{\perp}\}$ of the true wavevector $\bs{k}$  to the elements $\{q_{\parallel}\, , q_{\perp}\}$  of the measured wavevector  $\bs{q}$ through,
\begin{align}
    k_{\parallel} = {{H^{\rm T}\over H^{\rm F}}}\,\,q_{\parallel}\equiv {q_{\parallel}\over a_{\parallel}}\,,\hspace{1cm}
    k_{\perp} = {{d_{\rm A}^{\rm F}\over d_{\rm A
}^{\rm T}}}\,\,q_{\perp}\equiv {q_{\perp}\over a_{\perp}}\,,
\end{align}
where $ d_{\rm A}$ is the angular diameter distance and the subscripts T and F denote the true and fiducial quantities, respectively. 
The distortion can then be translated to the wavevector amplitude $k$ as follows (noting that $q_\parallel \equiv (\bshat{q}\cdot \bshat{r}) \,q\equiv  \mu_{q}\, q$) %
\begin{equation}
k(q,\mu_q)     ={q\over a_{\perp}}\sqrt{1- {(a^2_{\parallel}-a^2_{\perp})\over a_{\parallel}^2}\,\mu_q^2}\,.
\end{equation}
\begin{figure}
    \centering
    \includegraphics[width=0.7\textwidth]{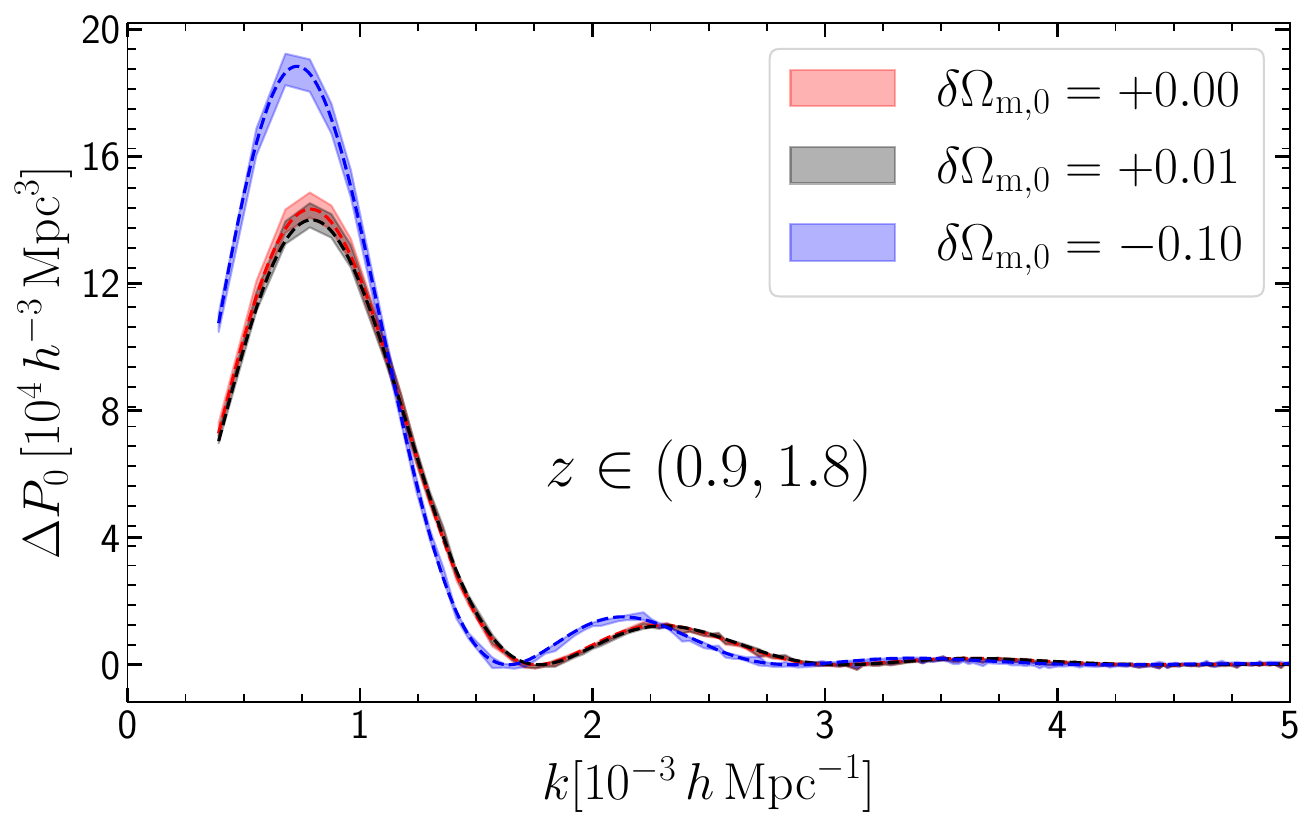}
    \caption{AP distortions on the FOTO signal from the \peuc survey. The expression in Eq.~(\ref{eq:AP_mt_DIP}, dashed lines) is in excellent agreement with the measurements from the mocks (shaded regions). The quantity $\delta \Omega_{\rm m,0}$ indicates the difference between the fiducial and true present-day values of the matter density parameter in a flat $\Lambda$CDM model.%
    }
    \label{fig:AP_t}
\end{figure}
The impact of the AP effect on the measured power spectrum multipoles $\widehat{P}_{\ell}(q)$ can then be written as \citep[e.g.,][]{Ballinger_1996,Beutler+14}
\begin{equation}
    \label{eq:AP_mt_multi}
    \widehat{P}_{\ell}(q) ={(2\ell + 1)\over \,a^2_\perp\,a_{\parallel}} \int P_{\rm obs}(\bs{k})\mc{L}_{\ell}(\mu_q){\dif^2 \Omega_{r}\over 4\pi}\,,
\end{equation}
where $P_{\rm obs}(\bs{k})$ is the measured power spectrum assuming the true cosmological parameters.
Therefore, for the FOTO signal, we have
\begin{equation}
    \label{eq:AP_mt}
    \widehat{P}_{0,\mathrm{dip}}(q) ={1\over \,a^2_\perp\,a_{\parallel}} \int P_{0,\mathrm{dip}}[k(q,\mu_q)]{\dif^2 \Omega_{r}\over 4\pi}\,,
\end{equation}
and, 
assuming that ${(a^2_{\parallel}-a^2_{\perp})\ll a_{\parallel}^2}$, we can write %
\begin{equation}
    \label{eq:AP_mt_DIP}
     \widehat{P}_{0,\mathrm{dip}}(q) \approx {1\over \,a^2_\perp\,a_{\parallel}}  P_{0,\mathrm{dip}}\left(\frac{q}{a_{\perp}}\right)\,.
\end{equation}
To evaluate the accuracy of Eq.~\eqref{eq:AP_mt_DIP}, we measure the monopole of the power spectra from our mocks using three distinct sets of fiducial cosmological parameters. %
 \begin{figure}
    \centering
    \includegraphics[width=0.5\textwidth]{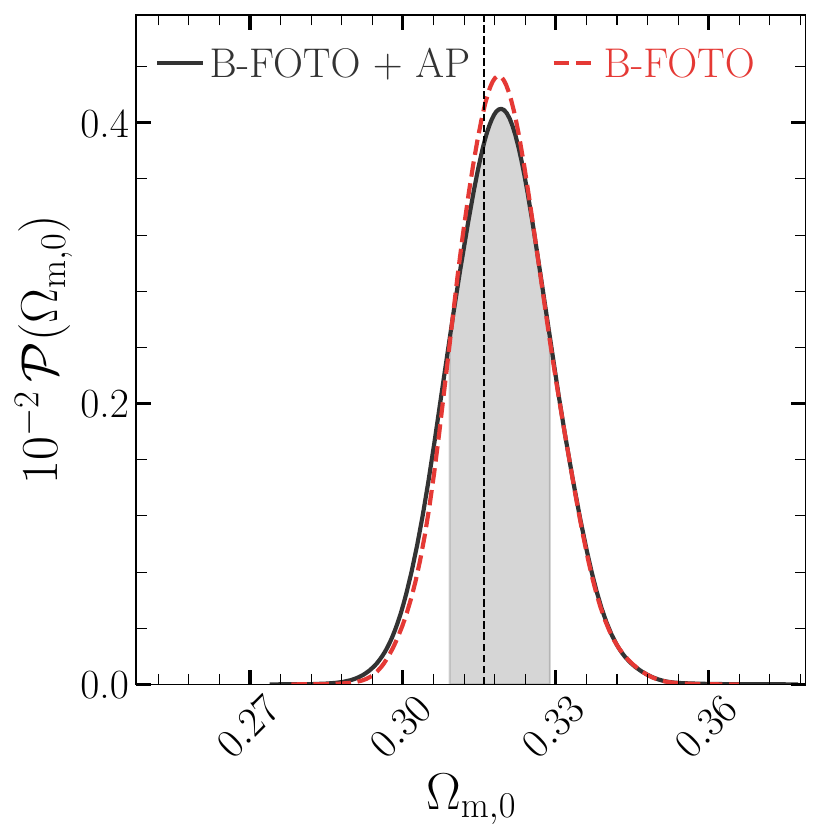}%
    \caption{ As in Fig.~\ref{fig:Corner_m} but including AP distortions (black). The posterior
    distribution shows only minimal changes with respect to Fig.~\ref{fig:Corner_m} (red). }
    \label{fig:Corner_m_AP}
\end{figure}
Fig.~\ref{fig:AP_t}
shows that the AP corrections alter 
the frequency and the amplitude of the FOTO signal. However, in all cases,
we find remarkable agreement between the theoretical prediction and the mock results. 
Moreover, we repeat the density inference by  varying the true cosmology in Eq.~\eqref{eq:AP_mt_DIP} %
and find the same posteriors shown in the main text where we ignored the AP corrections as shown in Fig~\ref{fig:Corner_m_AP}. %
\section{Derivation of the \RBFS signal.}
\label{app:alpha_deriv}
In this section, we 
compute the overdensity of the galaxies whose
redshifts have been corrected using
Eq.~\eqref{eq:redshift_boost}.
The correction induces the  radial shift 
\begin{align}
\delta r = c\,\int_{z_\mathrm{hel}}^{z_\mathrm{art}}
\frac{\mathrm{d}z}{H(z)}&\simeq 
c\,\frac{z_\mathrm{art}-z_\mathrm{hel}}{H(z_\mathrm{hel})}\,=(1+z_\mathrm{art})\,\frac{-v_{\parallel,\mathrm{art}}}{H(z_\mathrm{art})}\;.
\end{align}
Consequently, a galaxy at the 
redshift-space position $\bs{x}$ in the heliocentric rest frame
is moved to the `boosted'  position $\bs{r}$,  
such that
\begin{equation}\bs{x} = \left({r} +{v_{\parallel,\mathrm{art}}\over a\,H}\, \right)\bshat{r} \equiv \bs{r} - \delta {r}\,\bshat{r}\label{eq:app_rad_shift}\,,
\end{equation}
where $\bshat{r} = \bshat{x}$.

Requiring that the number of galaxies is conserved gives
\begin{align}
    n_{\rm corr}(\bs{r})\;\dif^3 r &=n_{\rm obs}(\bs{x})\;\dif^3 x \,,\\
    \bar{n}_{\rm corr}({r})[1+\delta_{\rm corr}(\bs{r})]\;\dif^3 r &=\bar{n}_{\rm obs}(x)[1+\delta_{\rm obs}(\bs{x})]\;\dif^3 x \,,\\
    \delta_{\rm corr}(\bs{r}) & = \frac{\bar{n}_{\rm obs}({x})}{\bar{n}_{\rm corr}({r})}[1+\delta_{\rm obs}(\bs{x})] \frac{\dif^3 x}{\dif^3 r} -1\,.
    \label{eq:coserv_append}
    \end{align}
    The last expression shows that
we can break down the derivation into three steps: \textit{i)} rewriting  $\delta_{\rm obs}(\bs{x})$ in the `boosted' coordinate $\bs{r}$, \textit{ii)} the change in the comoving background density, and \textit{iii)} the Jacobian of the coordinate transformation. 

The first ingredient is trivially $\delta_{\rm obs}(\bs{x}) = \delta_{\rm obs}(\bs{r})$ at linear order. As for the second piece, we first define the  comoving background density through the luminosity function $\phi(x,L)$ (assuming threshold luminosity $L_{\rm min}$) as
\begin{equation}
    \bar{n}(x)\equiv n(L_{\rm min}(x),x) = \int_{L_{\rm min}(x)}^{\infty} \phi(x,L) \dif L\,.
\end{equation}
We then compute calculate the impact of the transformation as follows
\begin{align}
    \bar{n}(x) &= \bar{n}(r-\delta r) = \int_{L_{\rm min}({r}-\delta{r})}^{\infty} \phi({r}-\delta{r},L) \,\dif L\,,\nonumber\\
     &=\textcolor{black}{ \int_{L_{\rm min}({r}) }^{\infty} \left[\phi(r,L) - \delta r \frac{\partial \phi(r,L)}{\partial r}\right ] \dif L} \textcolor{black}{-\frac{\dif L_{\rm min}}{\dif r} \delta{ r}\left[\phi(r,L_{\rm min}) - \delta {r} \frac{\partial \phi(r,L_{\rm min})}{\partial r}\right ]}   \,\nonumber\\
    &=\textcolor{black}{\bar{n}({r}) - \delta {r} \frac{\partial\, \bar{n}(r)}{\partial r}}
    \,\textcolor{black}{-\phi(r,L_{\rm min})\frac{\dif L_{\rm min}}{\dif r} \delta{ r}}  \nonumber\\
    \label{eq:app_numb_den}&=\bar{n}(r)+ {{v_{\parallel,\mathrm{art}}\over a\,H\,r}}\frac{\dif\, \bar{n}(r)}{\dif\ln r}.
\end{align}

For the final ingredient, we can write the Jacobian as
\begin{align}
\label{eq:app_jacob}
        \left|\frac{\dif^3 x}{\dif^3 r}\right| &=\left[1 - \frac{\partial \delta r^i}{\partial r^i}\right] = \left[1 -\nabla_{\parallel}\delta\bs{r}_{\parallel} - \delta\bs{r}_{\parallel} (\bs{\nabla}\cdot\bshat{r}) - \bs{\nabla}_{\perp}\delta \bs{r}_\perp  \right]\,,
\end{align}
where  $\nabla$ as the gradient operator w.r.t. the $r$ coordinates and 
 \begin{align}
    \delta\bs{r}_{\parallel} &= \delta\bs{r}\cdot \bshat{r} \,\,\,,
    &\delta\bs{r}_{\perp} = \delta\bs{r} -(\delta \bs{r}\cdot \bshat{r}) \bshat{r}\,,\\
    \nabla_{\parallel} &= \bshat{r}\cdot \bs{\nabla} \,\,\,,
    &\bs{\nabla}_{\perp} = \bs{\nabla} -\bshat{r}(\bshat{r}\cdot \bs{\nabla})\,.
\end{align}
The last term on the RHS of the Jacobian vanishes by construction as the shift is only along the line of sight. The third term is given by $\delta\bs{r}_{\parallel} (\bs{\nabla}\cdot\bshat{r}) =- 2{v_{\parallel,\mathrm{art}}/(r\,a\,H)} $.  As for the first term, it is given by
\begin{align}
    \nabla_{\parallel}\delta\bs{r}_{\parallel} = \frac{\dif z}{\dif r}{\partial(\delta\bs{r}_{\parallel}) \over\partial z} 
    ={-a\,H r  \over c }\left[1-  {\dif \ln H \over\dif \ln(1+ z)} \right]{v_{\parallel,\mathrm{art}}\over a \,H\, r}\,.
\end{align} 
Putting all terms together, the Jacobian can be written as 
\begin{equation}
\label{eq:app_jacob2}
    \left|\frac{\dif^3 x}{\dif^3 r}\right|  = 1 +  {v_{\parallel,\mathrm{art}}\over a\,H\, r} \left\{2 +{a H r  \over c }\left[1-  {\dif \ln H \over\dif \ln(1+ z)} \right]\right\}\,.
\end{equation}

Plugging in Eqs.~\eqref{eq:app_numb_den} and~\eqref{eq:app_jacob2} into Eq.~\eqref{eq:coserv_append}, gives%
    \begin{align}
    \delta_{\rm corr}(\bs{r}) &=[1+\delta_{\rm obs}(\bs{r})] \left[1 +{ v_{\parallel,\mathrm{art}}\over a\, H}\frac{\dif\, \ln\bar{n}_{\rm obs}(\bs{r})}{\dif r}  \right] \text{\LARGE $\times$ }\nonumber \\
    &\hspace{2cm}\left\{1+\left[2+{a H r  \over c }\left(1-   {\dif \ln H \over\dif \ln(1+ z)} \right)\right]{v_{\parallel,\mathrm{art}}\over a\,H\, r}\right\} -1\,\nonumber\\
         &=  \delta_{\rm obs}(\bs{r})  + {v_{\parallel,\mathrm{art}}\over a\, H}\frac{\dif\, \ln\bar{n}_{\rm obs}(\bs{r})}{\dif r} +\left\{2+\textcolor{black}{{ a H r  \over c}\left[1- {\dif \ln H \over\dif \ln(1+ z)} \right]}\right\}{v_{\parallel,\mathrm{art}}\over a\, H\,r} -1\,\nonumber\\
        &=  \delta_{\rm obs}(\bs{r})+{v_{\parallel,\mathrm{art}}\over a\, H\, r}\left\{\frac{\dif\, \ln\bar{n}_{\rm obs}(\bs{r})}{\dif\ln r}  +\, \frac{\dif\, \ln r^2}{\dif\ln r}+{a H r  \over c}\left[1- {\dif \ln H \over\dif \ln(1+ z)} \right]\right\}\,\nonumber\\
        &\equiv   \delta_{\rm obs}(\bs{r})+{v_{\parallel,\mathrm{art}}\over a\,H\,r}\alpha_c\,.
\end{align}
Finally, we obtain  Eq.~\eqref{eq:corrected_overdns} by setting $\bv{\mathrm {art}} = -\bv{\odot}$.
\subsection{Correcting relativistic aberration}
\label{app:aberration}
    In this subsection, we derive the impact of additionally  correcting for the relativistic aberration \footnote{This correction is not employed in the main text, as the B-FOTO signal is obtained with  only the radial transformation. }  using 
\begin{equation}
\bshat{r}\simeq \bshat{x} + 
{1\over c}\left[\bs{v}_\mathrm{\mathrm{art}} - {v}_{\parallel,\mathrm{art}}\,\bshat{x}  \right]\,\equiv
\bshat{x} + \frac{\bv{\perp,{\rm \mathrm{art}}}}{c} = \bshat{x} + {\delta \bs{r}_{\perp}\over r}.
\end{equation}
The coordinate map then becomes
\begin{equation}
    \bs{x} =\bs{r} +{\bs{v}_{\parallel,\mathrm{art}}\over a\,H}\, \bshat{r} -{r\,\bv{\perp,{\rm \mathrm{art}}} \over c}\equiv\bs{r} -\delta r_{\parallel}\, \bshat{r} -\delta \bs{r}_{\perp} \,. 
\end{equation}
The angular correction does not impact the average comoving number density -- see Eq.~\eqref{eq:app_numb_den} -- however, it impacts the Jacobian. In that case, the orthogonal derivative $\bs{\nabla}_{\perp}\delta \bs{r}_\perp$ no longer vanishes, it results in 
   $\bs{\nabla}_{\perp}\delta \bs{r}_\perp  %
   =-2  (v_{\parallel,\mathrm{art}}/c) \,.$
The Jacobian is then given by 
\begin{align}
\label{eq:app_jacob_app}
        \left|\frac{\dif^3 x}{\dif^3 r}\right| & = \left[1 -\nabla_{\parallel}\delta\bs{r}_{\parallel} - \delta\bs{r}_{\parallel} (\bs{\nabla}\cdot\bshat{r}) - \bs{\nabla}_{\perp}\delta \bs{r}_\perp  \right]\,\\
        &=    1 +  {v_{\parallel,\mathrm{art}}\over a\,H\, r} \left\{2 +{a H r  \over c}\left[1-  {\dif \ln H \over\dif \ln(1+ z)} \right] \right\}+ {2\over c} {v}_{\parallel,\mathrm{art}}\\ 
        &=1 +  {v_{\parallel,\mathrm{art}}\over a\,H\, r} \left\{2 +{a H r  \over c}\left[3-  {\dif \ln H \over\dif \ln(1+ z)} \right] \right\}\,.
\end{align}
Consequently, the final  density contrast is then changed to 
\begin{align}
        \delta_{\rm aber}(\bs{r}) &= 
          \delta_{\rm obs}(\bs{r})+{v_{\parallel,\mathrm{art}}\over a \,H\, r}\left\{\frac{\dif\, \ln\,r^2\,\bar{n}_{\rm obs}(\bs{r})}{\dif\ln r} + {a H r  \over c}\left[3- {\dif \ln H \over\dif \ln(1+ z)} \right]\right\}\,.
\end{align}
From the previous relation, it is straightforward to see that correcting for both the radial shift and the relativistic aberration (by  setting $\bs{v}_{\rm art}=-\bv{\odot}$) results in 

 \begin{equation}
 \delta_\mathrm{aber} -   \delta_{\mathrm {com}}=
 \,\frac{2\mc{Q}(\bv{\odot}  \cdot \hat{\bs{r}})}{c}\;,
\,
 \end{equation}
which does not completely cancel the FOTO signal.

\newpage
\appendix
 \clearpage
\bibliographystyle{citation_style}
\bibliography{bibs}
\end{document}